%% file: paper.tex
\documentclass[submission, Phys]{SciPost}

\input{incl_settings.tex}

\input{incl_shortcuts.tex}

\graphicspath{{./figures/}}

\usepackage{glossaries}
\usepackage{cleveref}
\usepackage{placeins}

\newacronym{LHC}{LHC}{Large Hadron Collider}
\newacronym{HL-LHC}{HL-LHC}{High Luminosity LHC}
\newacronym{HEP}{HEP}{High Energy Physics}
\newacronym{MC}{MC}{Monte Carlo}
\newacronym{ML}{ML}{Machine Learning}

\begin{document}

\begin{center}{\Large \textbf{
Sequential SBI for Resonance Searches at the LHC}}\end{center}

\begin{center}
Henning Bahl\textsuperscript{$\dagger$}\footnotetext{$^\dagger$ Former address} and
Ranit Das
\end{center}

\begin{center}
Institut für Theoretische Physik, Universit\"at Heidelberg, Germany 
\end{center}

\begin{center}
\today
\end{center}

% For convenience during refereeing: line numbers
%\linenumbers

%%%%%%%%%%%%%%%%%%%%%%%%%%%%%%%%%%%%%%%%%%%%%%%%%%
\section*{Abstract}
{\bf
Neural simulation-based inference has been shown to extract significantly more information from LHC data than histogram-based analyses. So far, it has mostly been applied to couplings, for which the polynomial structure of the scattering amplitude can be exploited. Masses and widths instead move and deform the signal in phase space, and learning their likelihood requires many event samples covering the continuous theory parameter space. We show how sequential simulation-based inference can address this by iteratively concentrating the simulations where the observed data points.
Using a parton-level example with an exact reference likelihood and reconstruction-level analyses of $A\to ZH\to \ell\ell b\bar b$ and $Y\to HH\to (t\bar t)(t\bar t)$, we demonstrate that sequential training produces significantly more accurate and better calibrated likelihoods than single round training with the same amount of training events.}

\vspace{10pt}
\noindent\rule{\textwidth}{1pt}
\tableofcontents\thispagestyle{fancy}
\noindent\rule{\textwidth}{1pt}
\vspace{10pt}

\clearpage
%%%%%%%%%%%%%%%%%%%%%%%%%%%%%%%%%%%%%%%%%%%%%%%%%%
\section{Introduction}

One of the central challenges in particle physics is to exploit the full potential of the current and future LHC datasets. LHC events are high-dimensional, but traditional inference methods compress their information into few-dimensional histograms of summary statistics. This reduction of dimensionality and the binning can result in a loss of information. Modern machine learning methods can overcome this limitation~\cite{Plehn:2022ftl,Feickert:2021ajf,Butter:2022rso} by learning unbinned likelihood ratios~\cite{Cranmer:2015bka,Cranmer:2019eaq,Rizvi:2023mws} using the full high-dimensional information.

ML-based matrix-element method~\cite{Martini:2020rnb,Butter:2022vkj,Heimel:2023mvw} and neural simulation-based inference (SBI)~\cite{Brehmer:2020cvb,Brehmer:2018eca,Brehmer:2018kdj,Brehmer:2018hga,Chatterjee:2021nms,Chatterjee:2022oco,Kong:2022rnd,Schofbeck:2024zjo,Bahl:2024meb} has been successfully used to estimate likelihood ratios for statistical inference.. It was first implemented for LHC purposes in the \textsc{MadMiner} tool~\cite{Brehmer:2019xox}. The advantages of SBI for LHC processes have been demonstrated not only in a variety of phenomenological studies~\cite{Brehmer:2019gmn,Barman:2021yfh,Bahl:2021dnc,GomezAmbrosio:2022mpm,Barrue:2023ysk,Mastandrea:2024irf,Ghosh:2025fma,Silva:2025hzo,Benato:2025rgo,Bahl:2025mib}, but recently also in the first experimental SBI analysis at the LHC~\cite{ATLAS:2024jry,ATLAS:2025clx}. Most of these studies target couplings, for instance Wilson coefficients in the Standard Model Effective Field Theory (SMEFT)~\cite{Degrande:2012wf,Brivio:2017vri}. For couplings, the squared matrix element is a polynomial in the theory parameters. The likelihood ratio at any parameter point can then be reconstructed from a small set of learned basis ratios or from its derivatives, and the continuous parameter dependence does not have to be learned from simulations~\cite{Brehmer:2018eca,Chatterjee:2021nms,Chatterjee:2022oco,Chen:2023ind,Bahl:2024meb}.

Masses and widths are qualitatively different. They enter through propagators, so varying them moves and deforms resonant structures (or the effect of higher-order corrections) in phase space, and no finite polynomial basis exists. Searches for new resonances and the investigation of known particles are, at the same time, among the most important tasks of the LHC program, and their likelihoods are often high-dimensional in the parameters as well. Already a single cascade decay comes with two masses, two widths, and a rate. Per-event likelihoods have of course long been used to measure masses, in the classical matrix-element method applied to the top mass~\cite{D0:1998eiz,CDF:2006nne,Fiedler:2010sg}, but always at a fixed and fully specified final state. Applications of learned likelihood ratios to resonances have so far been rare~\cite{Hollingsworth:2020kjg, Cheng:2024yig}. A parameter-conditional likelihood-ratio estimator has to learn the continuous mass and width dependence from simulations spread over the full parameter space. Covering this space uniformly quickly becomes prohibitively expensive as its dimensionality and the required resolution increase, while most of it is irrelevant once the data are observed.

This problem is well known beyond particle physics, where sequential SBI has been used to address it by guiding a new round of simulations with the inference result from the previous round~\cite{Gutmann:2016bolfi,papamakarios2016fast,Lueckmann:2017snpe,Papamakarios:2018zoy,Greenberg:2019apt,Hermans:2019ioj,Durkan:2020snre,Miller:2020hua,Miller:2021hys,Deistler:2022tsnpe,Lyu:2025vqk}.  Sequential SBI has been benchmarked in Ref.~\cite{Lueckmann:2021sbibm} and applied to cosmology~\cite{Cole:2021gwr}, strong lensing~\cite{Montel:2022fhv}, and gravitational-wave physics~\cite{Bhardwaj:2023xph}. Ref.~\cite{Lyu:2025vqk} drops the discrete rounds altogether and grows the training set dynamically. These methods are almost exclusively Bayesian and target the posterior.  We adapt the idea to classifier-based likelihood-ratio estimation. In each round, we evaluate the learned likelihood on the observed data, use it to construct a proposal for the next set of simulated parameter points, and retrain on the accumulated simulations. Since signal and background use parameters from the same proposal, it cancels in the learned likelihood ratio, and the final likelihood can be used for frequentist limit setting in the same way as SBI without sequential rounds~\cite{Bahl:2025mib}. Complementary approaches to high-dimensional inference are discussed in Refs.~\cite{Amram:2025vqw,Cheng:2025ewj,Caron:2019xkx}.

We evaluate sequential SBI on one parton-level benchmark and two reconstruction-level analyses. For the parton-level cascade $u\bar u\to Z'\to X\bar X\to 4q$, we calculate the exact likelihood ratio from matrix elements and use it as a reference for comparison with sequential SBI. At the reconstruction level, we infer three parameters for the heavy-scalar cascade $A\to ZH\to\ell\ell b\bar b$ and five parameters for the boosted topology $Y\to HH\to 4t$. Across all three studies, we compare sequential SBI with single-round training at equal simulation budgets. We also assess the calibration of the inferred confidence regions using coverage tests.

The paper is organized as follows. In \cref{sec:sequential_sbi}, we present the sequential SBI algorithm, the likelihood construction, and the limit-setting procedure. We then study the parton-level cascade in \cref{sec:dijet}, the reconstruction-level $A\to ZH$ analysis in \cref{sec:zhllbb}, and the boosted four-top channel in \cref{sec:ytohh4t}. We present our conclusions in \cref{sec:conclusions}. We provide additional results for the toy model in \cref{sec:cascade_app}, for $A\to ZH$ in \cref{sec:zhllbb_app}, and for the four-top channel in \cref{sec:ytohh}.

%%%%%%%%%%%%%%%%%%%%%%%%%%%%%%%%%%%%%%%%%%%%%%%%%%
\section{Method}
\label{sec:sequential_sbi}

\subsection{Simulation Based Inference}

Simulation-based inference (SBI) is a method to constrain theory parameters when a forward simulator can generate events $x$ at a parameter point $\theta$, but the likelihood $p(x|\theta)$ cannot be evaluated~\cite{Cranmer:2019eaq}. At the LHC, the simulation chain includes hard scattering, parton showering, hadronization, and detector simulation. Evaluating the likelihood of a reconstructed event would require marginalizing over the unobserved intermediate states of this chain, which is intractable. Neural SBI avoids this by training on simulated parameter--event pairs $(\theta,x)$ to approximate the likelihood, the posterior $p(\theta|x)$, or a likelihood ratio~\cite{Papamakarios:2018zoy,Greenberg:2019apt,Hermans:2019ioj,Durkan:2020snre}.

For frequentist inference, estimating the likelihood-ratio from simulated events allows us to do statistics and get confidence regions. The optimal classifier separating events drawn from two hypotheses is a known monotonic function of the ratio of their probability densities. A classifier can estimate this ratio without evaluating either density~\cite{Cranmer:2015bka,Brehmer:2018eca,Brehmer:2018kdj}. A single classifier can be conditioned on $\theta$, and it can estimate the likelihood ratio across parameter space using multidimensional event information without explicit binning.

The accuracy of the likelihood ratio can depend on how the finite simulation budget is distributed across parameter space. For a broad parameter range, much of the expensive simulation budget can be wasted generating events at parameter points far from the confidence region. Sequential SBI addresses it by using the observed data to guide successive simulation rounds, concentrating training events in the regions relevant to the inference.

%%%%%%%%%%%%%%
\subsection{Sequential SBI}
\label{sec:method}
\begin{figure}
    \centering
    \includegraphics[width=0.8\linewidth]{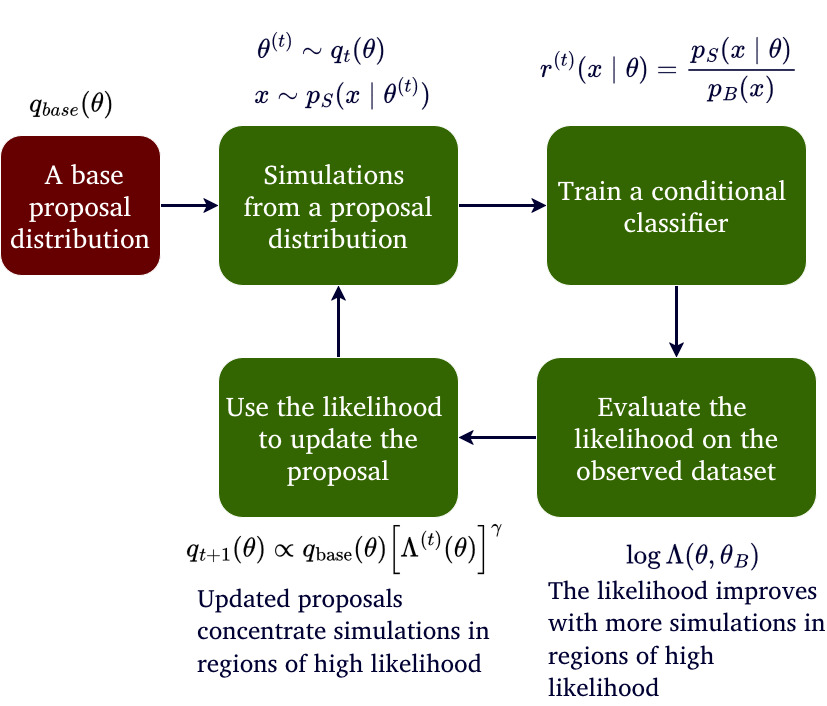}
    \caption{Overview of sequential SBI.}
    \label{fig:sequentialSBI_chart}
\end{figure}

We perform sequential simulation-based inference (SBI) over rounds
$t=1,2,\ldots$. In each round, we use the current likelihood estimate to
concentrate new MC simulations in parameter regions favored by the observed data.
This adaptively improves the likelihood-ratio estimate in the regions most
relevant for inference. \cref{fig:sequentialSBI_chart} illustrates this procedure.

\paragraph{Step 1: Define the proposal distribution.}
At round $t$, we draw model parameters from a proposal distribution
$q_t(\theta)$. For the first round, we choose
\begin{equation}
    q_1(\theta)=q_{\mathrm{base}}(\theta),
\end{equation}
where $q_{\mathrm{base}}(\theta)$ covers the overall parameter space of interest.
We typically take $q_{\mathrm{base}}(\theta)$ to be uniform over the chosen
parameter domain.

\paragraph{Step 2: Generate MC data.}
At round $t$, we sample parameter points
\begin{equation}
    \theta_i^{(t)} \sim q_t(\theta)
\end{equation}
and generate signal events according to
\begin{equation}
    x \sim p_S(x\mid\theta_i^{(t)}).
\end{equation}
Here, $p_S(x|\theta)$ is the normalized density of the observables $x$ for signal
events generated with the parameters $\theta$, after the full simulation chain and
the event selection. The
background density $p_B(x)$ is defined in the same way for the background process and does
not depend on the signal parameters $\theta$. We assign the background events random parameter values sampled from the same proposal
$q_t(\theta)$. The resulting signal and background training distributions are
\begin{align}
    p_{\mathrm{sig}}^{(t)}(x,\theta)
    &= q_t(\theta)\,p_S(x\mid\theta), \\
    p_{\mathrm{bkg}}^{(t)}(x,\theta)
    &= q_t(\theta)\,p_B(x).
    \label{eq:param_cond_llr}
\end{align}

\paragraph{Step 3: Learn the parameter-conditional likelihood}
We first train a parameter-conditional classifier $C^{(t)}(x,\theta)$ to
distinguish signal from background, and turn it into a likelihood ratio through the
standard calibrated-classifier construction~\cite{Cranmer:2015bka,Hermans:2019ioj,Rizvi:2023mws}.
For equal class priors, the optimal classifier converges towards
\begin{equation}
    r^{(t)}(x\mid\theta)
    =
    \frac{C^{(t)}(x,\theta)}
         {1-C^{(t)}(x,\theta)}
    =
    \frac{q_t(\theta)p_S(x\mid\theta)}
         {q_t(\theta)p_B(x)}
    =
    \frac{p_S(x\mid\theta)}
         {p_B(x)}.
\end{equation}
Using the same parameter proposal for both classes ensures that
$q_t(\theta)$ cancels in $r^{(t)}(x\mid\theta)$. The proposal distribution therefore determines where the classifier receives
training information, but does not modify the likelihood ratio being learned. The likelihood ratio can then be used to construct the likelihood $\Lambda^{(t)}(\theta)$ as described in \cref{sec:limits} (see~\cref{eq:full_likelihood_expected,eq:mcsplit}).

% After training, we evaluate the learned likelihood ratio on the fixed observed
% dataset
% \begin{equation}
%     \mathcal{D}_{\mathrm{obs}}
%     =
%     \{x_k\}_{k=1}^{N_{\mathrm{data}}}.
% \end{equation}
% Assuming statistically independent events, the overall log-likelihood
% ratio is 
% \begin{equation}
%     \log \Lambda^{(t)}(\theta)
%     =
%     \sum_{k=1}^{N_{\mathrm{data}}}
%     \log r^{(t)}(x_k\mid\theta).
% \label{eq:sequential_loglr_freq}
% \end{equation}

\paragraph{Step 4: Update the proposal and repeat.}
We use the likelihood estimate from round $t$ to construct the proposal for
the next round $t+1$, 
\begin{equation}
    q_{t+1}(\theta)
    \propto
    q_{\mathrm{base}}(\theta)
    \left[\Lambda^{(t)}(\theta)\right]^{\gamma},
\label{eq:sequential_proposal_freq}
\end{equation}
in the spirit of the sequential schemes of
Refs.~\cite{Gutmann:2016bolfi,papamakarios2016fast,Greenberg:2019apt,Durkan:2020snre,Miller:2021hys,Deistler:2022tsnpe,Lyu:2025vqk},
 where $0<\gamma\leq1$ controls the degree of concentration near high likelihood regions. Smaller values of $\gamma$ allow broader coverage of the parameter space, while larger values place more simulations in regions favored by the observed data.

We then sample new parameter points from $q_{t+1}(\theta)$, generate additional
simulations, and retrain the classifier using the simulations accumulated up to
the new round.
Because signal and background samples in each round use the same proposal,
their accumulated parameter distributions remain matched, and the proposal
dependence continues to cancel in the likelihood ratio.

Repeating these steps reallocates the simulation budget
toward regions of parameter space favored by the observed data. The classifier gains increasingly dense training coverage in the high-likelihood
region while avoiding unnecessary simulations in strongly disfavored regions.
The final likelihood-ratio estimate can then be used to determine
maximum-likelihood estimates, construct profile-likelihood test statistics,
and derive confidence regions.

\subsection{Likelihood construction}
\label{sec:limits}
The trained classifier provides an estimate of the parameter-conditional
signal-to-background likelihood ratio,
\begin{equation}
    r(x|\theta)
    =
    \frac{p_S(x|\theta)}{p_B(x)} \;,
\end{equation}
where $p_S(x|\theta)$ and $p_B(x)$ denote the normalized signal and background
densities, respectively. For expected signal and background yields
$S(\theta)$ and $B$, the normalized event density under the
signal-plus-background hypothesis is
\begin{equation}
    p(x|\theta)
    =
    \frac{
        S(\theta)\,p_S(x|\theta)
        +
        B\,p_B(x)
    }{
        S(\theta)+B
    } \;.
\end{equation}
Following Refs.~\cite{Bahl:2024meb,Bahl:2025mib}, we can combine the
event-level shape information with the total event rate through the extended
likelihood,
\begin{align}
    \mathcal{L}_{\mathrm{full}}(\{x\}|\theta)
    &=
    \operatorname{Pois}
    \bigl(n\,|\,S(\theta)+B\bigr)
    \prod_{i=1}^{n} p(x_i|\theta)
    \notag\\
    &=
    \frac{e^{-[S(\theta)+B]}}{n!}
    \prod_{i=1}^{n}
    \left[
        S(\theta)\,p_S(x_i|\theta)
        +
        B\,p_B(x_i)
    \right].
    \label{eq:pfull}
\end{align}
Using the classifier likelihood ratio, we can rewrite the extended likelihood
as
\begin{equation}
    \mathcal{L}_{\mathrm{full}}(\{x\}|\theta)
    =
    \frac{e^{-[S(\theta)+B]}}{n!}
    \prod_{i=1}^{n}
    p_B(x_i)
    \left[
        B + S(\theta)\,r(x_i|\theta)
    \right].
\end{equation}
The log-likelihood ratio between two parameter points $\theta$ and $\theta_0$
is therefore
\begin{equation}
    \log \Lambda_{\mathrm{full}}(\theta,\theta_0)
    =
    -\left[S(\theta)-S(\theta_0)\right]
    +
    \sum_{i=1}^{n}
    \log
    \frac{
        B+S(\theta)\,r(x_i|\theta)
    }{
        B+S(\theta_0)\,r(x_i|\theta_0)
    } \;.
    \label{eq:full_lr_sb}
\end{equation}
For the background-only hypothesis $\theta_0 = \theta_B$ (i.e., $\theta_B$ represents the SM), with $S(\theta_B)=0$, this reduces to
\begin{equation}
    \log \Lambda_{\mathrm{full}}(\theta,\theta_B)
    =
    -S(\theta)
    +
    \sum_{i=1}^{n}
    \log
    \left[
        1+\frac{S(\theta)}{B}\,
        r(x_i|\theta)
    \right].
    \label{eq:full_lr_bonly}
\end{equation}
The first term contains the total-rate information, while the event sum
contains the kinematic shape information learned by the classifier.

If the total event rate is very sensitive to systematic uncertainties, we can instead use a
shape-only likelihood. Writing the expected yields as
\begin{equation}
    S(\theta)
    =
    \mathcal{L}\,\sigma_S(\theta),
    \qquad
    B
    =
    \mathcal{L}\,\sigma_B,
\end{equation}
where $\mathcal{L}$ denotes the integrated luminosity, the luminosity cancels
from the relative signal and background contributions. We define the
per-event shape likelihood ratio with respect to the background-only
hypothesis as
\begin{equation}
    \lambda_{\mathrm{shape}}(x|\theta,\theta_B)
    =
    \frac{
        1+\dfrac{\sigma_S(\theta)}{\sigma_B}\,r(x|\theta)
    }{
        1+\dfrac{\sigma_S(\theta)}{\sigma_B}
    }.
    \label{eq:shape_likelihood_event}
\end{equation}
The corresponding dataset-level shape-only log-likelihood ratio is
\begin{equation}
    \log \Lambda_{\mathrm{shape}}(\theta,\theta_B)
    =
    \sum_{i=1}^{n}
    \log \lambda_{\mathrm{shape}}(x_i|\theta,\theta_B),
    \label{eq:shape_likelihood}
\end{equation}
or, explicitly,
\begin{equation}
    \log \Lambda_{\mathrm{shape}}(\theta,\theta_B)
    =
    \sum_{i=1}^{n}
    \log\left[
        1
        +
        \frac{\sigma_S(\theta)}{\sigma_B}
        r(x_i|\theta)
    \right]
    -
    n\log\left[
        1
        +
        \frac{\sigma_S(\theta)}{\sigma_B}
    \right].
\end{equation}
The denominator in \cref{eq:shape_likelihood_event} ensures the
normalization of the signal-plus-background density after conditioning on the
observed event count. The shape-only likelihood therefore removes the
information carried by the absolute event yield while retaining the relative
signal-to-background contribution and the event-level kinematic information.

For the parton-level study in \cref{sec:dijet}, we use the corresponding
parton-level signal and background cross sections for $\sigma_S(\theta)$ and
$\sigma_B$. For the reconstruction-level analyses in \cref{sec:zhllbb,sec:ytohh}, we instead use fiducial cross sections evaluated after the event selection. For an unweighted Monte Carlo sample, we calculate the fiducial cross section as
\begin{equation}
    \sigma_{\mathrm{fid}}
    =
    \sigma_{\mathrm{gen}}\,
    \frac{N_{\mathrm{pass}}}{N_{\mathrm{gen}}} \;,
\end{equation}
where $\sigma_{\mathrm{gen}}$ denotes the generated cross section,
$N_{\mathrm{gen}}$ the number of generated events, and $N_{\mathrm{pass}}$
the number of events that pass the reconstruction-level event selection.

\paragraph{Expected limits from Monte Carlo.}
We quote expected limits throughout this work. Rather than evaluating
\cref{eq:shape_likelihood} on a random realization containing
$N_{\mathrm{exp}}$ events, we evaluate the expectation of the per-event
log-likelihood ratio using the full Monte Carlo sample~\cite{Bahl:2025mib}.
We therefore make the replacement
\begin{align}
    \sum_{i=1}^{N_{\mathrm{exp}}}
    \log \lambda_{\mathrm{shape}}(x_i|\theta,\theta_B)
    \;\longrightarrow\;
    \frac{N_{\mathrm{exp}}}{N_{\mathrm{MC}}}
    \sum_{x_i\in\{x\}_{\mathrm{MC}}}
    \log \lambda_{\mathrm{shape}}(x_i|\theta,\theta_B)
    \notag\\
    =
    N_{\mathrm{exp}}
    \left\langle
        \log \lambda_{\mathrm{shape}}(x|\theta,\theta_B)
    \right\rangle_{\{x\}_{\mathrm{MC}}}.
    \label{eq:mcsum}
\end{align}
This uses all available simulated events rather than an arbitrary subset of
size $N_{\mathrm{exp}}$, thereby removing fluctuations associated with a
finite random realization of the expected dataset. The residual noise from the
finite simulated sample, and when it matters, is discussed in
\cref{sec:app_bkgfluct}. \Cref{eq:mcsum} is the
straightforward version of an unbinned Asimov dataset, and its cost grows with the
Monte Carlo sample. More economical constructions, in which a small weighted
reference sample is built such that the generating parameters maximize the
weighted likelihood, have recently been
proposed~\cite{deSa:2026hch,deSa:2026awa}.

For a signal-plus-background generating hypothesis, the signal and background
samples contribute independently, such that
\begin{align}
    N_{\mathrm{exp}}
    \left\langle
        \log \lambda_{\mathrm{shape}}(x|\theta,\theta_B)
    \right\rangle_{\{x\}_{\mathrm{MC}}}
    \;\longrightarrow\;&
    N_S^{\mathrm{exp}}
    \left\langle
        \log \lambda_{\mathrm{shape}}(x|\theta,\theta_B)
    \right\rangle_{\{x\}_{\mathrm{MC},S}}
    \notag\\
    &+
    N_B^{\mathrm{exp}}
    \left\langle
        \log \lambda_{\mathrm{shape}}(x|\theta,\theta_B)
    \right\rangle_{\{x\}_{\mathrm{MC},B}}.
    \label{eq:mcsplit}
\end{align}
The expression generalizes directly to multiple statistically independent
background processes by adding their individual contributions. For
background-only expected limits, one sets $N_S^{\mathrm{exp}}=0$, such that only
the background expectation contributes.

For the full likelihood in \cref{eq:full_lr_bonly}, we apply the same
replacement to the event sum while retaining the term $-S(\theta)$:
\begin{align}
    \log \Lambda_{\mathrm{full}}^{\mathrm{exp}}(\theta,\theta_B)
    ={}&
    -S(\theta)
    +
    N_S^{\mathrm{exp}}
    \left\langle
        \log\left[
            1+\frac{S(\theta)}{B}\,r(x|\theta)
        \right]
    \right\rangle_{\{x\}_{\mathrm{MC},S}}
    \notag\\
    &+
    N_B^{\mathrm{exp}}
    \left\langle
        \log\left[
            1+\frac{S(\theta)}{B}\,r(x|\theta)
        \right]
    \right\rangle_{\{x\}_{\mathrm{MC},B}}.    \label{eq:full_likelihood_expected}
\end{align}
Here, $N_S^{\mathrm{exp}} = S(\theta_{\text{true}})$ and $N_B^{\mathrm{exp}} = B$ are the expected yields of the observed dataset. This retains both rate and shape information while averaging over fluctuations in the
event count and kinematics.

%%%%%%%%%%%%%%
\subsection{Calibration and confidence regions}
\paragraph{Likelihood-ratio calibration.}

Before evaluating \cref{eq:shape_likelihood,eq:full_lr_bonly}, we calibrate the learned
signal-to-background likelihood ratio following Ref.~\cite{Brehmer:2018eca}. An exact likelihood ratio satisfies
\begin{align}
    \left\langle r(x|\theta)\right\rangle_{p_B}
    &=
    \int dx\,
    p_B(x)
    \frac{p_S(x|\theta)}{p_B(x)}
    \notag\\
    &=
    \int dx\,p_S(x|\theta)
    =
    1
\end{align}
for every parameter point $\theta$. A finite-capacity classifier does not
necessarily satisfy this normalization exactly, particularly near the
boundaries of the parameter space where the training coverage is sparse.

We therefore normalize the raw classifier likelihood-ratio estimate
$\hat r(x|\theta)$ according to
\begin{equation}
    r_{\mathrm{cal}}(x|\theta)
    =
    \frac{
        \hat r(x|\theta)
    }{
        \left\langle
            \hat r(x|\theta)
        \right\rangle_{p_B}
    } \;,
    \label{eq:ratio_calibration}
\end{equation}
or equivalently,
\begin{equation}
    \log r_{\mathrm{cal}}(x|\theta)
    =
    \log \hat r(x|\theta)
    -
    \log
    \left\langle
        \hat r(x|\theta)
    \right\rangle_{p_B}.
\end{equation}
We estimate the background expectation using an independent background Monte
Carlo sample at each parameter point. By construction,
\begin{equation}
    \left\langle
        r_{\mathrm{cal}}(x|\theta)
    \right\rangle_{p_B}
    =
    1.
\end{equation}
We then use $r_{\mathrm{cal}}(x|\theta)$ in
\cref{eq:shape_likelihood,eq:full_lr_bonly}. This normalization is particularly
important for the likelihood because an overall misnormalization
of the learned likelihood ratio can mimic a change in the signal rate.
Without this correction, the likelihood can reinterpret background events as
signal and spuriously favor poorly constrained regions near the boundaries of
the parameter space.

\paragraph{Profiling and confidence regions}

We use either the full likelihood in \cref{eq:full_lr_bonly} or the shape-only
likelihood in \cref{eq:shape_likelihood}, written generically as $\Lambda$, to
construct confidence regions in the model parameters. For a parameter point
$\theta$, we define the likelihood-ratio test statistic
\begin{equation}
    q(\theta)
    =
    -2
    \left[
        \log \Lambda(\theta,\theta_B)
        -
        \log \Lambda(\hat{\theta},\theta_B)
    \right],
    \label{eq:teststat}
\end{equation}
where
\begin{equation}
    \hat{\theta}
    =
    \operatorname*{arg\,max}_{\theta}
    \log \Lambda(\theta,\theta_B)
\end{equation}
denotes the maximum-likelihood estimate. Since both terms in
\cref{eq:teststat} are evaluated relative to the same background
reference, this reference cancels in $q(\theta)$.

When only a subset of the model parameters is of interest, we profile over
the remaining parameters. We write
\begin{equation}
    \theta
    =
    \left(
        \theta_{\mathrm{int}},
        \theta_{\mathrm{prof}}
    \right),
\end{equation}
where $\theta_{\mathrm{int}}$ denotes the parameters for which we construct
the confidence region and $\theta_{\mathrm{prof}}$ denotes the parameters
profiled over. The profile-likelihood test statistic is
\begin{equation}
    q(\theta_{\mathrm{int}})
    =
    -2
    \left[
        \log
        \Lambda
        \left(
            \theta_{\mathrm{int}},
            \hat{\theta}_{\mathrm{prof}}(\theta_{\mathrm{int}}),
            \theta_B
        \right)
        -
        \log
        \Lambda
        \left(
            \hat{\theta},
            \theta_B
        \right)
    \right],
    \label{eq:profile_teststat}
\end{equation}
where $\hat{\theta}_{\mathrm{prof}}(\theta_{\mathrm{int}})$ denotes the
values of the profiled parameters that maximize the likelihood at fixed
$\theta_{\mathrm{int}}$, while $\hat{\theta}$ denotes the global
maximum-likelihood estimate.

We denote $\psi$ as the full parameter vector $\theta$ or the subset $\theta_{\mathrm{int}}$, depending on whether we use \cref{eq:teststat} or \cref{eq:profile_teststat}. We denote the corresponding test statistic evaluated on the observed data by $q_{\mathrm{obs}}(\psi)$, distinguishing it from the random variable $q$ across repeated experiments. In the plots, we denote the observed likelihood-ratio test statistic by $\Delta\chi^2(\psi)\equiv q_{\mathrm{obs}}(\psi)$. The $p$-value associated with a parameter point $\psi$ is approximated by
\begin{equation}
    p_{\psi}
    =
    \int_{q_{\mathrm{obs}}(\psi)}^{\infty}
    dq\,
    p(q|\psi)
    \simeq
    1-
    F_{\chi^2}
    \bigl(q_{\mathrm{obs}}(\psi)|k\bigr),
    \label{eq:pvalue}
\end{equation}
where $p(q|\psi)$ is the probability density of the test statistic under repeated experiments with $\psi$ as the true parameter point. Under the Wilks approximation~\cite{Wilks:1938dza} and in the asymptotic limit~\cite{Cowan:2010js}, this density follows a $\chi_k^2$ distribution, with $k$ the number of parameters tested. Here, $F_{\chi^2}(q|k)$ denotes its cumulative distribution function. A confidence region with confidence level $1-\alpha$
contains all parameter points satisfying
\begin{equation}
    p_{\psi}>\alpha,
\end{equation}
or equivalently,
\begin{equation}
    q_{\mathrm{obs}}(\psi)
    <
    F_{\chi^2}^{-1}(1-\alpha|k).
\end{equation}

\paragraph{Empirical coverage.}

Both the asymptotic approximation in \cref{eq:pvalue} and the learned
likelihood ratio can lead to incorrect confidence regions. The former can
break down for finite event samples or when the regularity conditions of
Wilks' theorem are violated, while the latter can become inaccurate in
regions with limited simulation coverage. We therefore test the frequentist
coverage of the inferred confidence regions empirically~\cite{Bahl:2025mib}.

For a fixed truth point $\theta_{\mathrm{true}}$, we generate
$N_{\mathrm{toys}}$ independent pseudo-datasets by resampling the simulated
events, which requires the available pool to be much larger than the
pseudo-dataset; sampling instead from a learned reference density avoids this
limitation~\cite{deSa:2026awa}. For each pseudo-dataset
$j$, we construct the confidence region and determine whether it contains
$\theta_{\mathrm{true}}$. The empirical coverage of a nominal
$1-\alpha$ confidence region is
\begin{equation}
    c_{1-\alpha}
    =
    \frac{1}{N_{\mathrm{toys}}}
    \sum_{j=1}^{N_{\mathrm{toys}}}
    \Theta
    \left[
        p_{\theta_{\mathrm{true}}}^{(j)}
        -\alpha
    \right],
    \label{eq:coverage}
\end{equation}
where $\Theta$ denotes the Heaviside step function, so the sum counts the pseudo-experiments whose confidence regions contain $\theta_{\mathrm{true}}$. A correctly calibrated
inference procedure should give
\begin{equation}
    c_{1-\alpha}\simeq 1-\alpha
\end{equation}
up to fluctuations from the finite number of pseudo-experiments.
If $c_{1-\alpha}>1-\alpha$, the confidence regions overcover and are
conservative. If $c_{1-\alpha}<1-\alpha$, they undercover, indicating that the
reported confidence regions are too small.

%%%%%%%%%%%%%%%%%%%%%%%%%%%%%%%%%%%%%%%%%%%%%%%%%%
\section{Parton-level example: boosted resonance \texorpdfstring{$Z'\to X \bar X\to 4q$}{Z'->XX->4q}}
\label{sec:dijet}

We first consider a parton-level benchmark for which the exact likelihood ratio
is available. This allows us to test whether the sequential procedure
introduced in \cref{sec:limits} reproduces the confidence regions obtained
from the true likelihood.

%%%%%%%%%%%%%%
\subsection{Target Process and the exact likelihood}
\label{sec:cascade_exact}

We use as an example signal a heavy resonance $Z'$ decaying into a pair of lighter charged scalars $X$, each of which subsequently decays into two light quarks,
\begin{align}
    u\bar u \to Z' \to X\bar X \to (u\bar d)(\bar u d) \;.
    \label{eq:cascade}
\end{align}
As background, we consider the QCD process $u\bar u \to u\bar d\bar u d$. The final state contains four distinct
partons, so we can fully reconstruct both resonances. The signal model contains four shape parameters,
\begin{equation}
    \theta_{\mathrm{shape}}
    =
    \left(
        m_{Z'},\,
        \Gamma_{Z'},\,
        m_X,\,
        \Gamma_X
    \right)\;.
\end{equation}
We assume the $Z'$ boson to have a flavour-universal coupling to quarks, which we denote as $g$. The full theory parameter
vector is therefore
\begin{equation}
    \theta
    =
    \left(
        \theta_{\mathrm{shape}},\,g
    \right).
\end{equation}
We denote the signal matrix element for an event $x$ by $\mathcal{M}_S(x|\theta)$. The coupling $g$ appears four times in the signal amplitude,
\begin{equation}
    \mathcal{M}_S(x|\theta)
    =
    g^4\,
    \mathcal{M}_S(x|\theta_{\mathrm{shape}},g=1),
    \qquad
    \sigma_S(\theta)
    =
    g^8\,
    \sigma_S(\theta_{\mathrm{shape}}),
    \label{eq:theta8}
\end{equation}
where $\mathcal{M}_S(x|\theta_{\mathrm{shape}}) \equiv \mathcal{M}_S(x|\theta_{\mathrm{shape}},g=1)$ is the signal matrix
element at unit coupling and $\sigma_S(\theta_{\mathrm{shape}})$ is the
corresponding signal cross-section. The parameters $\theta_{\mathrm{shape}}$ control the signal kinematics,
while $g$ controls its relative normalization.

For the parton-level process, signal and background have the same initial and
final states. Their common flux and phase-space factors therefore cancel
point-by-point in the ratio of differential cross sections. We define
\begin{equation}
    r_{\mathrm{ME}}(x|\theta_{\mathrm{shape}})
    =
    \frac{
        |\mathcal{M}_S(x|\theta_{\mathrm{shape}})|^2
    }{
        |\mathcal{M}_B(x)|^2
    },
    \qquad
    r_{\sigma}(\theta_{\mathrm{shape}})
    =
    \frac{
        \sigma_S(\theta_{\mathrm{shape}})
    }{
        \sigma_B
    }.
    \label{eq:cascade_ratios}
\end{equation}
Since the classifier learns the ratio of normalized signal and background
densities, we can relate it to $r_{\mathrm{ME}}$ as
\begin{equation}
    r(x|\theta_{\mathrm{shape}})
    =
    \frac{
        p_S(x|\theta_{\mathrm{shape}})
    }{
        p_B(x)
    }
    =
    \frac{
        r_{\mathrm{ME}}(x|\theta_{\mathrm{shape}})
    }{
        r_{\sigma}(\theta_{\mathrm{shape}})
    }.
    \label{eq:cascade_matrix_ratio}
\end{equation}
The normalized
signal-plus-background density reads
\begin{equation}
    p_{S+B}(x|\theta)
    =
    \frac{
        p_B(x)
        +
        g^8 r_{\sigma}(\theta_{\mathrm{shape}})
        p_S(x|\theta_{\mathrm{shape}})
    }{
        1+
        g^8 r_{\sigma}(\theta_{\mathrm{shape}})
    }.
    \label{eq:cascade_mixture}
\end{equation}
The corresponding per-event shape likelihood ratio with respect to the
background-only hypothesis is
\begin{align}
    \lambda_{\mathrm{shape}}(x|\theta)
    &=
    \frac{
        p_{S+B}(x|\theta)
    }{
        p_B(x)
    }
    \notag\\
    &=
    \frac{
        1+
        g^8 r_{\sigma}(\theta_{\mathrm{shape}})
        r(x|\theta_{\mathrm{shape}})
    }{
        1+
        g^8 r_{\sigma}(\theta_{\mathrm{shape}})
    }
    \notag\\
    &=
    \frac{
        1+
        g^8 r_{\mathrm{ME}}(x|\theta_{\mathrm{shape}})
    }{
        1+
        g^8 r_{\sigma}(\theta_{\mathrm{shape}})
    }.
    \label{eq:cascade_perevent}
\end{align}
The shape-only log-likelihood ratio for a dataset
$\{x_i\}_{i=1}^{n}$ is therefore
\begin{align}
    \log \Lambda_{\mathrm{shape}}(\theta)
    &=
    \sum_{i=1}^{n}
    \log
    \lambda_{\mathrm{shape}}(x_i|\theta)
    \notag\\
    &=
    \sum_{i=1}^{n}
    \log\left[
        1+
        g^8 r_{\mathrm{ME}}
        (x_i|\theta_{\mathrm{shape}})
    \right]
    -
    n\log\left[
        1+
        g^8 r_{\sigma}(\theta_{\mathrm{shape}})
    \right].
    \label{eq:cascade_lr_cond}
\end{align}
The first term contains the event-level kinematic information, while the
second normalizes the signal-plus-background mixture. Since we condition on
the observed event count, only the relative cross section
$r_{\sigma}=\sigma_S/\sigma_B$ enters, and the likelihood contains no
information about the absolute event rate.

For the expected likelihood, we replace the event sum by its Monte Carlo
expectation using \cref{eq:mcsum,eq:mcsplit},
\begin{align}
    -2\log L_{\mathrm{shape}}(\theta)
    &=
    2\,N_{\mathrm{tot}}^{\mathrm{exp}}
    \log\left[
        1+
        g^8 r_{\sigma}(\theta_{\mathrm{shape}})
    \right]
    \notag\\
    &\quad
    -2
    \sum_{j=S,B}
    N_j^{\mathrm{exp}}
    \left\langle
        \log\left[
            1+
            g^8 r_{\mathrm{ME}}
            (x|\theta_{\mathrm{shape}})
        \right]
    \right\rangle_{\{x\}_{\mathrm{MC},j}},
    \label{eq:cascade_lr}
\end{align}
where
\begin{equation}
    N_{\mathrm{tot}}^{\mathrm{exp}}
    =
    N_S^{\mathrm{exp}}
    +
    N_B^{\mathrm{exp}}.
\end{equation}
We use \cref{eq:cascade_lr} throughout the parton-level study. For the
exact reference, we evaluate
$r_{\mathrm{ME}}(x|\theta_{\mathrm{shape}})$ directly from the matrix
elements. For the learned likelihood, the classifier instead estimates
$r(x|\theta_{\mathrm{shape}})$, from which
\begin{equation}
    r_{\mathrm{ME}}(x|\theta_{\mathrm{shape}})
    =
    r_{\sigma}(\theta_{\mathrm{shape}})
    r(x|\theta_{\mathrm{shape}})
\end{equation}
follows using the known cross-section ratio. This allows us to compare the
classifier-based inference directly to the exact matrix-element result.

The $g^8$ dependence makes the likelihood highly sensitive to the
coupling. In addition, the likelihood depends explicitly on
$r_{\sigma}(\theta_{\mathrm{shape}})$, so inaccuracies in the signal cross
section can distort the profiled likelihood.

As truth point for the observed signal dataset, we use
\begin{align}
    \theta_\text{true}
    = \left( m_{Z'},\, \frac{\Gamma_{Z'}}{m_{Z'}},\, m_X,\, \frac{\Gamma_X}{m_X},\, g \right)
    = \left( 2600~\text{GeV},\, 0.035,\, 300~\text{GeV},\, 0.15,\, 0.445 \right) \;,
    \label{eq:cascade_theta_true}
\end{align}
i.e., $\Gamma_{Z'} = 91$~GeV, $\Gamma_X = 45$~GeV and $g^8 = 1.55\times10^{-3}$. Within the fiducial cuts the signal-cross section at unit coupling is $\sigma_s = 0.0845$~pb and the background cross-section is $\sigma_b = 3.24\times10^{-3}$~pb. We fix the observed background yield to $N^{\text{exp}}_{B}=1000$ events, corresponding to
an integrated luminosity of $\mathcal{L}=309~\mathrm{fb}^{-1}$. The true coupling $g=0.445$ corresponds to the expected signal yield $N^{\text{exp}}_{S}=g^8\,\sigma_s\,\mathcal{L}=40.4$.

%%%%%%%%%%%%%%
\subsection{Simulations and event selection}
\label{sec:cascade_sim}
Both samples are generated with \textsc{MadGraph5\_aMC@NLO}~\cite{Alwall:2014hca} at $\sqrt{s}=13$~TeV with the \texttt{NNPDF2.3LO} parton densities~\cite{Ball:2012cx}, using a UFO model~\cite{Degrande:2011ua,Alloul:2013bka} in which the $Z'$ couples to a complex charged scalar. Signal events are generated as a decay-chain process (see \cref{eq:cascade}), which makes phase-space integration and event generation more efficient by using the resonant structure. For reweighting, the phase-space points are already fixed, so we evaluate the full $2\to4$ matrix element directly. This retains the more accurate off-shell dependence for the matrix element ratio $r_{\text{ME}}$. The decay-chain and full $2\to4$ descriptions agree at the percent level.

Signal and background events are subject to the same selection cuts,
$p_{T,j}>40$~GeV, $|\eta_j|<2.8$, $\Delta R_{jj}>0.4$, and
$H_T>1500$~GeV. We additionally fix the renormalization scale to
$\mu_R=2600$~GeV. Since the QCD background is
$\mathcal{O}(\alpha_s^4)$, evaluating the matrix-element ratio requires the
background coupling $\alpha_s(\mu_R)$ at every phase-space point, including
events drawn from the signal sample. For a dynamical scale, this requires the
exact event-dependent scale prescription to be reproducible during
reweighting. Using a fixed $\mu_R$ avoids this difficulty and ensures a
consistent evaluation of the matrix-element likelihood ratio for all events.

In each sequential SBI round, we add $800$ parameter points with $500$ signal events per point, and the networks are retrained on all accumulated simulations. The first round uses $800$ Sobol points~\cite{SOBOL} in the range
\begin{align}
    m_{Z'} &\in [2000,4500]~\text{GeV} \;, &
    \Gamma_{Z'}/m_{Z'} &\in [0.01,0.3] \;, \notag \\
    m_X/m_{Z'} &\in [1/22,\, 5/11] \;, &
    \Gamma_X/m_X &\in [0.01,0.3] \;.
    \label{eq:cascade_prior}
\end{align}
\cref{sec:grid_designs} shows why this choice ensures more uniform coverage in the parameter space.In subsequent rounds, parameter points are drawn from the proposal in \cref{eq:sequential_proposal_freq} with $\gamma=0.5$, constructed from the previous round's likelihood. We describe how we sample it in \cref{sec:cascade_proposals}. We perform six rounds in total. A fixed pool of $2.4\times10^6$ background events is generated beforehand, and in each round we use an equal number of signal and background events to maintain class balance.

We compare sequential SBI with two single-round baselines, each using $2.4\times10^6$ signal events and an equal number of background events over the full range of \cref{eq:cascade_prior}. The coarse grid uses the same $800$ round-1 points with $3000$ signal events per point; the fine grid uses the first $4800$ points of the same Sobol sequence with $500$ events per point.

%%%%%%%%%%%%%%
\subsection{Network architecture and hyperparameters}

The classifier is a fully connected network with a 24-dimensional input comprising the four final-state parton momenta, the invariants \(\{s,m_{01}^2,m_{23}^2,t\}\), and the parameters \((m_{Z'},\Gamma_{Z'},m_X,\Gamma_X)\). Kinematic inputs are standardised, while the parameters are scaled by their median and interquartile range. The network has three hidden layers of 128 units with GELU activations and \(0.1\) dropout, followed by a sigmoid output. We train with binary cross entropy and AdamW for 250 epochs using a batch size of 1024, with the learning rate cosine-annealed from \(10^{-3}\) to \(10^{-5}\). At each parameter point, the signal sample is split equally between training and validation, and we retain the epoch with the lowest validation loss. Background events are assigned parameter values by reshuffling the signal parameters at every epoch, ensuring that both classes share the same parameter distribution. In each round, we train an ensemble of five networks from independent random initializations on all accumulated simulations and average their predictions at the likelihood level.

%%%%%%%%%%%%%%%%%% figure %%%%%%%%%%%%%%%%%%%
\begin{figure}[t]
    \centering
    \includegraphics[width=\textwidth]{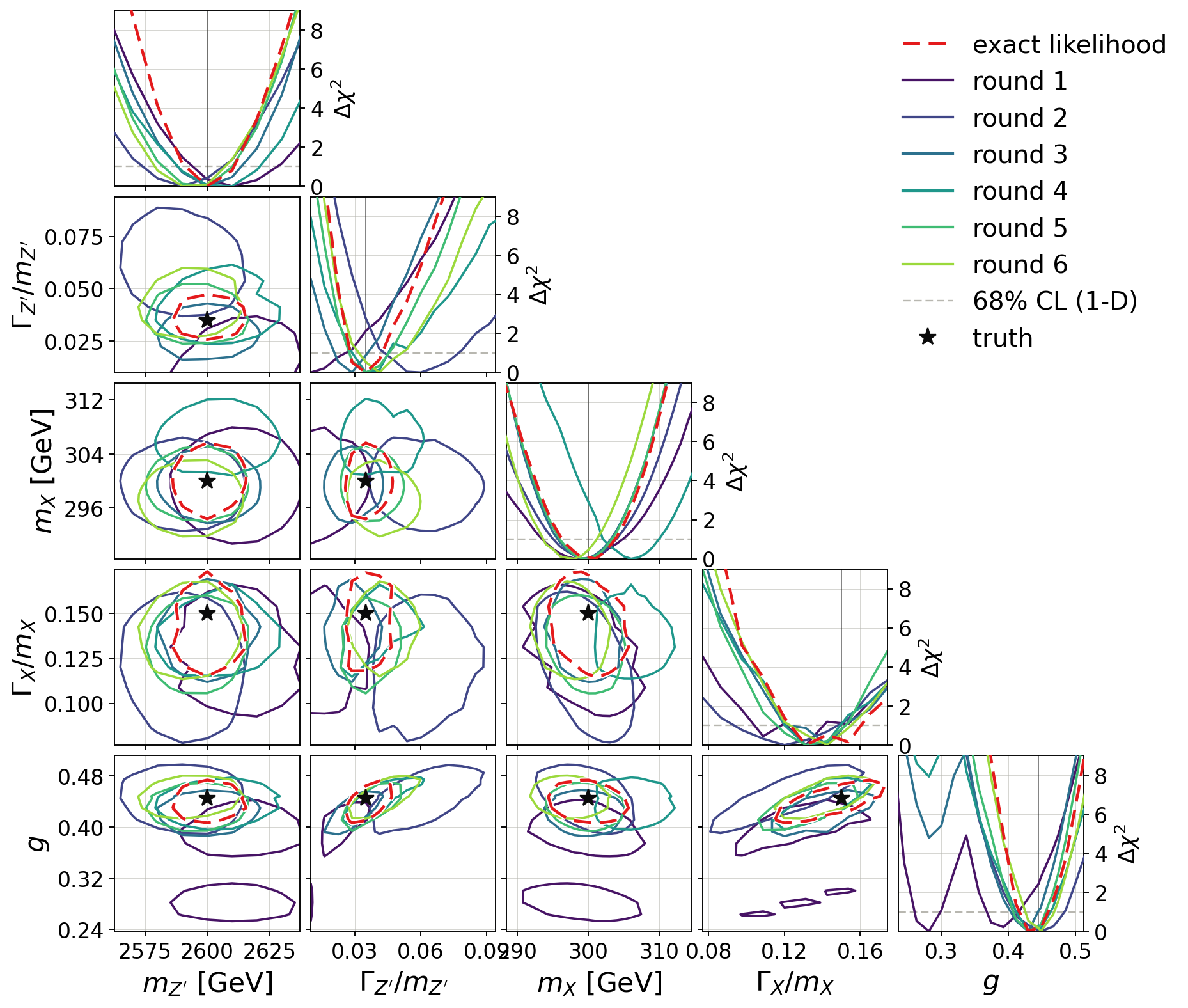}
    \caption{Sequential evolution of the inferred likelihood over six rounds: one-dimensional profiles on the diagonal and $68\%$ CL regions below it. The learned contours progressively converge to the exact likelihood (red dashed). The true parameter point is indicated by the black star.}
    \label{fig:cascade_rounds_68}
\end{figure}
%%%%%%%%%%%%%%%%%% figure %%%%%%%%%%%%%%%%%%%

\subsection{Evaluation}

\begin{figure}[t]
    \centering
    \includegraphics[width=\textwidth]{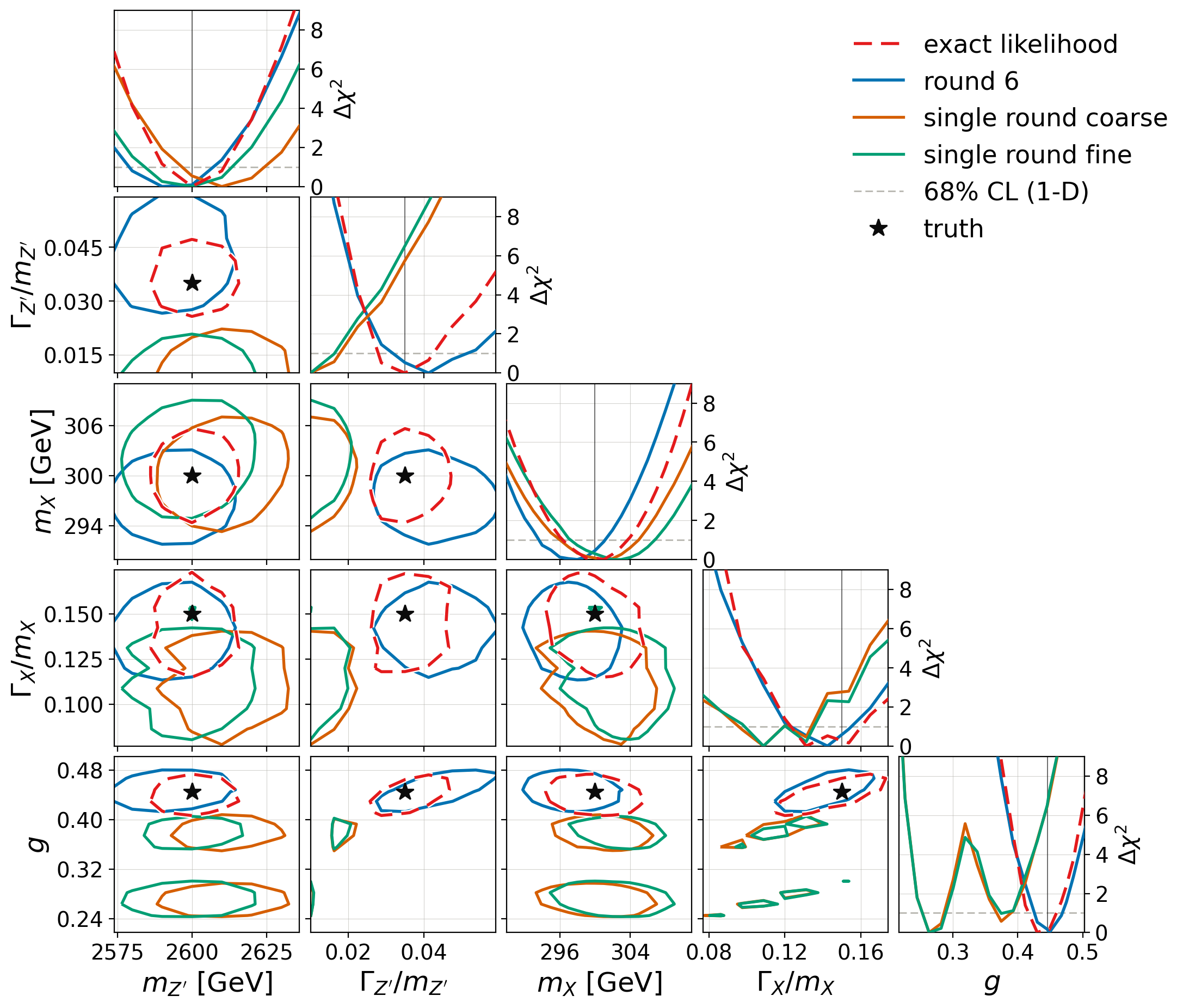}
    \caption{Comparison of the final sequential inference with the coarse ($800$ points with $3000$ events each) and fine ($4800$ points with $500$ events each) single rounds at fixed simulation budget. The sequential strategy most closely reproduces the exact likelihood contours, showing the benefit of adaptively concentrating simulations in the high-likelihood region.}
    \label{fig:cascade_single_68}
\end{figure}

We evaluate the expected likelihood of \cref{eq:cascade_lr} with
$N_\text{MC}=5\times10^4$ signal events generated at $\theta_\text{true}$ and
$N_\text{MC}=5\times10^4$ background events. The likelihoods shown in
\cref{sec:cascade_rounds,sec:cascade_budget} are evaluated and profiled on the grid
\begin{align}
    m_{Z'} &\in [2520,2680]~\text{GeV} \;\; (17~\text{points}) \;, &
    \Gamma_{Z'}/m_{Z'} &\in [0.01,0.16] \;\; (25~\text{points}) \;, \notag \\
    m_X &\in [285,320]~\text{GeV} \;\; (36~\text{points}) \;, &
    \Gamma_X/m_X &\in [0.03,0.3] \;\; (25~\text{points}) \;,
    \label{eq:cascade_grid}
\end{align}
with all axes evenly spaced, which gives $382{,}500$ points in the shape-parameter space. For each point, we additionally evaluate $42$ values of the coupling, $g\in[0.097,0.8]$, including $g=0.445$. The single-round results of \cref{fig:cascade_single_68}, trained on the coarse and fine grids of \cref{sec:cascade_sim}, are evaluated on the same grid. The coverage tests of \cref{sec:cascade_coverage} use a coarser evaluation grid with the same relative-width axes, $m_{Z'}\in[2400,2780]$~GeV and $m_X\in[200,390]$~GeV with $20$ points each, and $28$ values $g\in[0.356,0.8]$, i.e.\ $250{,}000$ shape points.

We measure the calibration $k$ (see \cref{{sec:limits}}) once on the union of the $250{,}000$ nodes of the coarser grid and an initial $20{,}000$-point Sobol set, and use this common calibration, interpolated locally linearly between the measured points, for sampling and for both evaluation grids.

%%%%%%%%%%%%%%

%%%%%%%%%%%%%%
\subsection{Results}

%%%%%%%%%%%%%%%%%% figure %%%%%%%%%%%%%%%%%%%
%%%%%%%%%%%%%%%%%% figure %%%%%%%%%%%%%%%%%%%
\subsubsection{Sequential SBI}
\label{sec:cascade_rounds}

\Cref{fig:cascade_rounds_68} shows the evolution of the $68\%$ confidence regions from the inferred likelihood across the six sequential rounds. The first round gives broad and several small displaced localized contours. This is a result of the limited coverage of the initial proposal. For the later rounds, as the proposal is iteratively concentrated in the high-likelihood region, the learned likelihood approaches the exact result. By rounds $4$--$6$, the contours closely track the exact likelihood across all one- and two-parameter projections. The true parameter point remains within the inferred high-likelihood region throughout, and the final-round contours are in reasonable agreement with the exact likelihood. In \cref{sec:contours_parton_app},~\cref{fig:cascade_rounds_95} shows how the $95 \%$ confidence region changes with rounds.

%%%%%%%%%%%%%%%%%% figure %%%%%%%%%%%%%%%%%%%
\begin{figure}[t]
    \centering
    \includegraphics[width=0.75\textwidth]{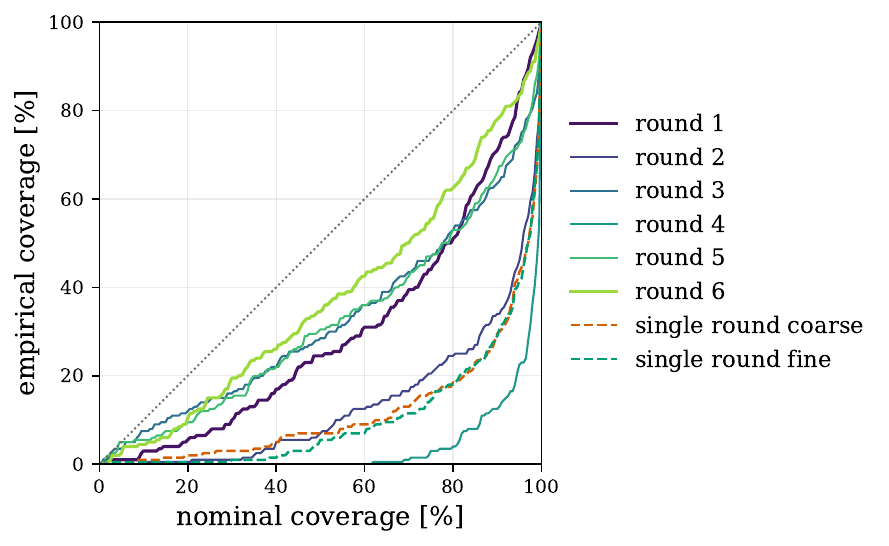}
    \caption{Empirical against nominal coverage of the five-parameter region, for the six sequential rounds and the coarse ($800$ points with $3000$ events each) and fine ($4800$ points with $500$ events each) single rounds. Coverage tests use $200$ pseudo-experiments each. Results below the dotted diagonal are overconfident.}
    \label{fig:cascade_coverage}
\end{figure}
%%%%%%%%%%%%%%%%%% figure %%%%%%%%%%%%%%%%%%%

%%%%%%%%%%%%%%
\subsubsection{Comparing sequential SBI to single rounds}
\label{sec:cascade_budget}

\Cref{fig:cascade_single_68} compares $68\%$ confidence regions of the final sequential SBI result with two single-round baselines using the same simulation budget of $2.4\times10^6$ signal events. Sequential SBI reproduces the exact likelihood substantially better across the parameter projections. Both single-round baselines develop displaced or spurious contours, despite using the same total number of simulated events. This shows that increasing either the number of events per point or the density of the initial points is insufficient. The gain comes from adaptively relocating the simulation budget toward the region favored by the observed data as the inference proceeds. In \cref{sec:contours_parton_app},~\cref{fig:cascade_single_95} shows the $95 \%$ confidence region comparison.

%%%%%%%%%%%%%%
\subsubsection{Coverage}
\label{sec:cascade_coverage}

To get the coverage, we do $200$ pseudo-experiments, where we draw
$N_S\sim\operatorname{Pois}(N^{\text{exp}}_{S})$ and
$N_B\sim\operatorname{Pois}(N^{\text{exp}}_{B})$ with $N^{\text{exp}}_{S}=40.4$ and $N^{\text{exp}}_{B}=1000$. We bootstrap the corresponding numbers of
events from the signal and background evaluation samples, rebuild the
likelihood on the $250{,}000$-point parameter grid, and test whether the
five-dimensional confidence region contains the injected point at the
corresponding Wilks threshold.

\Cref{fig:cascade_coverage} shows the resulting empirical coverage as a function of
the nominal coverage. All methods exhibit some undercoverage, but the
sequential procedure improves substantially as the simulation budget is
adapted to the likelihood-supported region. Round~6 gives the best overall
calibration and lies closest to the diagonal over most of the range. In contrast, both
single-round baselines strongly undercover across almost the entire range,
despite using the same total simulation budget. Their confidence regions
are therefore systematically too narrow, whereas the later sequential
rounds recover a substantially better calibrated likelihood.

%%%%%%%%%%%%%%%%%%%%%%%%%%%%%%%%%%%%%%%%%%%%%%%%%%
\section{Reco-level example I: a heavy scalar cascade \texorpdfstring{$A\to ZH\to \ell\ell b\bar b$}{A -> ZH -> llbb}}
\label{sec:zhllbb}

 To test iterative SBI under more realistic conditions, we show a full reconstruction-level analysis with parton shower, hadronization, fast detector simulation, and the dominant SM background. Given the computational cost of reconstruction-level simulations, we infer three parameters in this example.

%%%%%%%%%%%%%%
\subsection{Target process}
\label{sec:zhllbb_signal}

We take our reconstruction-level signal to be a heavy BSM pseudoscalar decaying into a $Z$ boson as well as another scalar BSM resonance,
\begin{align}
    pp \to A \to Z H \to (\ell^+\ell^-)(b\bar b) \;,
    \qquad \ell = e,\mu \;.
    \label{eq:zhllbb}
\end{align}
Such a process can appear for example in a two-Higgs-doublet model~\cite{Branco:2011iw}, where a mass splitting $m_A > m_H$ is favoured by electroweak baryogenesis and a strong first-order phase transition~\cite{Biekotter:2023eil}. It has been searched for experimentally with both masses free in the $\ell\ell b\bar b$ and $\ell\ell WW$ final states~\cite{ATLAS:2020gxx} and, above the $t\bar t$ threshold, in $\ell\ell t\bar t$~\cite{CMS:2024yiy}. The two are complementary in $m_H$, since $H\to b\bar b$ dominates below $2m_t$ and $H\to t\bar t$ above it. The $\nu\nu b\bar b$ final state has been proposed but not yet covered~\cite{Biekotter:2023eil}.

Following actual experimental searches~\cite{ATLAS:2020gxx}, we infer
\begin{align}
    \theta = (m_A, m_H)
    \qquad\text{and}\qquad
    \mu \equiv \sigma\times\text{BR} \;,
\end{align}
where $\theta$ are the parameters the classifier is conditioned on and the sequential rounds of \cref{sec:method} proposes over, while $\mu\geq 0$ is the free rate parameter. As our truth benchmark point, we use
\begin{align}
    \theta_\text{true} = (m_A,\, m_H) = \left( 600~\text{GeV},\; 250~\text{GeV} \right) \;.
    \label{eq:zhllbb_theta_true}
\end{align}
Both widths are fixed to $5\%$ of the respective mass during generation and are not scanned. These values are well below the reconstruction resolutions. We describe the signal rate in the following section.

All four final-state particles are fully resolved rather than boosted. With $m_A - m_H = 350$~GeV the $Z$ is produced with moderate momentum and its decay leptons are well separated, and the two $b$-jets from the $H$ boson are resolved as separate $R=0.4$ jets. The two parameters therefore enter through two distinct invariant masses built from the same four objects,
\begin{align}
    m_{b\bar b} = \sqrt{(p_{b_1}+p_{b_2})^2} \;, \qquad
    m_{\ell\ell b\bar b} = \sqrt{(p_{\ell_1}+p_{\ell_2}+p_{b_1}+p_{b_2})^2} \;.
    \label{eq:zhllbb_obs}
\end{align}
Loosely following Ref.~\cite{ATLAS:2020gxx}, one can improve the $m_{\ell\ell b\bar b}$ resolution by using the known $Z$ boson mass. I.e., we use
\begin{align}
    m_A^\text{corr} = m_{\ell\ell b\bar b} - m_{\ell\ell} + m_Z \;,
    \label{eq:mAcorr}
\end{align}
instead of $m_{\ell\ell b\bar b}$.

The experimental searches proceed one mass hypothesis at a time~\cite{ATLAS:2020gxx}. After a cut-based selection , each tested $(m_A, m_H)$ defines its own signal region through a window in $m_{b\bar b}$ around the assumed daughter mass $m_H$. Inside these SRs the binned $m_{\ell\ell b\bar b}$ distribution is the discriminant, peaking at the assumed parent mass for the signal and falling smoothly for the background.

%%%%%%%%%%%%%%%%%% figure %%%%%%%%%%%%%%%%%%%
\begin{figure}[t]
    \centering
    \includegraphics[width=\textwidth]{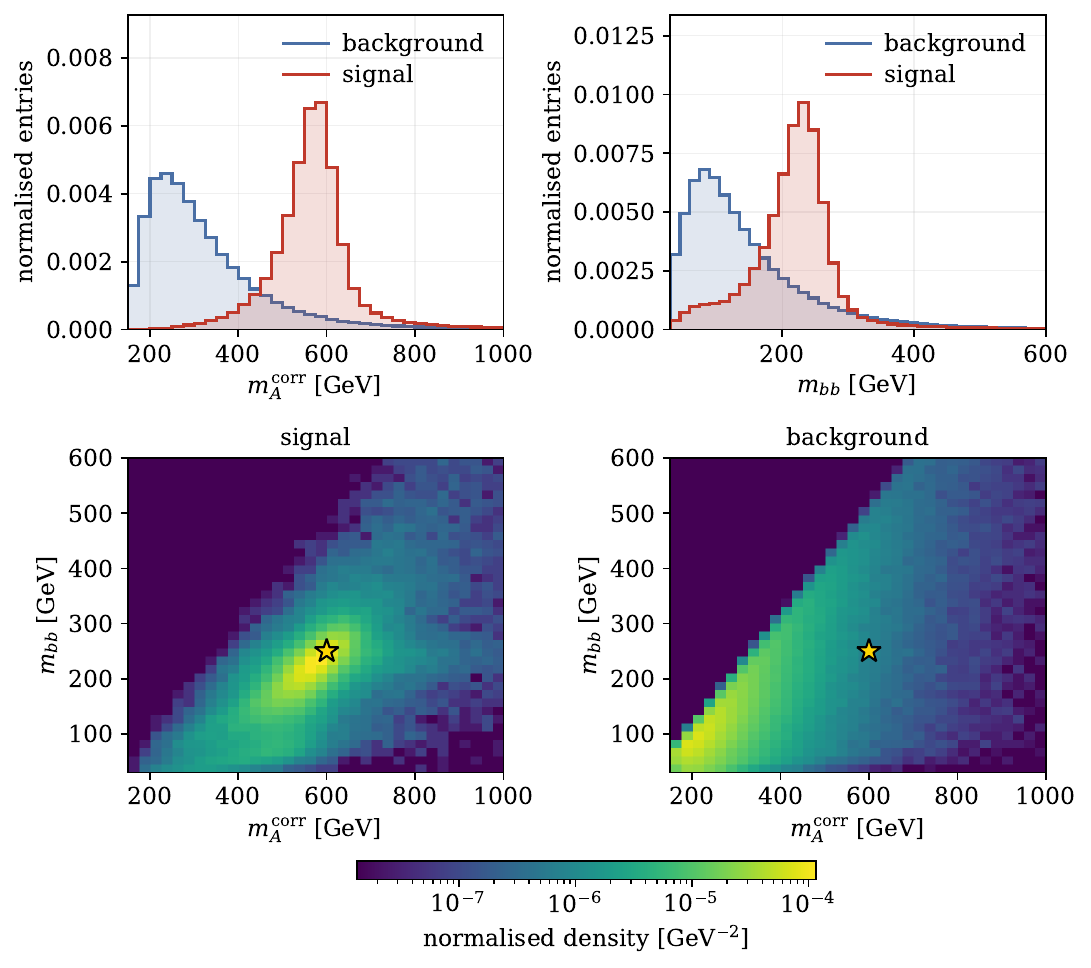}
    \caption{The two mass observables of the $A\to ZH\to\ell\ell b\bar b$ final state after the selection, for the benchmark signal $(m_A, m_H) = (600, 250)$~GeV and for the $Z(\to\ell\ell)+b\bar b$ background. Top: the corrected four-object mass $m_A^{\rm corr}$, which provides information about the parent resonance, and $m_{bb}$, which constraints the daughter mass. Bottom: their joint density on a common logarithmic colour scale with the benchmark point marked.}
    \label{fig:zhllbb_observables}
\end{figure}
%%%%%%%%%%%%%%%%%% figure %%%%%%%%%%%%%%%%%%%

\Cref{fig:zhllbb_observables} shows the two observables. The parent mass $m_A^{\rm corr}$ separates the two classes well but says nothing about $m_H$, while $m_{bb}$ constrains the daughter mass and overlaps the background far more strongly. Both distributions are also broad and asymmetric relative to the injected masses. The joint density in the lower panels shows that the two observables are strongly correlated, bounded by the kinematic edge $m_{bb} \lesssim m_A^{\rm corr}$. The background piles up along that edge at low mass while the signal is clustered around the truth point.
 
%%%%%%%%%%%%%%
\subsection{Fitting the signal rate}
\label{sec:zhllbb_rate}

In the parton-level cascade study of \cref{sec:dijet}, the signal rate is controlled by the coupling of the new resonance, so rate and shape follow from the same underlying parameters and the coupling can be inferred directly. At reconstruction level we do not want the result to depend on a specific coupling structure: the signal kinematics is fixed by the two masses alone, while the signal rate depends on the full model, through the production mechanism and through every branching ratio of the cascade.

In this section use the extended likelihood \cref{eq:full_likelihood_expected}. The learned likelihood ratio $r(x|\theta)$ is independent of $\mu$, but it enters the extended likelihood of \cref{eq:pfull} only through the expected signal yield,
\begin{align}
    S(\theta,\mu) = \mu \, \kappa(\theta) \;,
    \qquad
    \kappa(\theta) = \mathcal{L} \, \epsilon(\theta) \;.
    \label{eq:kappa}
\end{align}
$\mathcal{L}$ is the integrated luminosity and $\epsilon(\theta)$ is the selection efficiency, the fraction of simulated signal events at the parameter point $\theta$ that survive the reconstruction-level selection of \cref{sec:zhllbb_sim}. The yield coefficient $\kappa$ is the expected number of selected signal events. We measure $\epsilon$ directly as the ratio of selected and generated MC events  at every generated parameter point. Between these points, we interpolate it linearly in the $(m_A,m_H)$ plane on a Delaunay triangulation of the generated points, and outside their convex hull we use the value of the nearest generated point. We interpolate $\epsilon$ to every point at which we evaluate the likelihood, since $S(\theta,\mu)$ is needed there but signal is only simulated at the generated points. During the sequential rounds, the proposal for each round uses $\epsilon$ interpolated from the points generated so far, so the interpolation becomes denser near the best fit as the rounds proceed. For the final confidence regions, $\epsilon$ is interpolated from all $234$ generated points.

The parameter $\mu$ we infer is
\begin{align}
    \mu = \sigma(pp\to A)\times\text{BR}(A\to ZH)\times\text{BR}(Z\to\ell\ell)\times\text{BR}(H\to b\bar b). \;
\end{align}
Known branching fractions can be included in either $\mu$ or $\kappa$. For example, ATLAS~\cite{ATLAS:2020gxx} reports $\sigma\times\mathrm{BR}(A\to ZH)\times\mathrm{BR}(H\to b\bar b)$, with the leptonic $Z$ branching fraction treated as a known factor. In our notation, this convention sets $\mathrm{BR}(Z\to\ell\ell)\simeq 6.7\%$ in $\kappa$ rather than $\mu$.
%

%%%%%%%%%%%%%%
\subsection{Simulation and event selection}
\label{sec:zhllbb_sim}

The hard process is generated with \textsc{MadGraph5\_aMC@NLO}~\cite{Alwall:2014hca} using the UFO implementation~\cite{Degrande:2011ua,Alloul:2013bka} of the $A$--$Z$--$H$ vertex. Events are produced for $pp$ collisions at $\sqrt{s}=13.6$~TeV with the \texttt{PDF4LHC15\_nlo\_30\_pdfas} parton densities~\cite{Butterworth:2015oua} accessed through \textsc{LHAPDF6}~\cite{Buckley:2014ana}, showered and hadronized with \textsc{Pythia\,8}~\cite{Sjostrand:2014zea}, and passed through \textsc{Delphes\,3}~\cite{deFavereau:2013fsa} with the ATLAS card, with jets clustered by the anti-$k_T$ algorithm~\cite{Cacciari:2008gp} at $R=0.4$ using \textsc{FastJet}~\cite{Cacciari:2011ma}. The dominant background is $Z(\to\ell\ell)+b\bar b$, generated with the same shower and detector configuration. Events are selected by requiring exactly two opposite-sign same-flavour electrons or muons. with $p_T > 15$~GeV and $|\eta| < 2.6$, the leading one above $27$~GeV, a $Z$ window $76 < m_{\ell\ell} < 106$~GeV, at least two $b$-tagged jets with $m_{b\bar b} > 30$~GeV, and $E_T^\text{miss} < 50$~GeV to suppress the $t\bar t$ contribution (and other backgrounds) that we do not consider for this prototype study.

The background cross section after the generator-level cuts is $\sim 3.6$~pb, which with a flat $K$-factor of $\sim 1.4$~\cite{FebresCordero:2009xzo} and a selection efficiency of $\sim 8\%$ measured on $5.0\times10^7$ generated events gives
\begin{align}
    N^{\text{exp}}_{B} \simeq 1.2\times10^{5}
    \qquad\text{at}\qquad
    \mathcal{L} = 300~\text{fb}^{-1} \;.
\end{align}
The signal selection efficiency at the benchmark point, measured on $6.3\times10^5$ generated events, is $\epsilon \sim 0.17$, so that $\kappa = 5.2\times10^{4}$~pb$^{-1}$. We inject
\begin{align}
    N^{\text{exp}}_{S} = 500 \;,
    \qquad\text{i.e.}\qquad
    \sigma\times\text{BR} = 9.57\times10^{-3}~\text{pb}
\end{align}
for the full visible rate including $\text{BR}(Z\to\ell\ell)$. In the convention in which ATLAS quotes $\sigma\times\text{B}(A\to ZH)\times\text{B}(H\to b\bar b)$ this corresponds to $0.14$~pb. This is a comparatively large signal compared to ATLAS limits, chosen so that there is enough sensitivity to constrain the BSM masses.

We run five sequential SBI rounds, adding $40$ parameter points per round on top of the $85$-point grid from round 1. The points of the later rounds are drawn from the proposal in \cref{eq:sequential_proposal_freq} with $\gamma = 0.8$, constructed from the previous round's likelihood, as described in \cref{sec:zhllbb_proposals}. For each parameter point, we add $10^3$ signal events and $10^3$ background events to the training dataset. The round 1 grid is uninformed and covers the full parameter range,
\begin{align}
    m_A \in [400,1000]~\text{GeV} \;, \qquad
    m_H \in [100,550]~\text{GeV} \;.
    \label{eq:zhllbb_prior}
\end{align}
It is a regular grid with a spacing of 50~GeV in both masses, with nodes at $m_A=400,\dots,1000$~GeV and $m_H=150,\dots,450$~GeV, restricted to $m_A-m_H\geq100$~GeV so that the decay $A\to ZH$ is open. This gives $85$ parameter points. The proposals of the later rounds are drawn over the full range of \cref{eq:zhllbb_prior}.

%%%%%%%%%%%%%%%%%%% figure %%%%%%%%%%%%%%%%%%%
\begin{figure}[t]
    \centering
    \includegraphics[width=0.75\textwidth]{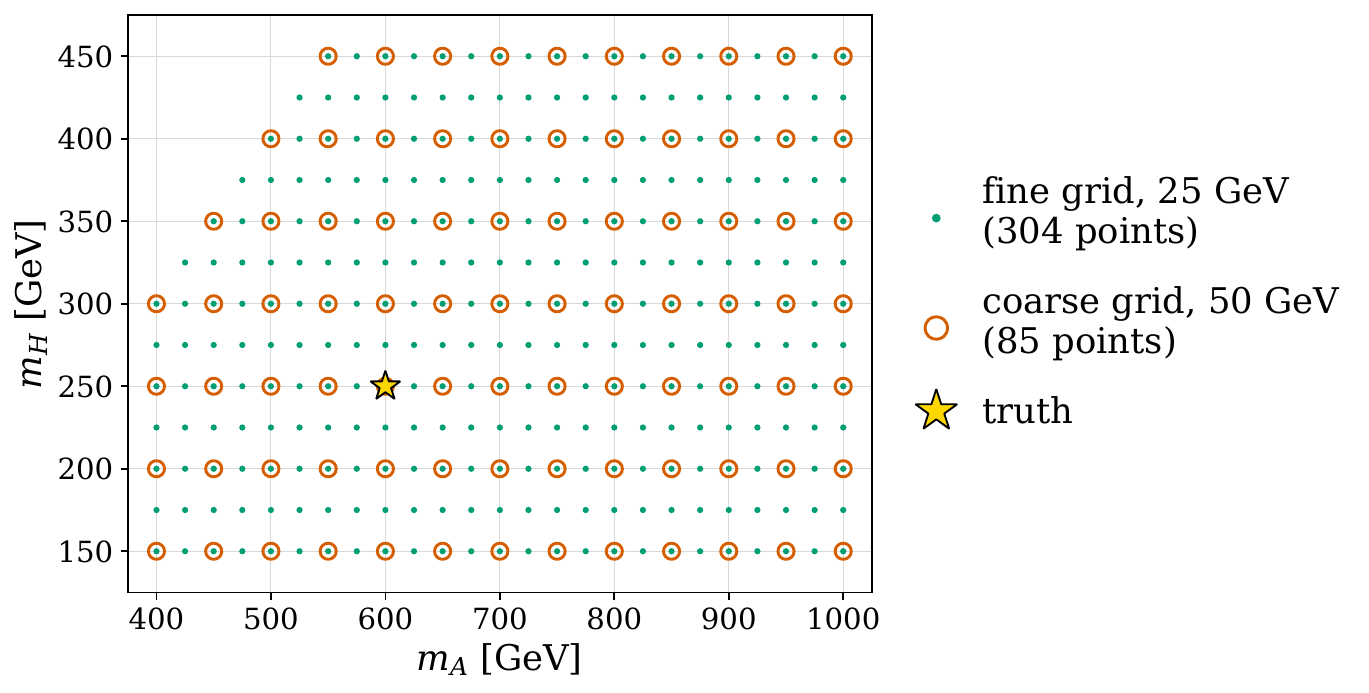}
    \caption{The two fixed $(m_A, m_H)$ grids the single shots train on, drawn on the same axes.}
    \label{fig:zhllbb_ssgrids}
\end{figure}
%%%%%%%%%%%%%%%%%%% figure %%%%%%%%%%%%%%%%%%%

For the single-round comparisons, we use two fixed grids over the range of \cref{eq:zhllbb_prior}, shown in \cref{fig:zhllbb_ssgrids}. The coarse grid uses the $85$ round-1 points at $50$~GeV spacing, with $3000$ signal and $3000$ background events per point. The fine grid contains $304$ points at $25$~GeV spacing, including the coarse grid, with $769$ events of each type per point. The total simulation used are $5.1\times10^5$ and $4.7\times10^5$ events, respectively, comparable to the $4.8\times10^5$ events used across the five sequential rounds.

%%%%%%%%%%%%%%
\subsection{Network architecture}
\label{sec:lgatr}

For the parton-level cascade study the conditional classifier is a fully connected network acting on the four-momenta and the conditioning parameters. At reconstruction level the observables $x$ are a set of reconstructed objects, each of which is a Lorentz four-vector, and the likelihood ratio we want to learn is an approximately Lorentz-invariant function of them. Rather than leaving the network to learn this approximate invariance from data, we build it into the architecture and use the Lorentz-equivariant geometric algebra transformer L-GATr~\cite{Spinner:2024hjm,Brehmer:2024yqw}, the Lorentz-group version of the geometric algebra transformer of Ref.~\cite{Brehmer:2023gatr}.

Each reconstructed object becomes one token of a transformer, embedded as an element of the geometric (Clifford) algebra of Minkowski space rather than as a plain list of numbers. We use the slim variant that keeps the scalar and vector grades, so that a token consists of one four-momentum plus a set of Lorentz scalars~\cite{Petitjean:2025zjf}. All network operations --- linear layers, the geometric product providing the non-linearity, and attention --- are equivariant under the proper orthochronous Lorentz group, so that a boost or rotation of the input event transforms the internal representations covariantly and an invariant output stays invariant by construction. Combined with the permutation equivariance of attention over tokens, this means neither symmetry has to be learned from training data, which is the reason equivariant networks are more sample-efficient than generic ones~\cite{Spinner:2024hjm}. Sample efficiency is precisely what we aim at with the sequential SBI approach.

The scalar channels of each token carry the non-kinematic information: the particle type and the conditioning parameters $\theta$ broadcast to every token, so that a single network represents the entire family of hypotheses. We also append constant reference vectors, so-called spurions --- the beam direction and the time direction --- as additional tokens to break exact Lorentz invariance~\cite{Spinner:2024hjm,Brehmer:2024yqw}. The network can then construct frame-dependent quantities such as rapidities while remaining equivariant relative to these references. The per-token scalar outputs are averaged over tokens into a single logit, which is trained with a binary cross-entropy loss. The inputs to the classifier are the four-momenta of the two leptons and the two $b$-jets.

\subsection{Evaluation}
We train the networks independently on the same data, and average their likelihood values during evaluation. The combined likelihood is profiled and evaluated on a lattice around the benchmark point,
\begin{align}
    m_A &\in [520,680]~\text{GeV} \;\; (50~\text{points}) \;, \notag \\
    m_H &\in [185,315]~\text{GeV} \;\; (50~\text{points}) \;, \notag \\
    \sigma\times\text{BR} &\in [0,\, 5\times10^{-2}]~\text{pb} \;\; (251~\text{points}) \;,
    \label{eq:zhllbb_grid}
\end{align}
all evenly spaced, i.e.\ at a spacing of $3.27\times2.65$~GeV in the mass plane. For the expectations in \cref{eq:mcsum}, we use $N_{\text{MC}}=8.8\times10^{4}$ signal events generated at $\theta_\text{true}$ and $N_{\text{MC}}=1.3\times10^{6}$ background events. The background sample is split into three disjoint parts, $35\%$ for training and $32.5\%$ each for calibration and evaluation. The $1.3\times10^{6}$ evaluation events are taken from the evaluation part, and the calibration $k$ in \cref{eq:ratio_calibration} is computed at every lattice point from another $4\times10^{5}$ background events of the calibration part. Neither set is used in training.

%%%%%%%%%%%%%%
\subsection{Results}
\subsubsection{Sequential SBI rounds}
\label{sec:zhllbb_rounds}

%%%%%%%%%%%%%%%%%% figure %%%%%%%%%%%%%%%%%%%
\begin{figure}[t]
    \centering
    \includegraphics[width=\textwidth]{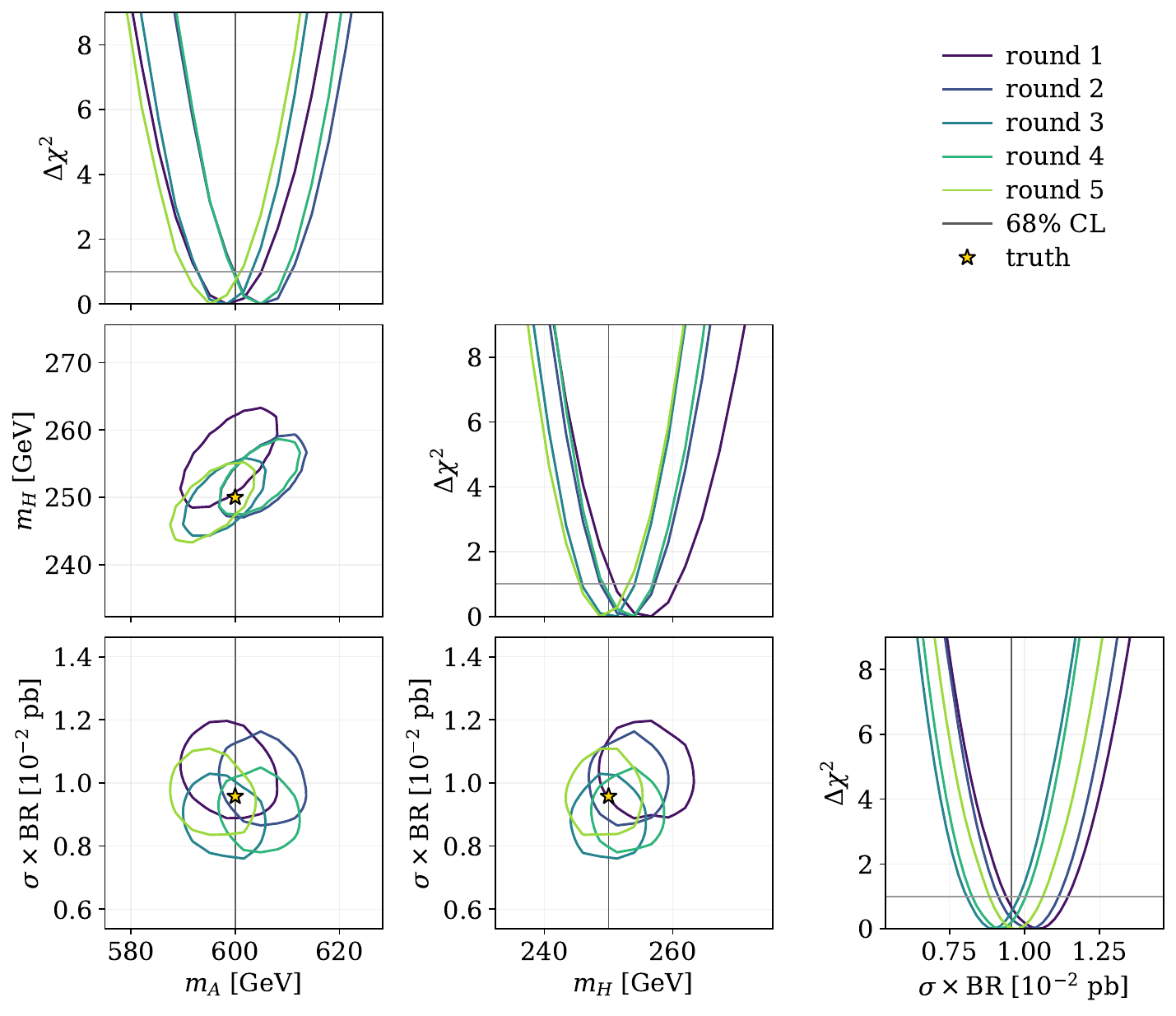}
    \caption{Five rounds of iterative SBI in the $A\to ZH\to\ell\ell b\bar b$ channel, shown over the full parameter set $(m_A, m_H, \sigma\times$BR$)$. Diagonal: the one-dimensional profile likelihood of each parameter, the other two profiled out. Below the diagonal: the $68\%$ CL region for each pair, the third parameter profiled out. Only the $68\%$ level is drawn, the $95\%$ regions are shown in \cref{fig:zhllbb_rounds95} in \cref{sec:zhllbb_95}. The star and the vertical lines mark the benchmark point.}
    \label{fig:zhllbb_rounds}
\end{figure}
%%%%%%%%%%%%%%%%%% figure %%%%%%%%%%%%%%%%%%%

\Cref{fig:zhllbb_rounds} shows how the $68\%$ confidence regions change with sequential rounds. The panels in the diagonal show the 1-D profile likelihoods and the lower panels show the three pairwise confidence regions of the parameters.  Round 1, trained on points spanning the full parameter range, gives the broadest confidence regions and is visibly shifted in $m_H$. Rhe $m_A-m_H$ and $m_H-\sigma\times$BR confidence regions exclude the truth in round 1. For the later rounds, the regions tighten around the benchmark point. A binned two-dimensional template analysis of the same two observables leads to much weaker limits as we demonstrate in \cref{sec:zhllbb_hist}. 

As an additional check, we also run a sequential SBI analysis using $N_S^\text{exp} = 0$ to test whether the algorithm potentially amplifies round~1 statistical fluctuations. We find this to not be the case. As expected, the proposals and the final likelihood are flat in the $m_A-m_H$ plane.

%%%%%%%%%%%%%%
\subsubsection{Sequential SBI versus single round}
\label{sec:zhllbb_single}

%%%%%%%%%%%%%%%%%% figure %%%%%%%%%%%%%%%%%%%
\begin{figure}[t]
    \centering
    \includegraphics[width=\textwidth]{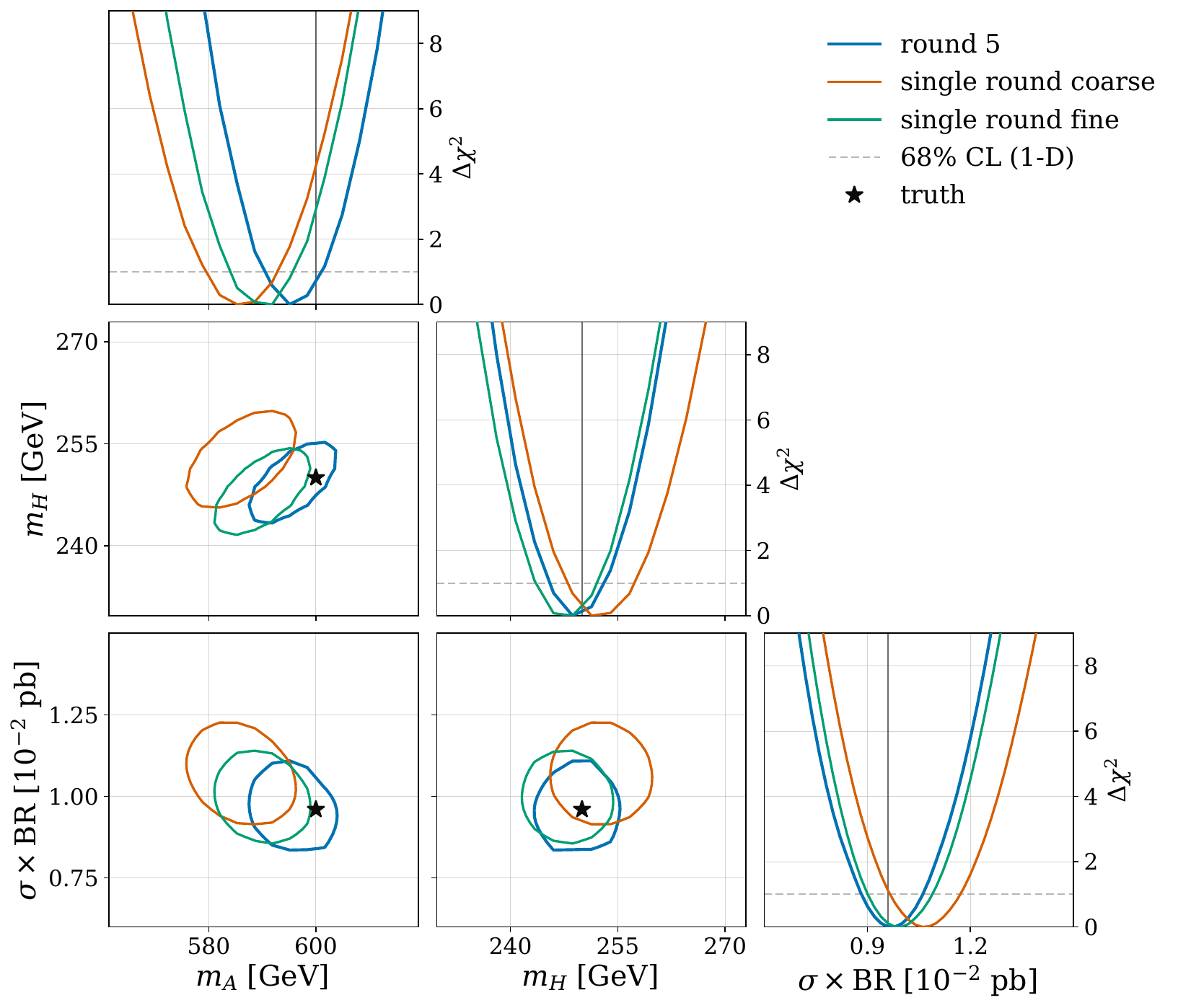}
    \caption{Same as \cref{fig:zhllbb_rounds} but the round-5 sequential rounds result is compared against single round training on a fixed grid. Only the $68\%$ confidence level region is drawn; the $95\%$ regions are shown in \cref{fig:zhllbb_single95} in \cref{sec:zhllbb_95}.}
    \label{fig:zhllbb_single}
\end{figure}
%%%%%%%%%%%%%%%%%% figure %%%%%%%%%%%%%%%%%%%

\begin{figure}[t]
    \centering
    \includegraphics[width=\textwidth]{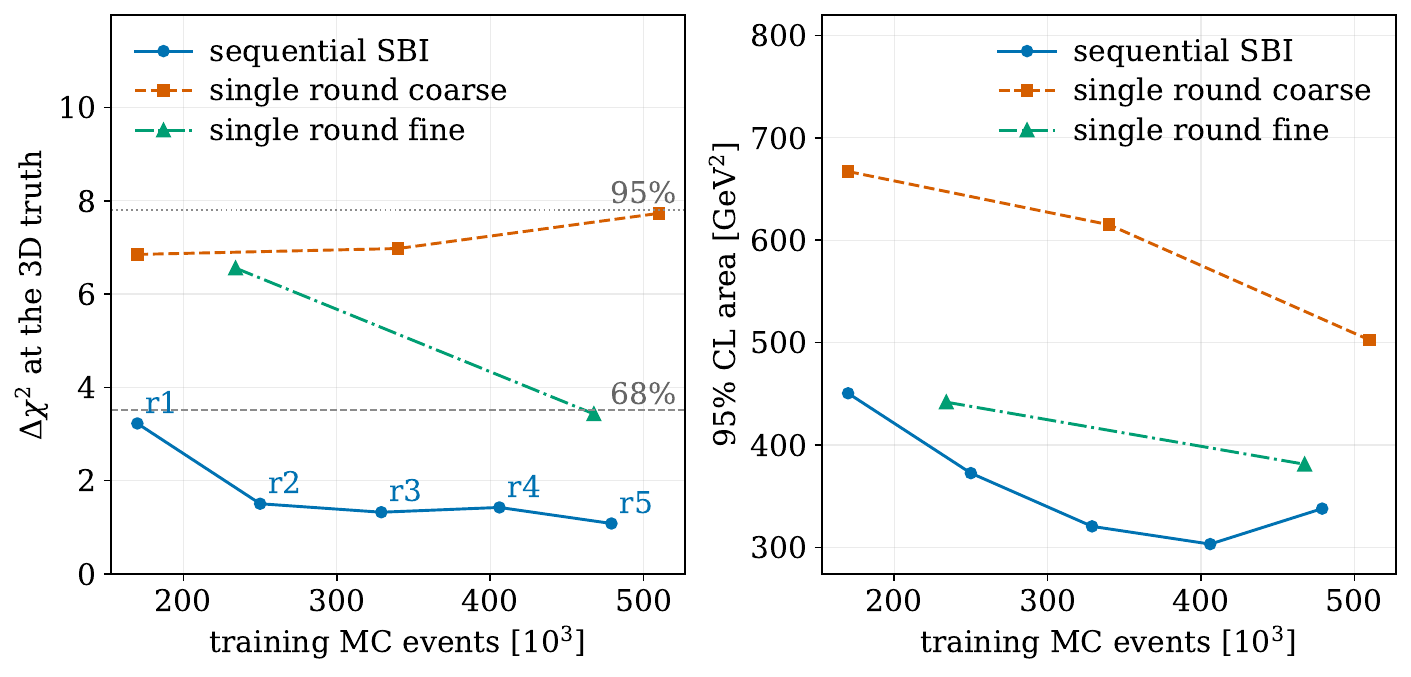}
    \caption{Left: $\Delta\chi^2$ at the truth point in the three-dimensional $(m_A,m_H,\sigma\times\text{BR})$ space as a function of the amount of MC events used in training comparing the various likelihood ratio estiamtors. Right: the area of the $95\%$ CL region in the $(m_A, m_H)$ plane as a function of the amount of MC events used in training.}
    \label{fig:zhllbb_budget}
\end{figure}

Next, we compare the results from the sequential rounds with two single round configurations both using a comparable number of simulated events. In \cref{fig:zhllbb_budget}, we show the comparison of the $68\%$ confidence regions of the sequential rounds against these single rounds. We observe that the single-round results struggle to accurately recover the truth point. Both are outperformed by the sequential SBI result.

%%%%%%%%%%%%%%%%%% figure %%%%%%%%%%%%%%%%%%%
%%%%%%%%%%%%%%%%%% figure %%%%%%%%%%%%%%%%%%%

We further quantify this comparison in \cref{fig:zhllbb_budget}, which compares the achieved precision both against the used simulation budget. In the left panel, we show the $\Delta\chi^2$ at the truth point in the three-dimensional $(m_A,m_H,\sigma\times\text{BR})$ space as a function of the amount of MC events used in training. The right panel shows the area of the 95\% CL area in the $(m_A,m_H)$ which we use as a measure of the constraining power of the results. Here, we in addition to the single round results discussed above also show single round results trained on the same parameter grid but with less MC events per parameter point. As visible in the left panel, every sequential SBI round lies inside the $68\%$ region of the three-dimensional truth, while all three coarse grid single round results lie outside it.  For the coarse grid, more simulation events on the grid does not move the fit towards the injected point. The fine grid single round results perform better. Both the $\Delta\chi^2$ at the truth point  and the confidence region area get better with the number of training events. The sequential rounds consistently provide a stronger constraint  in the $(m_A,m_H)$ plane in comparison to the single round results.

%%%%%%%%%%%%%%%%%% figure %%%%%%%%%%%%%%%%%%%
\begin{figure}[t]
    \centering
    \includegraphics[width=0.75\textwidth]{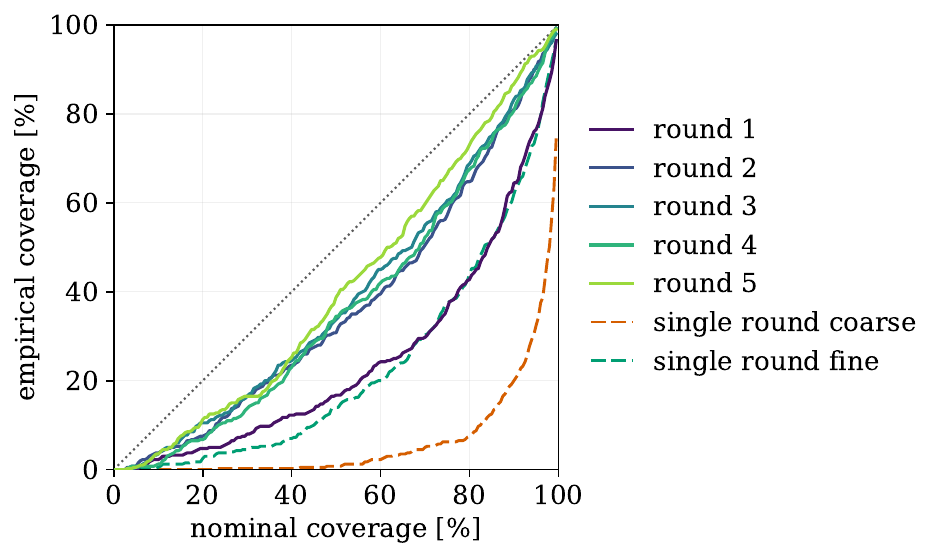}
    \caption{Empirical against nominal coverage of the three-parameter region in $(m_A, m_H, \sigma\times\text{BR})$, for the five sequential rounds and for the two single round runs. The dotted diagonal corresponds to perfect calibration; results below it are overconfident.}
    \label{fig:zhllbb_coverage}
\end{figure}
%%%%%%%%%%%%%%%%%% figure %%%%%%%%%%%%%%%%%%%

%%%%%%%%%%%%%%
\subsubsection{Coverage}
\label{sec:zhllbb_coverage}

We evaluate the empirical coverage as defined in \cref{eq:coverage} by creating pseudo datasets with numbers of signal and background events drawn from Poisson distributions with means $N^{\text{exp}}_{S}$ and $N^{\text{exp}}_{B}$. \cref{fig:zhllbb_coverage} shows the emperical coverage versus nominal coveral based on $400$ pseudo datasets.

We find the full three-dimensional confidence region of round~1 of the chain to be overconfident . From round~2 the overconfidence is reduced and the coverage is close to its nominal value even though still slighlty overconfident. The sequential rounds therefore not merely tighten the contours but also help the likelihood ratio estimator to converge towards the true confidence region. In contrast, the single round runs are far worse, in particular the coarse grid result. The fine grid result is approximately as overconfident as round~1 of the iterative chain.

%%%%%%%%%%%%%%%%%%%%%%%%%%%%%%%%%%%%%%%%%%%%%%%%%%
%%%%%%%%%%%%%%%%%%%%%%%%%%%%%%%%%%%%%%%%%%%%%%%%%%
\section{Reco-level example II: boosted resonance cascade \texorpdfstring{$Y\to HH\to 4t$}{Y -> HH -> 4t}}
\label{sec:ytohh4t}

As a second reconstruction-level example, we consider the boosted cascade $pp \to Y \to HH \to 4t$, where each $H \to t\bar t$ decay is reconstructed as a single large-radius jet. We allow the masses and widths of both resonances to vary, so that the classifier is conditioned on four theory parameters. Along with the signal cross section, the likelihood depends on five parameters. Our results in this section show that sequential SBI can be extended to higher-dimensional parameter spaces at the reconstruction level.

\subsection{Target Process}
\label{sec:ytohh4t_signal}

The signal process is a resonant cascade producing four top quarks,
\begin{align}
pp \to Y \to H H \to (t\bar t)(t\bar t) ;,
\label{eq}
\end{align}
where a heavy resonance $Y$ decays to a pair of heavy scalars $H$, each subsequently decaying to $t\bar t$. We take $Y$ to be the first Kaluza--Klein excitation of the graviton in a warped extra-dimensional model with Standard Model fields propagating in the bulk~\cite{Randall:1999ee,Agashe:2007zd}. We use the signal model and event-generation setup used in Ref.~\cite{Das:2026spe}. The same signal was also used in a recent CMS search~\cite{CMS:2024nsz}.

We infer the two masses and the two relative widths,
\begin{equation}
    \theta = \left( m_Y,\; \Gamma_Y/m_Y,\; m_H,\; \Gamma_H/m_H \right),
    \label{eq:yhh_theta}
\end{equation}
using an observed dataset generated at
\begin{equation}
    \theta_{\mathrm{true}}
    = \left( 3.5~\mathrm{TeV},\; 0.05,\; 400~\mathrm{GeV},\; 0.05 \right).
    \label{eq:yhh_theta_true}
\end{equation}
At $\theta_\text{true}$ we inject a signal rate of $\sigma\times\text{BR}=1.09\times10^{-2}$~pb, which corresponds to $N^{\text{exp}}_{S}=1409$ selected signal events at $\mathcal{L}=139~\text{fb}^{-1}$, see \cref{sec:likelihood}.

\subsection{Fitting the rate}
\label{sec:likelihood}

As in the $A\to ZH$ analysis, we infer the parameters $\theta$ in $\eqref{eq:yhh_theta}$
together with the signal rate,
\begin{equation}
    \mu = \sigma(pp\to Y)\times\mathrm{BR}(Y\to HH)
          \times\mathrm{BR}(H\to t\bar t)^2.
\end{equation}
The signal cross section after selection is
$\sigma_S(\theta,\mu)=\mu\,\epsilon(\theta)$, where
$\epsilon(\theta)$ is the selection efficiency which varies between $0.34$ and $0.98$ over the initial range of parameters.

For this process, the background rate is hard to predict theoretically implying that the total rate constraint is subject to large systematic uncertainties. So, we use the shape-only likelihood of \cref{eq:shape_likelihood}
rather than the extended likelihood of \cref{eq:full_lr_bonly}. So the inference uses the observed kinematic distributions
without using the predicted total event yield. We use $\sigma_B=4.91$~pb from the generated QCD cross section including its selection efficiency. For $\mathcal{L} = 139~\text{fb}^{-1}$, the background yield is
\begin{align}
    N^{\text{exp}}_{B} = \mathcal{L}\,\sigma_B = 6.82\times10^{5} \;.
\end{align}
At $\theta_{\text{true}}$, $\epsilon = 0.93$. We inject
\begin{align}
    N^{\text{exp}}_{S} = 1409 \;,
    \qquad\text{i.e.}\qquad
    \sigma\times\text{BR} = 1.09\times10^{-2}~\text{pb} \;.
\end{align}
At this yield, a simple cut-and-count analysis in the signal region
$3300<m_{JJ}<3700$~GeV would give a naive significance of $2.8\,\sigma$.

\subsection{Simulation and event selection}
\label{sec:ytohh4t_sim}
Signal events are generated for $pp$ collisions at $\sqrt{s}=13$~TeV
with \textsc{MadGraph5\_aMC@NLO}~\cite{Alwall:2014hca} using the
bulk-RS UFO model~\cite{Degrande:2011ua,Alloul:2013bka}, then showered
and hadronized with \textsc{Pythia\,8}~\cite{Sjostrand:2014zea}.
QCD dijet background events are generated with \textsc{Pythia\,8}
with $\hat{p}_T>500$~GeV. Both samples use
\textsc{Delphes\,3}~\cite{deFavereau:2013fsa} with the LHC Olympics
R\&D settings~\cite{Kasieczka:2021xcg}. Jets are clustered with
anti-$k_T$~\cite{Cacciari:2008gp} at $R=1.0$ using
\textsc{FastJet}~\cite{Cacciari:2011ma}. Following the LHC Olympics
selection, we require at least two jets with $|\eta|<2.5$ and
a leading jet with $p_T>1.2$~TeV. The two highest-$p_T$ jets
are then ordered by mass.

For sequential SBI, similar to~\cref{sec:dijet}, the first round uses $300$ Sobol points~\cite{SOBOL} distributed across the
parameter ranges.
\begin{align}
m_Y &\in [2.5,5]~\mathrm{TeV}, &
m_H &\in [350,600]~\mathrm{GeV}, \\
\Gamma_Y/m_Y &\in [0.015,0.25], &
\Gamma_H/m_H &\in [0.03,0.25].
\end{align}
We require $m_Y\geq 5m_H$, ensuring that the scalars $H$ are produced with a substantial boost. \cref{sec:grid_designs} compares this design with
independent uniform sampling and a regular lattice. Each subsequent round adds $300$ parameter points drawn from \cref{eq:sequential_proposal_freq} with $\gamma=0.5$. At each point,
we add $500$ signal and $500$ background events, giving a cumulative training sample of $3\times10^5\times t$ events at round~$t$. We describe how the new parameter points are selected in \cref{sec:ytohh4t_proposals}.

We compare sequential SBI with two single-round baselines, each using the same $1.8\times10^6$ simulated events as the six sequential rounds. Both grids cover the full parameter range and are shown in \cref{fig:ytohh4t_ssgrids}. The coarse grid uses the $300$ round-1 Sobol points with $3000$ signal and $3000$ background events per point; the fine grid uses the first $1800$ points of the same sequence with $500$ events of each type per point, including the coarse grid as its first $300$ points.

%%%%%%%%%%%%%%
\subsection{Network architecture and hyperparameters}
\label{sec:ytohh4t_net}

The L-GATr classifier of \cref{sec:lgatr} takes the two large-$R$
jets as input tokens, each conditioned on the four parameters
$\theta$. We train four classifiers independently on the same data
for $60$ epochs, using a cosine learning-rate schedule with an
initial learning rate of $10^{-3}$, and average their likelihood
surfaces for all results in this section. At each epoch, background
events are assigned parameter values sampled with replacement from
the signal parameter values. This gives both classes the same
parameter distribution, following step~2 of \cref{sec:method}.

%%%%%%%%%%%%%%%%%% figure %%%%%%%%%%%%%%%%%%%
\begin{figure}[h]
    \centering
    \includegraphics[width=\textwidth]{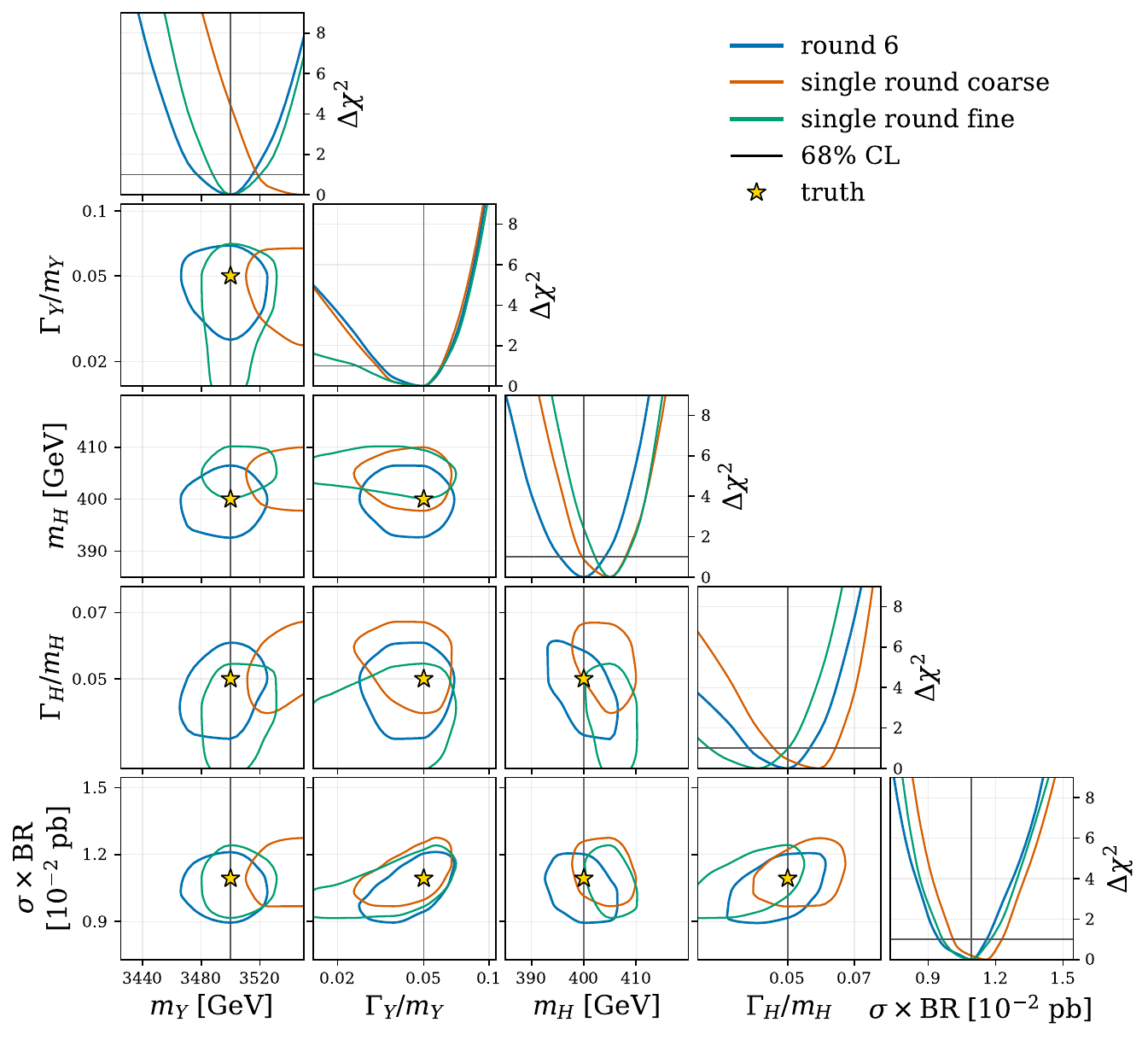}
    \caption{Round-6 sequential rounds result is compared against single round training on a fixed coarse and fine grid with the same simulation budget. We show only the $68\%$ CL regions.}
    \label{fig:ytohh4t_single68}
\end{figure}
%%%%%%%%%%%%%%%%%% figure %%%%%%%%%%%%%%%%%%%
\subsection{Evaluation}
\label{sec:ytohh4t_eval}
All likelihoods are evaluated and profiled on a common grid around $\theta_{\text{true}}$,
\begin{align}
    m_Y &\in [3425,3550]~\text{GeV} \;\; (6~\text{points}) \;, &
    \Gamma_Y/m_Y &\in [0.0154,0.134] \;\; (11~\text{points}) \;, \notag \\
    m_H &\in [375,425]~\text{GeV} \;\; (11~\text{points}) \;, &
    \Gamma_H/m_H &\in [0.0318,0.08] \;\; (7~\text{points}) \;, \notag \\
    \sigma\times\text{BR} &\in [0.546,2.19]\times10^{-2}~\text{pb} \;\; (25~\text{points}) \;.
    \label{eq:ytohh4t_grid}
\end{align}
The masses are evenly spaced in steps of $25$~GeV and $5$~GeV, while the relative widths and $\sigma\times$BR are spaced logarithmically, the latter from half to twice the injected rate. The injected parameter values are included in the grid. This gives $5{,}082$ points in the four-dimensional shape-parameter space.
As in the $A\to ZH$ analysis, event sums are replaced by their expectations under the injected hypothesis, \cref{eq:mcsum}. We use $N_\text{MC}=1.1\times10^5$ signal events generated at $\theta_\text{true}$ and $N_\text{MC}=1.5\times10^6$ background events withheld from training. The same background events are used to compute the calibration in \cref{eq:ratio_calibration} at each grid point.

%%%%%%%%%%%%%%
\subsection{Results}
\subsubsection{Comparing sequential SBI to single rounds}
\label{sec:ytohh4t_single}

%%%%%%%%%%%%%%%%%% figure %%%%%%%%%%%%%%%%%%%

%%%%%%%%%%%%%%%%%% figure %%%%%%%%%%%%%%%%%%%

We run six sequential SBI rounds. \Cref{fig:ytohh4t_single68} compares the one-dimensional profile
likelihoods and two-dimensional $68\%$ confidence regions from
sequential SBI round~6 to those from the coarse and fine single-round
analyses. The sequential SBI best fit coincides with the injected point,
and its contours contain the true parameter values. The single-round on a coarser grid with more data, instead favours
larger $m_Y$, with several contours extending beyond the evaluation
grid. The single-round on a finer grid constrains $m_Y$ well but
favours a larger $m_H$ and smaller relative widths. Its $68\%$
region in the mass plane is smaller than that of sequential SBI, but excludes the injected point.
Thus, the tighter single-round contours do not imply more accurate
inference. We assess the statistical reliability of these regions with the coverage tests in the next section.  \cref{sec:ytohh4t_rounds} shows how the $68\%$ and $95\%$ confidence regions change with sequential SBI rounds. \Cref{sec:ytohh4t_single_app} compares the $95\%$ confidence regions from sequential rounds to single rounds with coarse and fine grids. \Cref{sec:ytohh4t_hist} compares sequential SBI with a binned baseline.

%%%%%%%%%%%%%%
\subsubsection{Coverage}
\label{sec:ytohh4t_coverage}

%%%%%%%%%%%%%%%%%% figure %%%%%%%%%%%%%%%%%%%
\begin{figure}[t]
    \centering
    \includegraphics[width=0.75\textwidth]{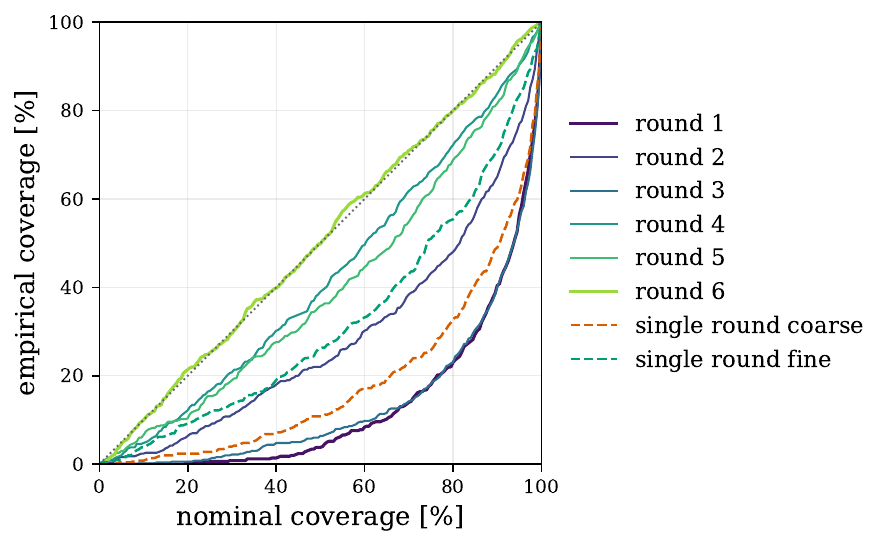}
    \caption{Empirical against nominal coverage of the full five-parameter region in $(m_Y, \Gamma_Y/m_Y, m_H, \Gamma_H/m_H, \sigma\times\text{BR})$, for rounds~1 to~6 of the chain and the coarse ($300$ points with $3000$ events each) and fine ($1800$ points with $500$ events each) single-round runs, from $500$ pseudo-experiments each. The dotted diagonal corresponds to perfect calibration; results below it are overconfident.}
    \label{fig:ytohh4t_coverage}
\end{figure}
%%%%%%%%%%%%%%%%%% figure %%%%%%%%%%%%%%%%%%%

We evaluate the joint coverage of all five parameters using
\cref{eq:coverage}, with $500$ pseudo-experiments for each analysis.
For each pseudo-experiment, we draw
$N_S\sim\operatorname{Pois}(N^{\text{exp}}_{S})$ and
$N_B\sim\operatorname{Pois}(N^{\text{exp}}_{B})$ with $N^{\text{exp}}_{S}=1409$ and $N^{\text{exp}}_{B}=6.82\times10^5$, sample the corresponding
signal and background events from the evaluation datasets, and
recompute the likelihood with the trained networks and calibration
held fixed.

\Cref{fig:ytohh4t_coverage} compares the empirical coverage with
the nominal confidence level. Rounds~1--3 substantially undercover:
their nominal $68\%$ regions contain the injected parameter point
in only $13\%$--$35\%$ of pseudo-experiments. Coverage improves
in rounds~4 and~5, although both remain below the diagonal.
By round~6, the empirical coverage closely follows the nominal
level across the range shown, reaching $69\%$ and $96\%$ at
nominal levels of $68\%$ and $95\%$, respectively. In contrast,
the fine single-round analysis reaches only $41\%$ and $83\%$,
and the coarse analysis $21\%$ and $61\%$. Sequential SBI therefore
substantially improves the calibration of the joint confidence
regions, with round~6 consistent with nominal coverage at the
injected point.

%%%%%%%%%%%%%%

\FloatBarrier
%%%%%%%%%%%%%%%%%%%%%%%%%%%%%%%%%%%%%%%%%%%%%%%%%%
\section{Conclusions}
\label{sec:conclusions}

Simulation-based inference is becoming an increasingly popular way of analysing LHC data, because it uses the information in the full high-dimensional phase space instead of a binned, low-dimensional summary statistic. So far, LHC applications have almost exclusively targeted couplings. There, the squared matrix element is a polynomial in the theory parameters, so the continuous parameter dependence of the likelihood ratio can be reconstructed from a small set of learned basis ratios or derivatives limiting the amount of simulated MC events needed. The same approach does not apply to masses and widths, which enter through propagators and shift or deform resonant structures in phase space. A parameter-conditional estimator must learn this dependence from simulations distributed across the parameter space. Covering that space uniformly is what makes inference on resonances expensive, and most of it is irrelevant once the data has been observed.

In this paper, we have adapted sequential SBI to classifier-based likelihood-ratio estimation at the LHC. In each round, we evaluate the current likelihood on the observed data, turn it into a proposal over the theory parameters, simulate the proposed parameters, and retrain on all the accumulated simulations. The final likelihood can be used for frequentist limit setting in the same way as a regular SBI analysis.

For the parton-level process $u\bar u\to Z'\to X\bar X\to 4q$, sequential training gave confidence regions similar to the exact matrix-element likelihood ratio. Both single-round analyses with the same simulation budget --- one using more events per parameter point, the other a finer grid --- recovered the two masses but failed to recover the relative $Z'$ width and the coupling. We then analysed reconstruction level, including parton shower, hadronization,
and a fast detector simulation. We first considered the heavy-scalar cascade $A\to ZH\to\ell\ell b\bar b$, where we jointly inferred three parameters: the masses $m_A$, $m_H$ and the signal rate. Using a Lorentz-equivariant transformer, we learned the likelihood ratio with only $\sim 2.4\times10^5$ simulated signal events. Then, we studied $Y\to HH\to 4t$, where we jointly inferred five parameters --- the masses $m_Y$ and $m_H$, the relative widths $\Gamma_Y/m_Y$ and $\Gamma_H/m_H$, and the signal rate --- with $\sim 9\times10^5$ simulated signal events. For reference, \cite{Cheng:2024yig} used $\sim 10^7$ simulations to infer three parameters for a single resonant mass window. 

For both reconstruction-level analyses, we show that sequential SBI gives tighter confidence regions than SBI at the same simulation budget. They also resulted in tighter confidence regions than a classical histogram based analysis used traditionally in experimental searches~\cite{ATLAS:2020gxx,CMS:2024yiy}. Furthermore we demonstrate through coverage tests that it gives better-calibrated confidence regions than single round SBI. Our results show that sequential SBI can be a practical approach to resonance-parameter inference for simulation based analyses.

%%%%%%%%%%%%%%%%%%%%%%%%%%%%%%%%%%%%%%%%%%%%%%%%%%
\section*{Acknowledgements}

We thank Tilman Plehn for useful discussions. RD acknowledges support from the Alexander von Humboldt Foundation. RD and HB acknowledge the use of ChatGPT and Claude for developing the code and in the writing of this paper.

%%%%%%%%%%%%%%%%%%%%%%%%%%%%%%%%%%%%%%%%%%%%%%%%%%
%%%%%%%%%%%%%%%%%%%%%%%%%%%%%%%%%%%%%%%%%%%%%%%%%%
\appendix
\crefalias{section}{appendix}
\crefalias{subsection}{appendix}

\section{Further results for toy model}
\label{sec:cascade_app}

\subsection{Proposal updates}
\label{sec:cascade_proposals}

From round~2 on, the parameter points of each round are drawn from the proposal in \cref{eq:sequential_proposal_freq} with $\gamma=0.5$, constructed from the previous round's likelihood. We sample it with the affine-invariant ensemble sampler \textsc{emcee}~\cite{Foreman-Mackey:2012any} over the box of \cref{eq:cascade_prior} together with $g\in[0.1,1]$, using $64$ walkers, $1500$ burn-in and $8000$ production steps, and evaluating the likelihood on $5\times10^4$ signal and $5\times10^4$ background events. From the resulting proposal, we keep $800$ randomly chosen samples and drop the coupling, since all signal events are generated at unit coupling. For $15\%$ of these points, both relative widths are redrawn log-uniformly over $[0.01,0.3]$, so that each round keeps generating over the full width range even if the previous round's likelihood is overconfident.

\subsection{$95 \%$ confidence regions}
\label{sec:contours_parton_app}
\begin{figure}[!htbp]
    \centering
    \includegraphics[width=\textwidth]{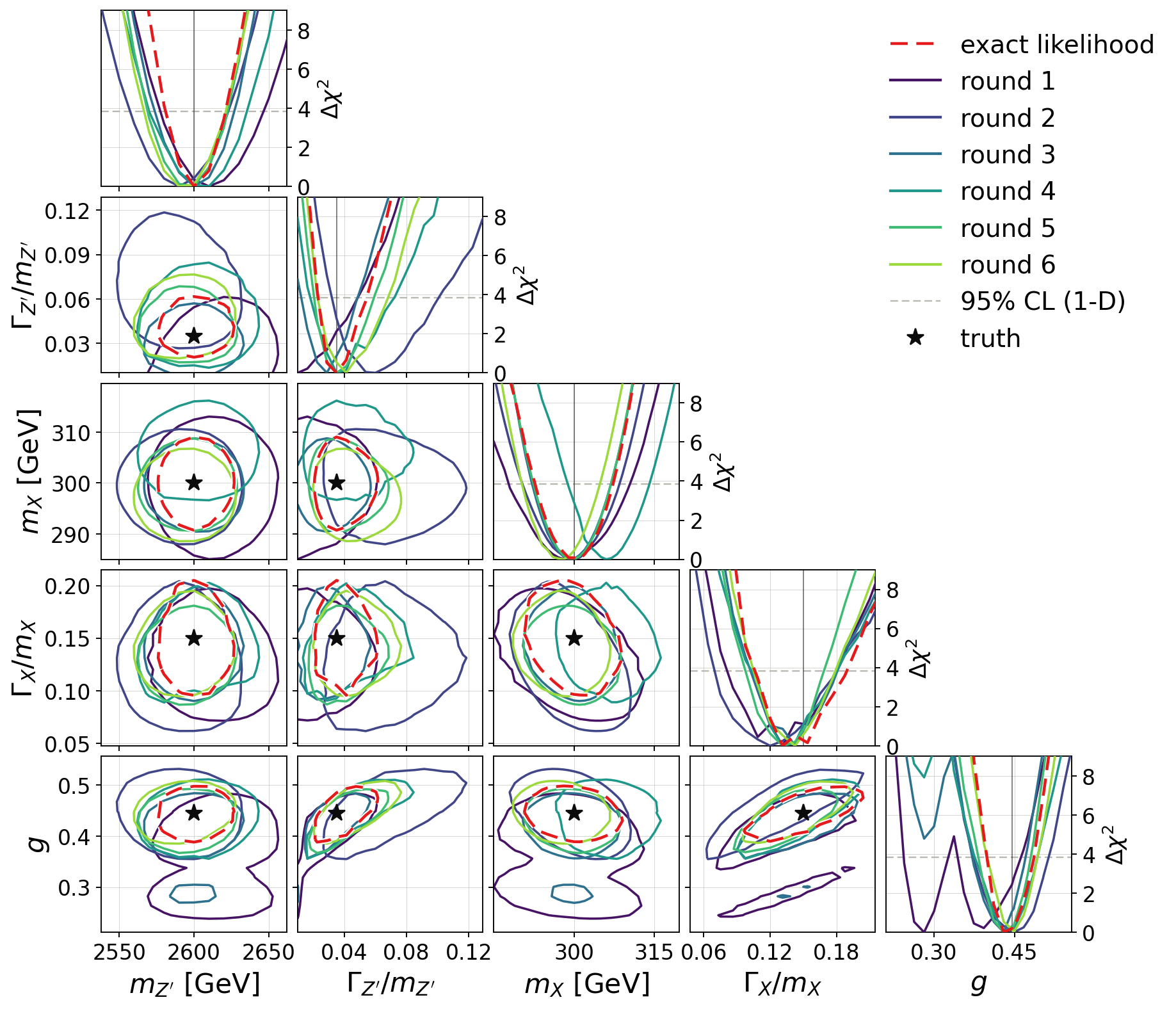}
    \caption{ $95\%$ confidence region versus sequential rounds for the parton-level study described in \cref{sec:dijet}. }
    \label{fig:cascade_rounds_95}
\end{figure}

\begin{figure}[!htbp]
    \centering
    \includegraphics[width=\textwidth]{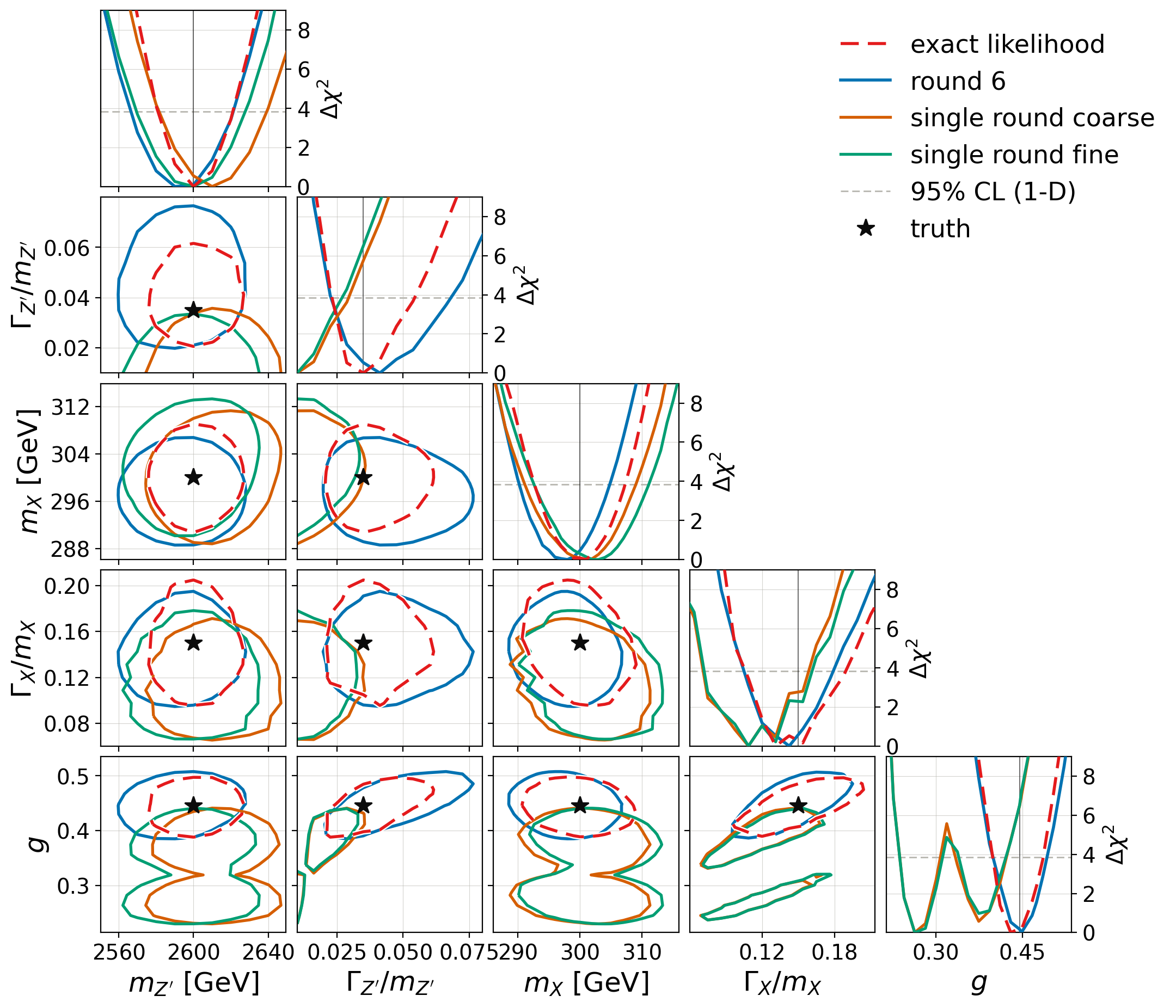}
    \caption{For the parton level study in \cref{sec:dijet}, this figure compares the $95\%$ CL region for sequential SBI to single rounds with the same amount of simulations.}
    \label{fig:cascade_single_95}
\end{figure}

In this subsection we show the $95\%$ confidence regions for the parton level study. ~\cref{fig:cascade_rounds_95} shows how the confidence regions change with sequential rounds. Similar to ~\cref{fig:cascade_rounds_68}, we see that the $95 \%$ confidence regions are broader and develop many suprious local minima in the earlier rounds, and regions become tighter in the later rounds. \cref{fig:cascade_single_95} shows how these confidence regions compare between sequential SBI and single rounds with the same amount of simulation. Similar to \cref{fig:cascade_single_68}, we see that the $95\%$ confidence regions are tighter and closer to the confidence regions from the exact likelihood, whereas the single rounds result in confidence regions that can exclude the benchmark point, as seen in many of the two-dimensional panels.

%%%%%%%%%%%%%%%%%% figure %%%%%%%%%%%%%%%%%%%
%%%%%%%%%%%%%%%%%% figure %%%%%%%%%%%%%%%%%%%
%%%%%%%%%%%%%%%%%% figure %%%%%%%%%%%%%%%%%%%
%%%%%%%%%%%%%%%%%% figure %%%%%%%%%%%%%%%%%%%

% \FloatBarrier
%%%%%%%%%%%%%%
\section{Further results for \texorpdfstring{$A\to ZH\to\ell\ell b\bar b$}{A -> ZH -> llbb}}
\label{sec:zhllbb_app}

%%%%%%%%%%%%%%
\subsection{Proposal updates}
\label{sec:zhllbb_proposals}

From round~2 onwards, we construct the proposal from the previous round's likelihood. We temper it with $\gamma=0.8$, following \cref{eq:sequential_proposal_freq}. We draw $4000$ candidate points by selecting cells with probability proportional to $L^\gamma$ and sampling uniformly within each cell. After rounding to integer GeV, we remove duplicates, the benchmark point, and points where $A\to ZH$ is kinematically forbidden. We retain the first $34$ valid points and add $6$ uniformly sampled points to maintain exploration across the full parameter range. We then remove duplicates and previously simulated points, slightly reducing the number of new points in later rounds.

\begin{figure}[H]
    \vspace{-0.6\intextsep}% fixed in place under fig:cascade_rounds_95; tighten so the caption clears the footer
    \centering
    \includegraphics[width=\textwidth]{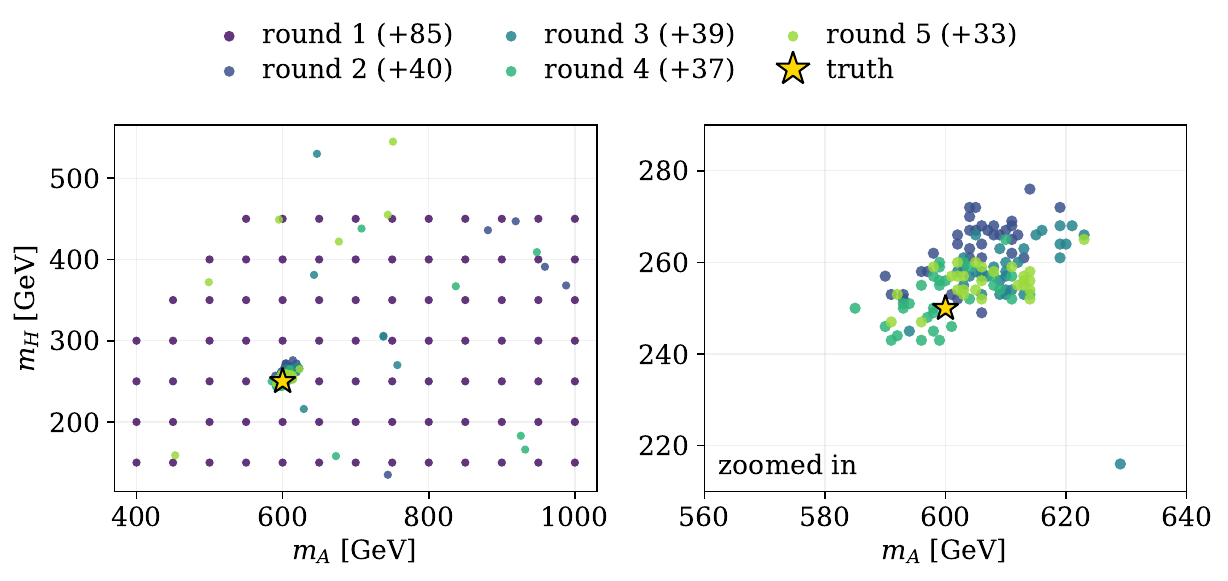}
    \caption{Left: The parameter points the sequential rounds trained on in \cref{sec:zhllbb}, coloured by the round that added them. Right: The same as the left panel, but zoomed on the benchmark point. }
    \label{fig:zhllbb_grids}
    \vspace{-0.6\intextsep}
\end{figure}

\Cref{fig:zhllbb_grids} shows the points added in each round. From round~2 onwards, the likelihood-based proposal concentrates them within a few tens of GeV of the injected point.

%%%%%%%%%%%%%%%%%% figure %%%%%%%%%%%%%%%%%%%
%%%%%%%%%%%%%%%%%% figure %%%%%%%%%%%%%%%%%%%

%%%%%%%%%%%%%%
\subsection{Binned baseline}
\label{sec:zhllbb_hist}

%%%%%%%%%%%%%%%%%% figure %%%%%%%%%%%%%%%%%%%
\begin{figure}[!htbp]
    \centering
    \includegraphics[width=\textwidth]{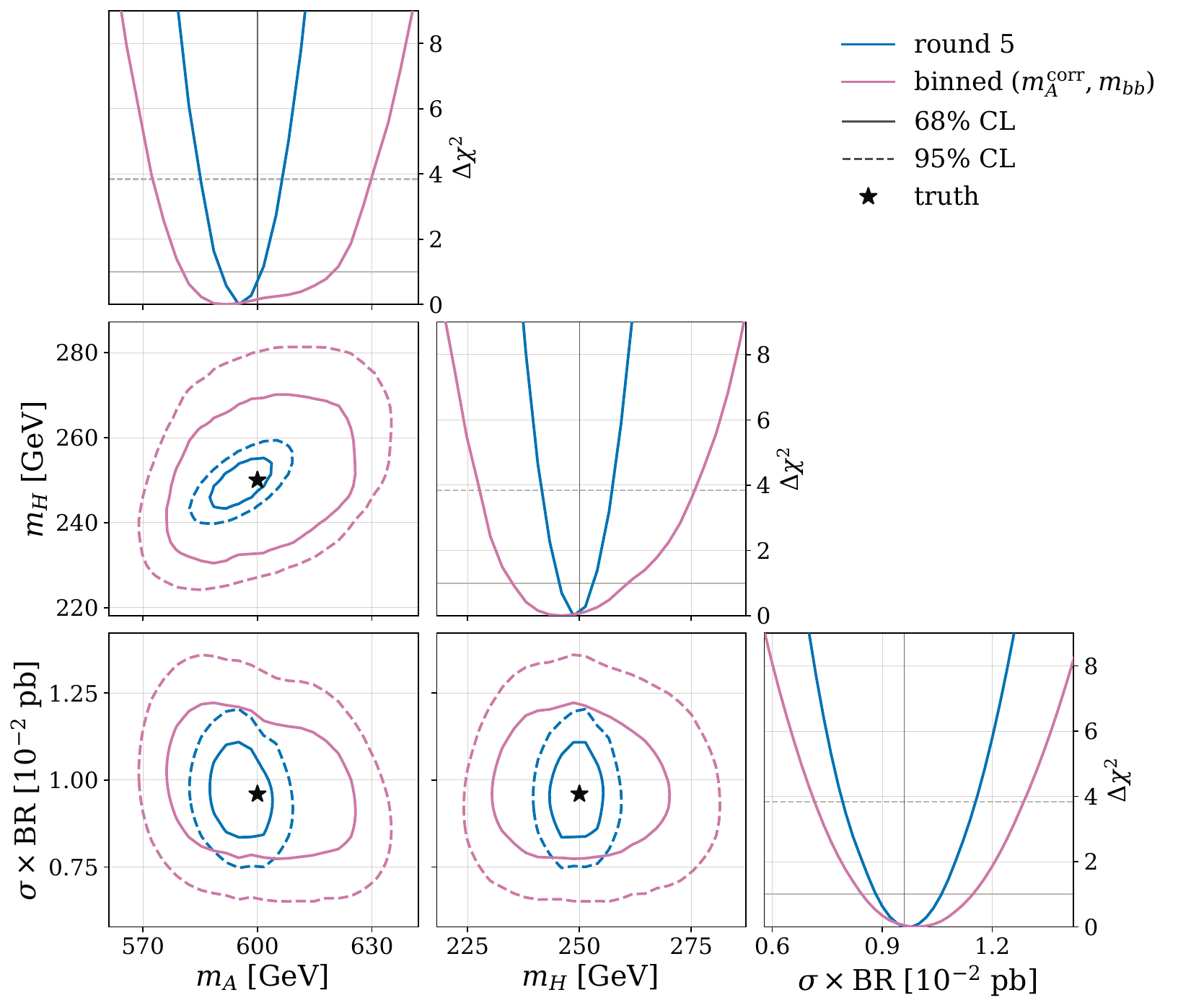}
    \caption{The learned ratio at round~5 compared against the binned $(m_A^{\rm corr}, m_{bb})$ template, in the same corner as \cref{fig:zhllbb_rounds,fig:zhllbb_single}. Solid and dashed lines correspond to the $68\%$ and $95\%$ CL regions, the star marks the injected point. Both analyses are evaluated on the same observed events and differ only in the per-event discriminant.}
    \label{fig:zhllbb_hist_corner}
\end{figure}
%%%%%%%%%%%%%%%%%% figure %%%%%%%%%%%%%%%%%%%

The strongest natural binned analysis for this channel is a two-dimensional $(m_A^{\rm corr}, m_{bb})$ template --- a discriminant similar to the one used in the experimental search~\cite{ATLAS:2020gxx}. We use two dimensions so that the information about the daughter mass is kept. We bin the two observables of \cref{fig:zhllbb_observables} into $20\times10$ bins of $50$~GeV, of the order of the reconstruction resolution, replace the NN-base likelihood ratio by the per-bin ratio of expected signal and background events. We insert it into the same extended likelihood of \cref{eq:full_lr_bonly}, on the same mass lattice and the same $\sigma\times$BR axis. The templates are built from the iterative chain's $234$ generated parameter points and are calibrated/evaluated on the same events as the learned ratio. Consequently, the two analyses differ only in the discriminant.

To avoid simulating even more MC events, we morph the signal template at a scanned $(m_A, m_H)$ rather than copied from the nearest generated point. I.e., we average the $k$ nearest template weighting by the distance to the $(k\!+\!1)$-th choosing $k=32$ to suppress noise in the profile likelihood. This plays the role that moment morphing~\cite{Baak:2014fta} plays in the experimental analyses, at the price of being purely local in the mass plane. We checked that the effect on the sensitivity is negligible.

\Cref{fig:zhllbb_hist_corner} compares the two analyses. The binned fit recovers the rate and both masses within its 68\% and 95\% confidence regions. The confidence region of the binned analysis is significantly wider in both mass directions. This shows that the SBI analysis is able to exploit the full phase-space information and does not loose information due to binning.

%%%%%%%%%%%%%%
\subsection{\texorpdfstring{$95\%$ CL regions}{95\% CL regions}}
\label{sec:zhllbb_95}

\begin{figure}[!htbp]
    \centering
    \includegraphics[width=\textwidth]{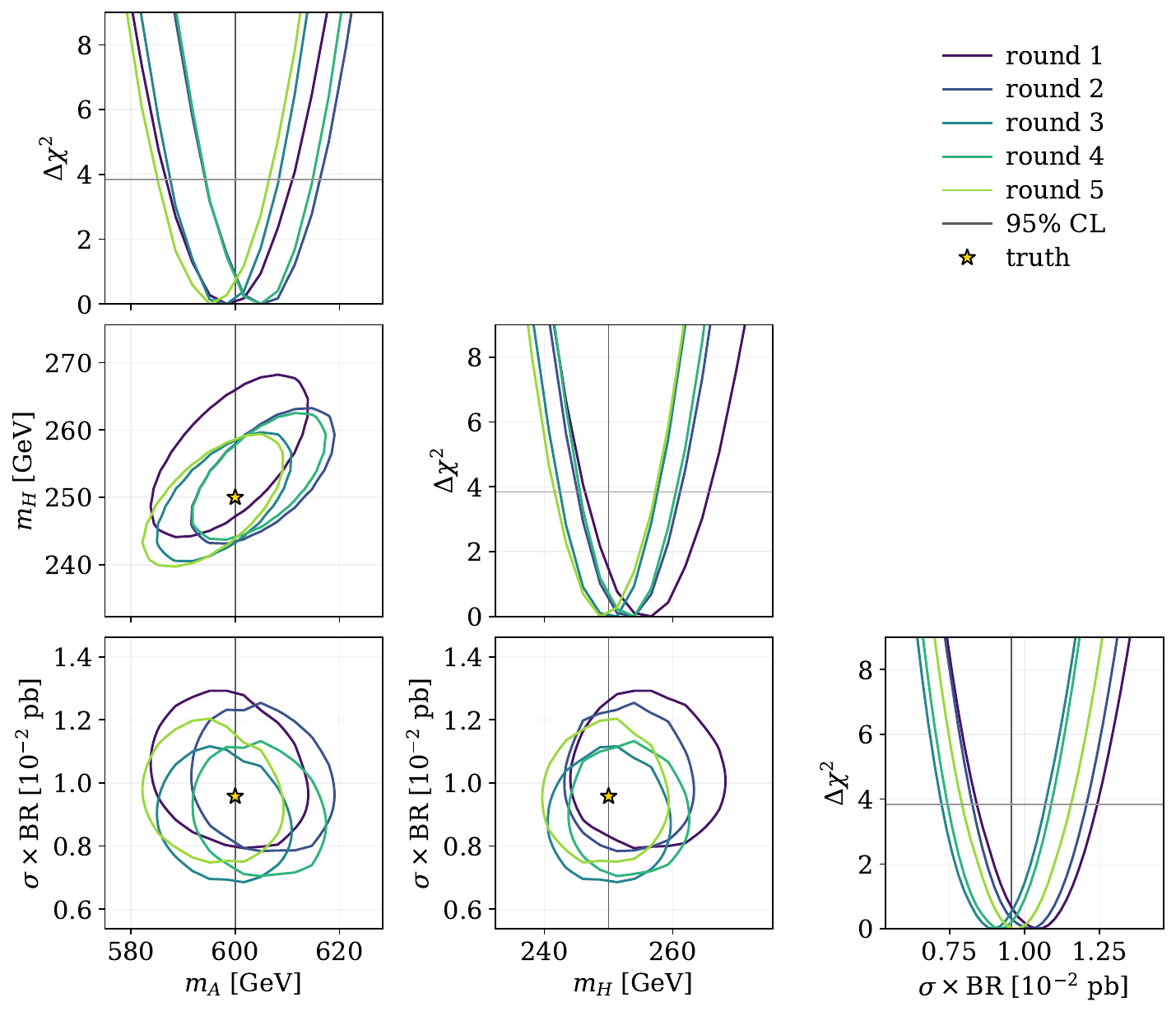}
    \caption{Similar to \cref{fig:zhllbb_rounds}, the five sequential SBI rounds showing the $95\%$ CL regions for the case described in \cref{sec:zhllbb}.}
    \label{fig:zhllbb_rounds95}
\end{figure}

\begin{figure}[!htbp]
    \centering
    \includegraphics[width=\textwidth]{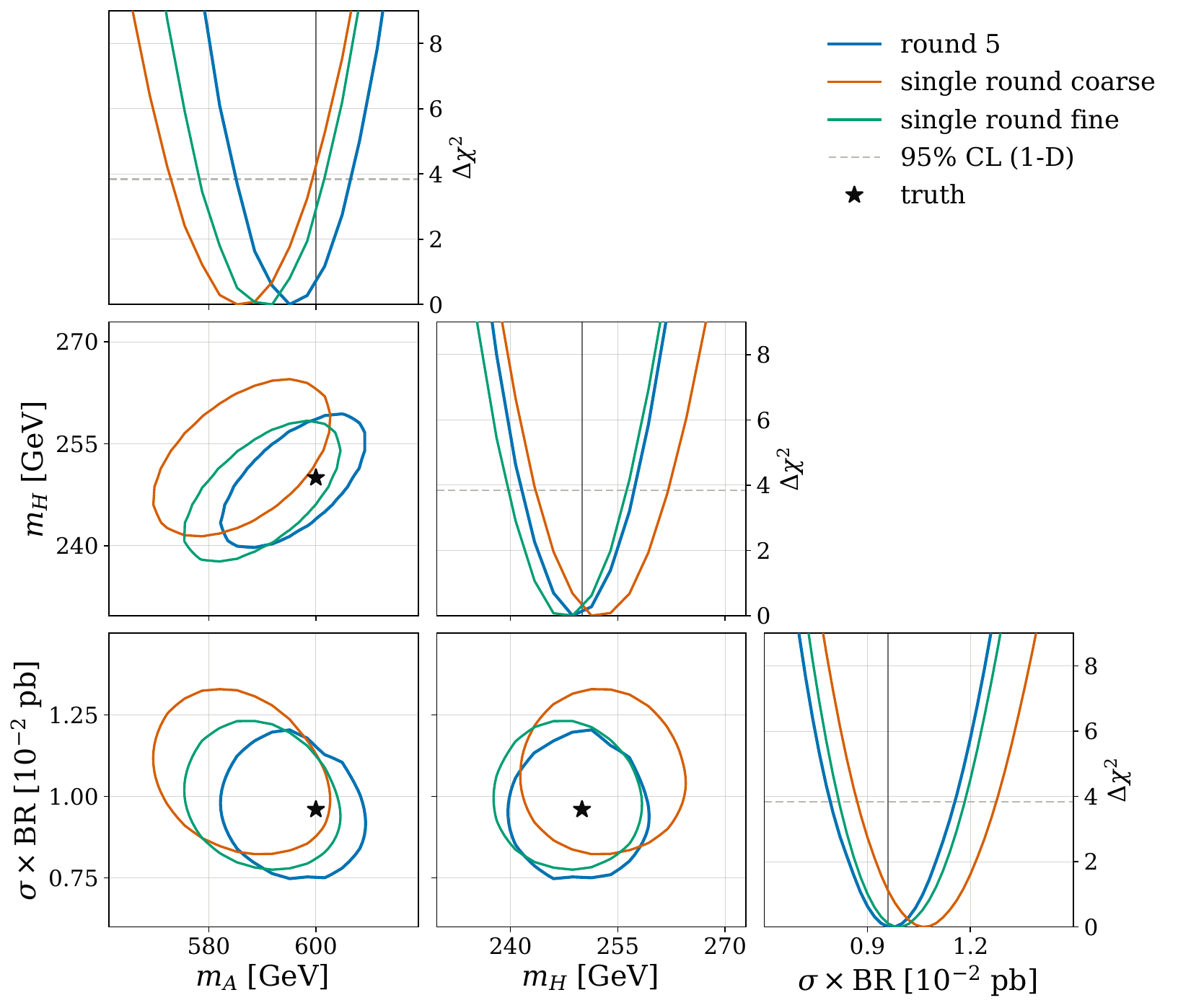}
    \caption{As \cref{fig:zhllbb_single}, the round-5 sequential SBI against the coarse and fine
    single rounds, but showing the $95\%$ CL regions for the case described in \cref{sec:zhllbb}.}
    \label{fig:zhllbb_single95}
\end{figure}

\Cref{fig:zhllbb_rounds,fig:zhllbb_single} show the $68\%$ CL regions. \Cref{fig:zhllbb_rounds95} shows how the $95\%$ CL regions change with sequential rounds. The confidence regions get tighter with rounds. \cref{fig:zhllbb_single95} compares the corresponding $95\%$ CL regions to the fixed fine and coarse grid with the same simulation budget. Similar to \cref{fig:zhllbb_single}, sequential rounds better constrains the parameters than single rounds with the same simulation budget.

%%%%%%%%%%%%%%%%%% figure %%%%%%%%%%%%%%%%%%%
%%%%%%%%%%%%%%%%%% figure %%%%%%%%%%%%%%%%%%%

%%%%%%%%%%%%%%%%%% figure %%%%%%%%%%%%%%%%%%%
%%%%%%%%%%%%%%%%%% figure %%%%%%%%%%%%%%%%%%%

\FloatBarrier
%%%%%%%%%%%%%%
\section{Further results for \texorpdfstring{$Y\to HH\to 4t$}{Y -> HH -> 4t}}
\label{sec:ytohh}

\subsection{Proposal updates}
\label{sec:ytohh4t_proposals}

From round~2 on, each round adds $300$ parameter points drawn from the proposal in \cref{eq:sequential_proposal_freq} with $\gamma=0.5$, built from the likelihood of the previous round.
We sample the proposal by importance resampling in the five-dimensional space $(\theta,\log(\sigma\times\mathrm{BR}))$. We draw $16{,}384$ candidates from a density $q$, comprising $90\%$ from a Gaussian fitted to the previous round's resampled points and broadened by a factor of $1.5$, and $10\%$ uniformly across the full parameter range. For round~2, we instead fit the Gaussian to $1350$ samples obtained from nine slice-sampling~\cite{Neal:2003slice} chains on the round-1 likelihood and broaden it by a factor of $2$.

We assign each candidate an importance weight $w=L^\gamma/q$, where $L$ is the previous round's likelihood averaged over the ensemble. We then obtain $3000$ samples by systematic resampling. To select generation points, we round masses to $1$~GeV and widths to $0.01$~GeV, discard points outside the support $m_Y\geq5m_H$, remove duplicates, and retain $300$ distinct points in random order.

%%%%%%%%%%%%%%%%%% figure %%%%%%%%%%%%%%%%%%%
\begin{figure}[!htbp]
    \centering
    \includegraphics[width=\textwidth]{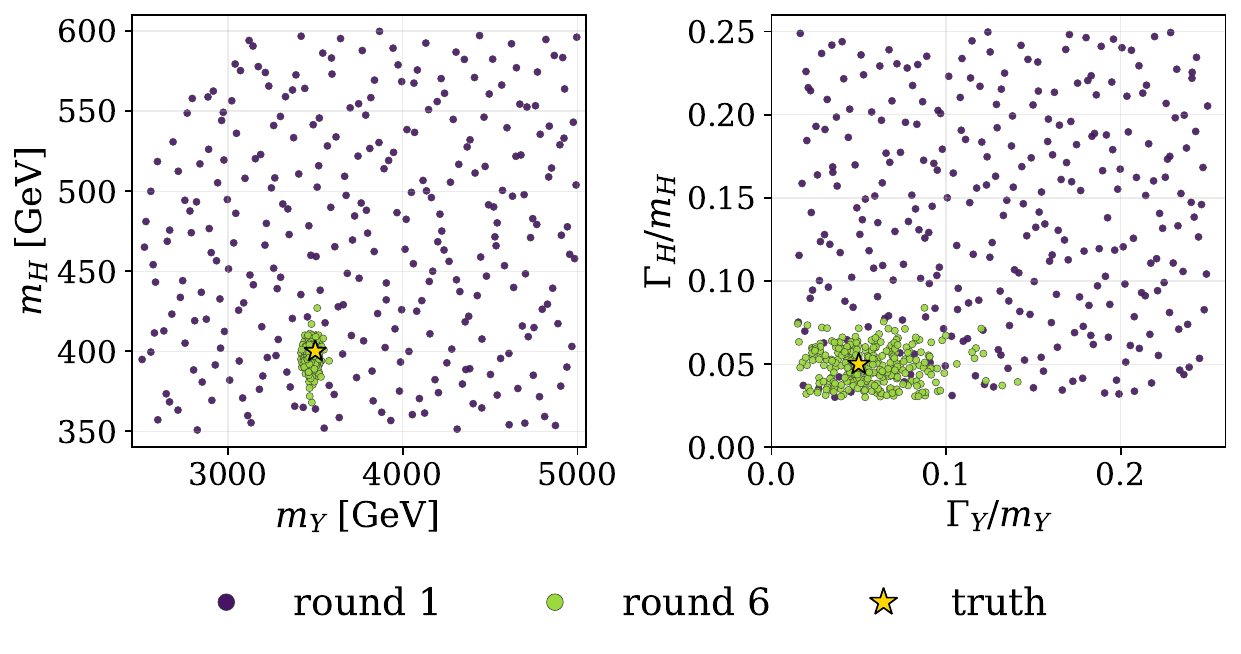}
    \caption{Parameter points used in the   of round 1, the $300$-point Sobol grid, and of round 6, drawn from the posterior of round 5. Left: the mass plane over the full prior. Right: the two relative widths.}
    \label{fig:ytohh4t_grids}
\end{figure}
%%%%%%%%%%%%%%%%%% figure %%%%%%%%%%%%%%%%%%%

\Cref{fig:ytohh4t_grids} shows the parameter points used in the training of rounds~1 and~6. The parameter points are spread across the full prior range in round~1, but concentrate around the injected point in round~6.

%%%%%%%%%%%%%%
\subsection{Sequential rounds}
\label{sec:ytohh4t_rounds}

 %%%%%%%%%%%%%%%%%% figure %%%%%%%%%%%%%%%%%%%

\begin{figure}[!htbp]
    \centering
    \includegraphics[width=\textwidth]{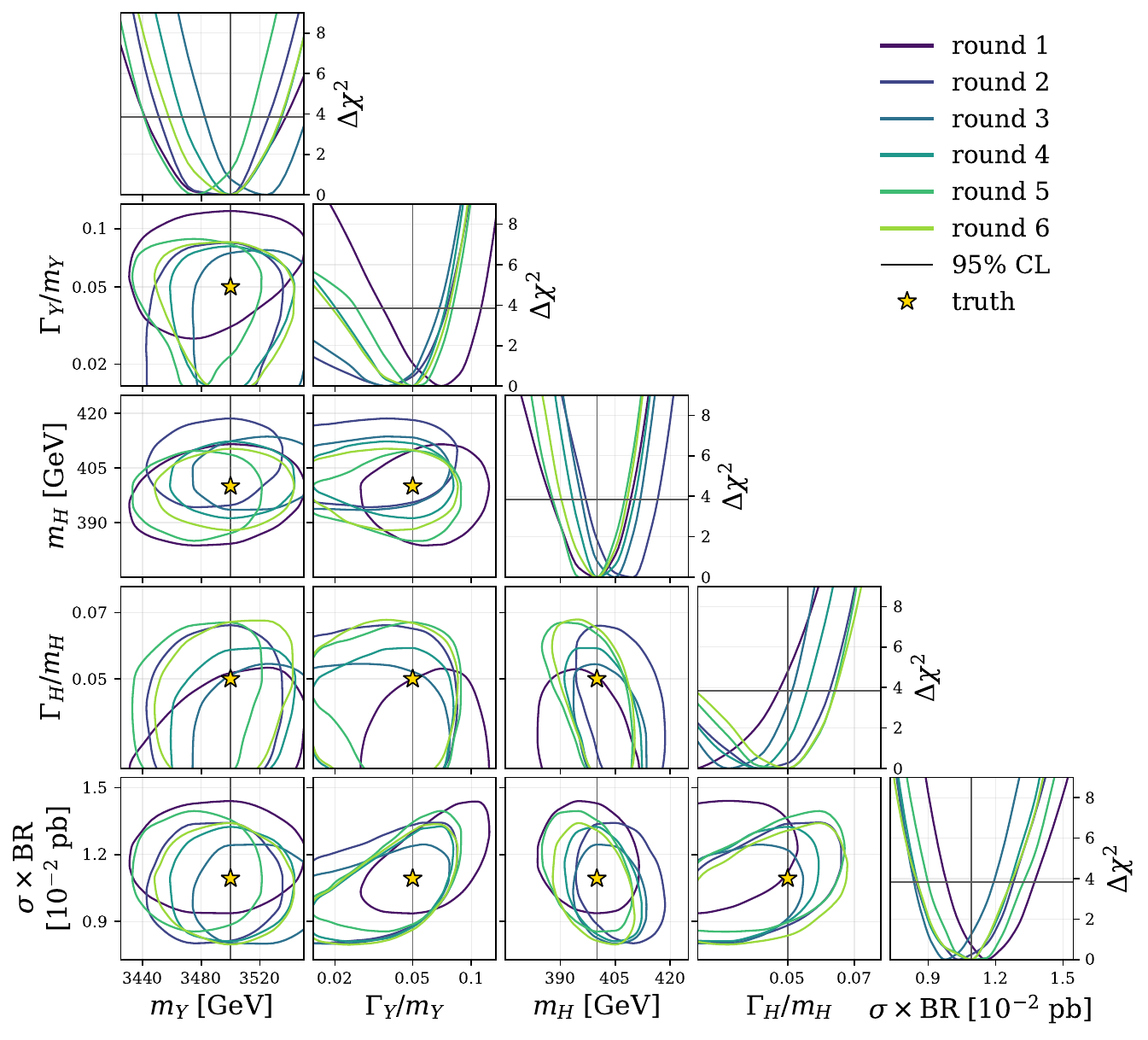}
    \caption{Rounds~1 to~6 of sequential SBI \cref{sec:ytohh4t}. Diagonal: one-dimensional profile likelihoods. Below the diagonal: $95\%$ CL regions of each pair, the other parameters profiled out. The star and the vertical lines mark the benchmark point.}
    \label{fig:ytohh4t_rounds}
\end{figure}

\begin{figure}[!htbp]
    \centering
    \includegraphics[width=\textwidth]{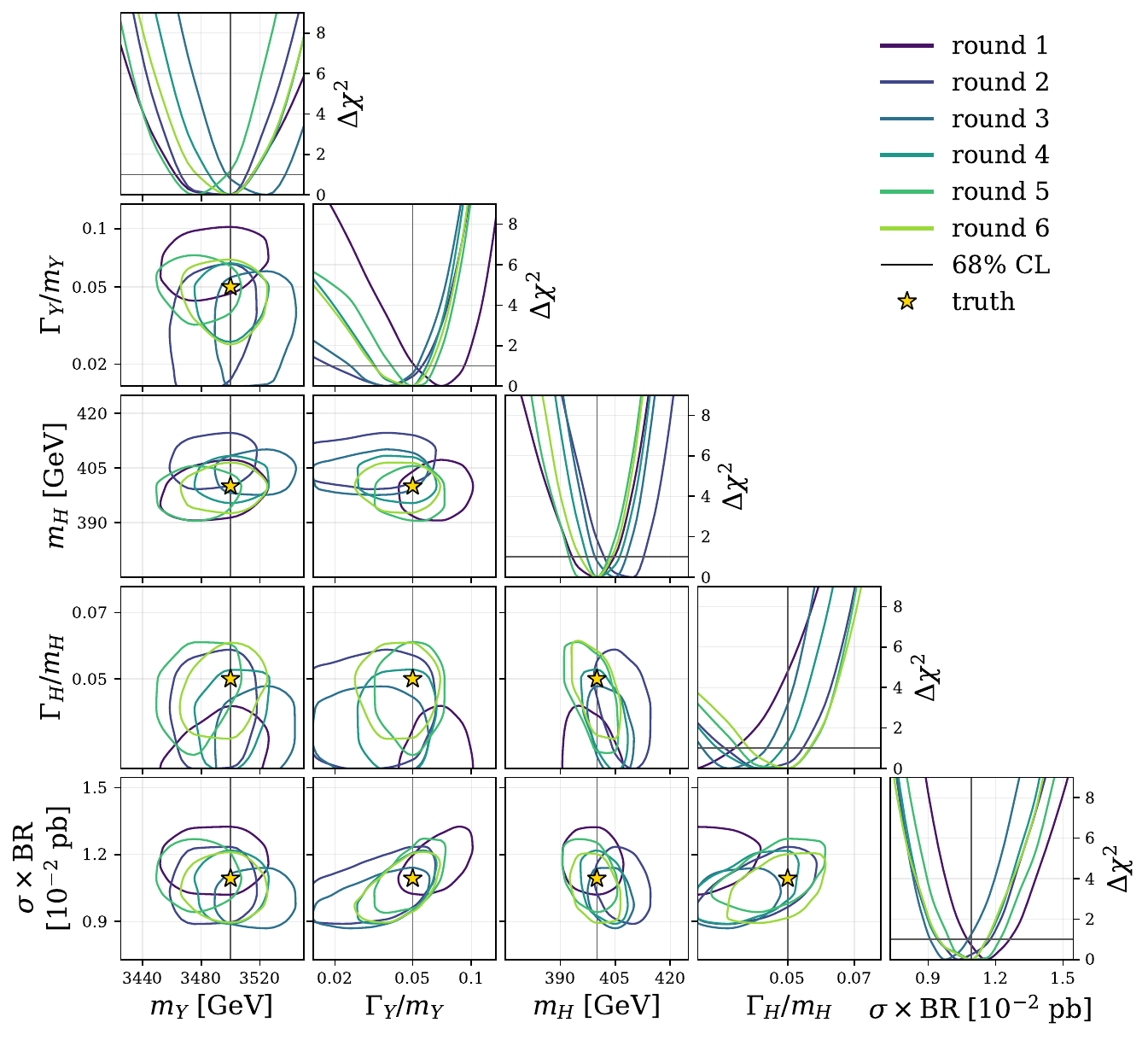}
    \caption{Rounds~1 to~6 for the case described in \cref{sec:ytohh4t}. Diagonal: one-dimensional profile likelihoods. Below the diagonal: $68\%$ CL regions of each pair, the other parameters profiled out. The star and the vertical lines mark the benchmark point.}
    \label{fig:ytohh4t_rounds68}
\end{figure}

\cref{fig:ytohh4t_rounds} (\cref{fig:ytohh4t_rounds68}) shows how the $68\%$ ($95\%$) confidence regions change from rounds 1 to 6. The diagonal panels display the one-dimensional profile likelihoods, while the panels below the diagonal show the pairwise confidence regions with the remaining parameters profiled out. The relative widths are less precisely determined, with the $95\%$ contours reaching the lower boundaries of the scan range. This is expected because detector smearing limits sensitivity to intrinsic widths below the experimental mass resolution. The masses and the cross sections are better constrained. The confidence regions for all the parameters get tighter for the later sequential rounds.

\FloatBarrier

%%%%%%%%%%%%%%
\subsection{Comparing $95\%$ confidence regions with single rounds}
\label{sec:ytohh4t_single_app}

%%%%%%%%%%%%%%%%%% figure %%%%%%%%%%%%%%%%%%%
\begin{figure}[!htbp]
    \centering
    \includegraphics[width=\textwidth]{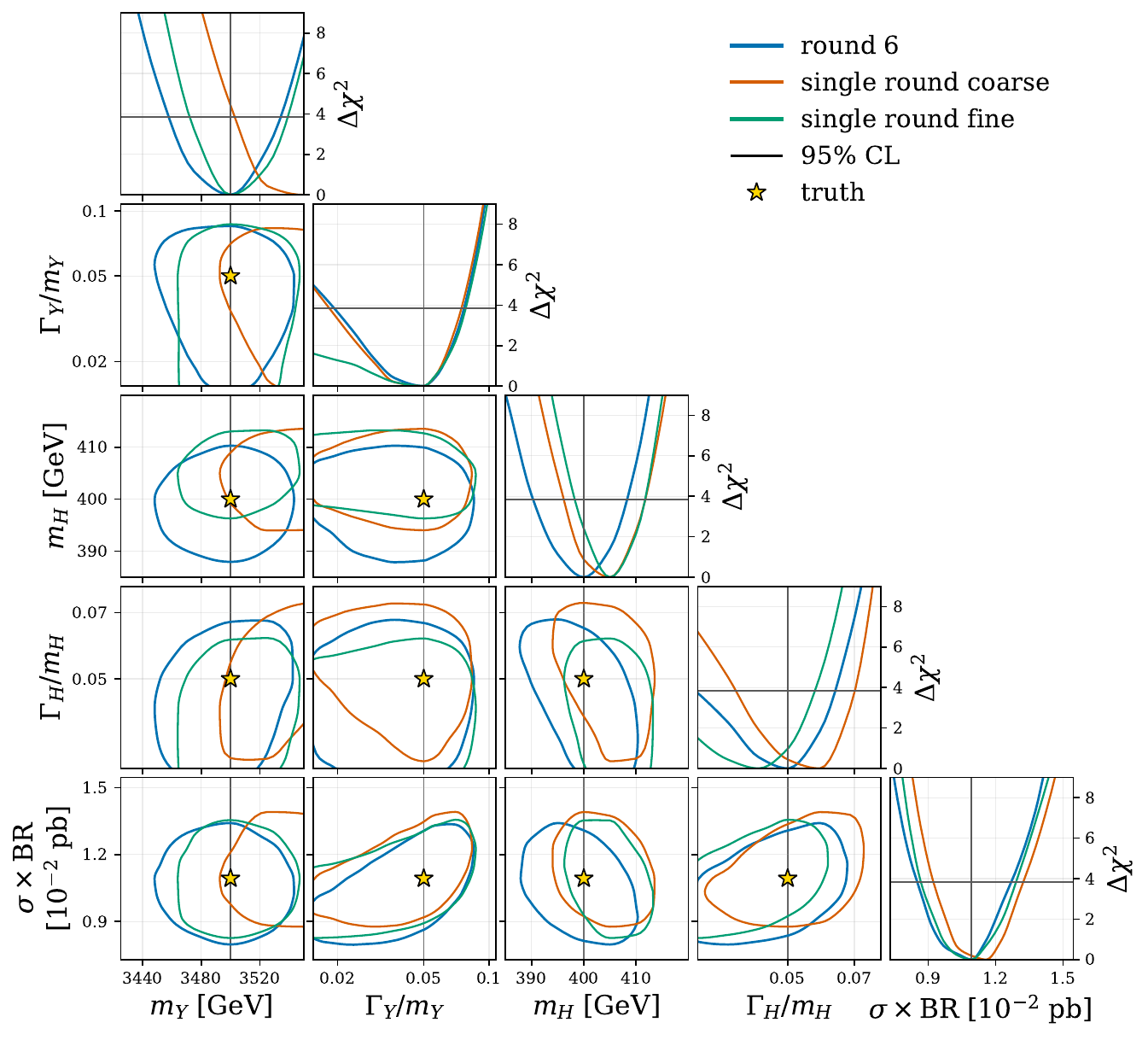}
    \caption{Round~6 compared to two single rounds of \cref{fig:ytohh4t_ssgrids}, all trained on $1.8\times10^{6}$ events for the case described in \cref{sec:ytohh4t}. The $68\%$ regions were shown in \cref{fig:ytohh4t_single68}.}
    \label{fig:ytohh4t_single}
\end{figure}

%%%%%%%%%%%%%%%%%% figure %%%%%%%%%%%%%%%%%%%

\cref{fig:ytohh4t_single} compares the $95\%$ confidence regions for sequential SBI to single round with the same amount of simulation, similar to \cref{sec:ytohh4t_single}. We see that the confidence regions from the final round of sequential SBI covers the benchmark point, unlike the single coarse round. While the fine grid results in a tighter contour, ~\cref{fig:ytohh4t_coverage}, we showed that it undercovers. 

% \FloatBarrier
%%%%%%%%%%%%%%
\subsection{Round-1 grid}
\label{sec:grid_designs}

\begin{figure}[!htbp]
    \centering
    \includegraphics[width=\textwidth]{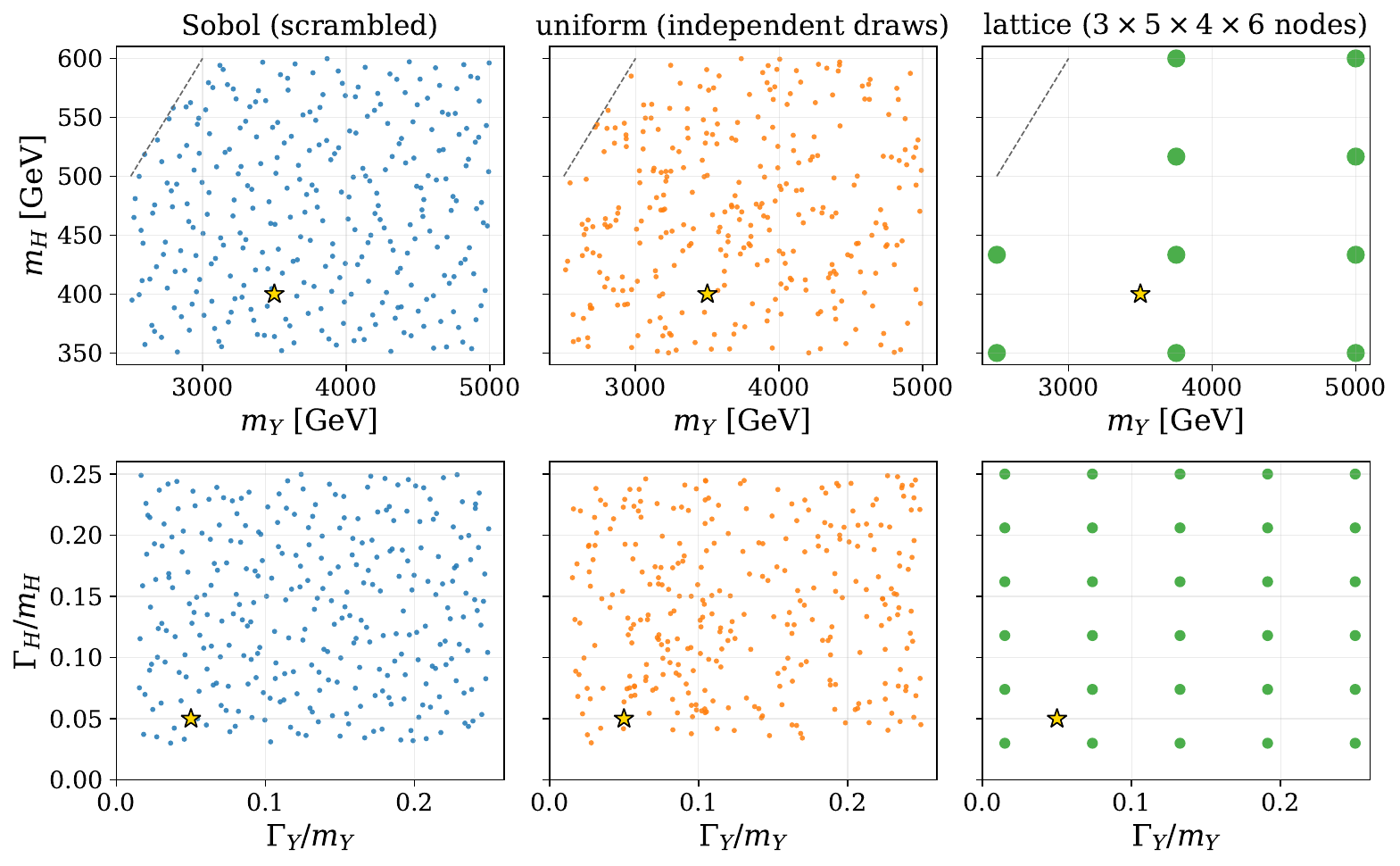}
    \caption{Three $300$-point round-1 designs over the prior: the scrambled Sobol grid we use (left), independent uniform draws (middle) and a $3\times5\times4\times6$ lattice (right). Top: the mass plane, with the boost requirement $m_Y\geq 5\,m_H$ dashed. Bottom: the two relative widths. The star marks the benchmark point.}
    \label{fig:grid_designs}
\end{figure}

For the four-parameter analysis, the first round must cover the parameter range with $300$ points. \Cref{fig:grid_designs} compares three designs uniform in the masses and relative widths: scrambled Sobol points (left), independent uniform draws (middle), and a balanced lattice (right), each restricted to $m_Y\geq5m_H$.

\begin{figure}[!htbp]
    \centering
    \includegraphics[width=0.85\textwidth]{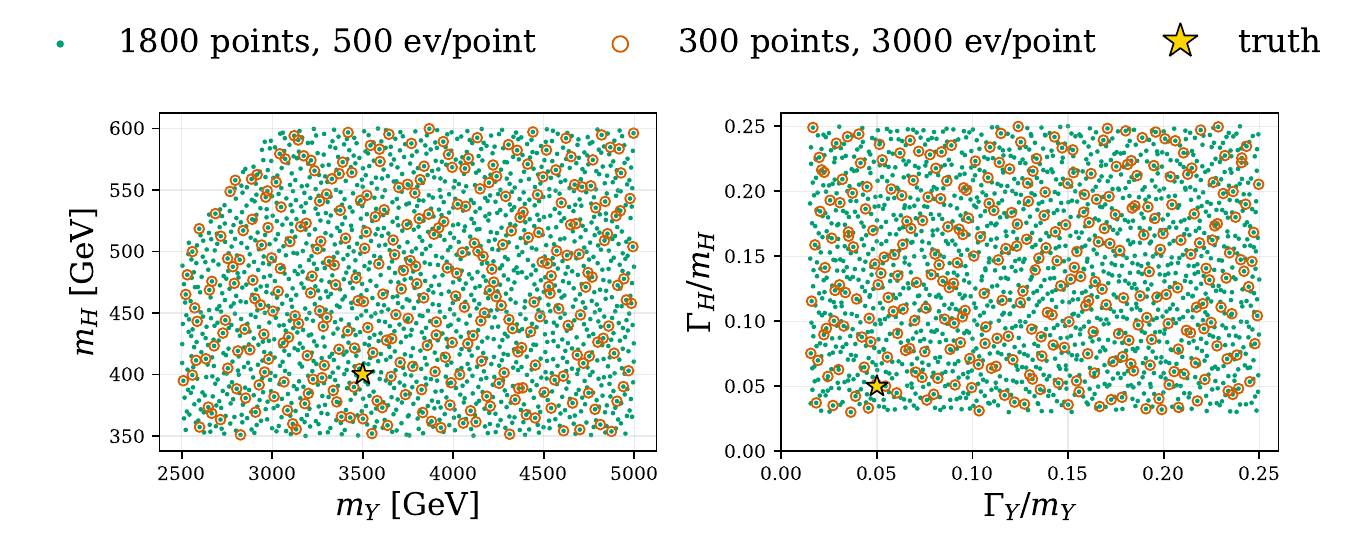}
    \caption{The two fixed grids the single rounds train on, drawn on the same axes: the mass plane (left) and the plane of the two relative widths (right). The $300$-point grid is the one used in round 1 of sequential rounds, and also in the coarse grid with fixed simulation budget. The $1800$-point grid is used as in the single round with fine grid with fixed simulation budget. }
    \label{fig:ytohh4t_ssgrids}
\end{figure}

We generate $2048$ scrambled Sobol points with a fixed seed using \texttt{scipy.stats.qmc.Sobol} (SciPy~1.16~\cite{Virtanen:2019joe}) and map them linearly to parameter ranges. We use the first $300$ Sobol points satisfying $m_Y\geq5m_H$ as the round-1 grid. The two single-round grids of \cref{sec:ytohh4t_sim}, shown in \cref{fig:ytohh4t_ssgrids}, come from the same sequence: the coarse grid is this $300$-point round-1 grid, and the fine grid consists of the first $1800$ points satisfying the constraint. The uniform design uses the same box and constraint, while the lattice starts from $3\times5\times4\times6=360$ equally spaced nodes, of which $300$ satisfy the constraint.

%%%%%%%%%%%%%%%%%% table %%%%%%%%%%%%%%%%%%%

%%%%%%%%%%%%%%%%%% table %%%%%%%%%%%%%%%%%%%

\Cref{tab:grid_designs} summarizes the differences between the three designs. The lattice samples each parameter at only a few values, e.g.\ $m_Y=2.5$, $3.75$, and $5$~TeV. Its $300$ points project onto just $10$ distinct points in the mass plane and $30$ in the relative-width plane, whereas the Sobol and uniform designs have $300$ distinct points in each projection.

We quantify uniformity using the centered discrepancy, which measures deviations between the fraction of points in a region and its fractional volume. For example, a region occupying $10\%$ of the volume should contain approximately $10\%$ of the points. The Sobol discrepancy is smaller than those of the uniform and lattice designs by factors of approximately $3.7$ and $76$, respectively, indicating the most uniform coverage.

\begin{table}[!htbp]
    \centering
    \begin{tabular}{lccc}
        \toprule
         & Sobol & uniform & lattice \\
        \midrule
        centred $L_2$ discrepancy & $1.2\times10^{-3}$ & $4.4\times10^{-3}$ & $9.1\times10^{-2}$ \\
        distinct values per parameter & $300$ & $300$ & $3,\,5,\,4,\,6$ \\
        distinct points in the mass plane & $300$ & $300$ & $10$ \\
        distinct points in the width plane & $300$ & $300$ & $30$ \\
        \bottomrule
    \end{tabular}
    \caption{Properties of the designs of \cref{fig:grid_designs}, with distances in the unit hypercube of the four prior ranges. The largest empty ball is the largest distance from any point of the support to its nearest design point. For Sobol and uniform, the first four rows are medians over $200$ realisations.}
    \label{tab:grid_designs}
\end{table}

%%%%%%%%%%%%%%%%%% figure %%%%%%%%%%%%%%%%%%%
%%%%%%%%%%%%%%%%%% figure %%%%%%%%%%%%%%%%%%%

%%%%%%%%%%%%%%%%%% figure %%%%%%%%%%%%%%%%%%%
%%%%%%%%%%%%%%%%%% figure %%%%%%%%%%%%%%%%%%%

%%%%%%%%%%%%%%
\subsection{Binned baseline}
\label{sec:ytohh4t_hist}

%%%%%%%%%%%%%%%%%% figure %%%%%%%%%%%%%%%%%%%
\begin{figure}[!htbp]
    \centering
    \includegraphics[width=0.95\textwidth]{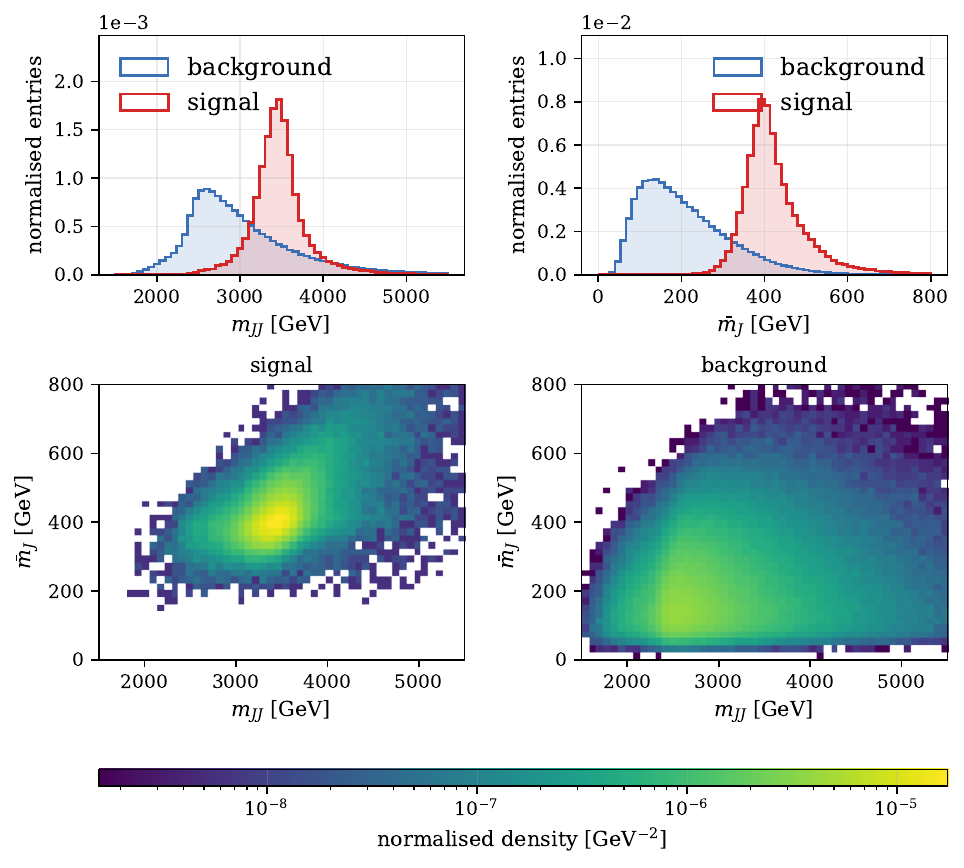}
    \caption{The two mass observables after the selection, for the benchmark signal $(m_Y, \Gamma_Y/m_Y, m_H, \Gamma_H/m_H) = (3500~\text{GeV}, 5\%, 400~\text{GeV}, 5\%)$ and the QCD di-jet background, normalised to unit area. Top: the two-jet mass $m_{JJ}$ and the mean large-$R$ jet mass $\bar m_J$ of \cref{eq:ytohh4t_obs}. Bottom: their joint density on a logarithmic colour scale.}
    \label{fig:ytohh4t_observables}
\end{figure}
%%%%%%%%%%%%%%%%%% figure %%%%%%%%%%%%%%%%%%%

For the binned baseline, we use two observables sensitive to the parent and daughter masses: the dijet invariant mass $m_{JJ}$ and the average jet mass $\bar m_J$. At $\theta_{\text{true}}$, the large boost of each daughter particle $H$ collimates its decay products, allowing the full hadronic decay $H\to t\bar t\to 6q$ to be contained within a single large-radius jet. The dijet invariant mass then probes $m_Y$, while the average jet mass probes $m_H$,
\begin{align}
    m_{JJ} &= \sqrt{\left(p_{J_1}+p_{J_2}\right)^2}\;, \\
    \bar m_J &= \frac{1}{2}\left(\sqrt{p_{J_1}^2}+\sqrt{p_{J_2}^2}\right)\;.
    \label{eq:ytohh4t_obs}
\end{align}
The distributions of these observables are shown in \cref{fig:ytohh4t_observables}.

Similar to $A\to ZH$ (\cref{sec:zhllbb_hist}), we compare the learned ratio with a binned template in the two mass observables of \cref{eq:ytohh4t_obs}. We use $40\times40$ bins spanning $m_{JJ}\in[1.5,5.5]$~TeV and $\bar m_J\in[0,800]$~GeV, with bin widths of $100$ and $20$~GeV, respectively. The signal-to-background template ratio replaces the network prediction in the shape-only likelihood of \cref{eq:shape_likelihood}, using the same efficiency, scan grid, events, and calibration (\cref{eq:ratio_calibration}).

\begin{figure}[!htbp]
    \centering
    \includegraphics[width=\textwidth]{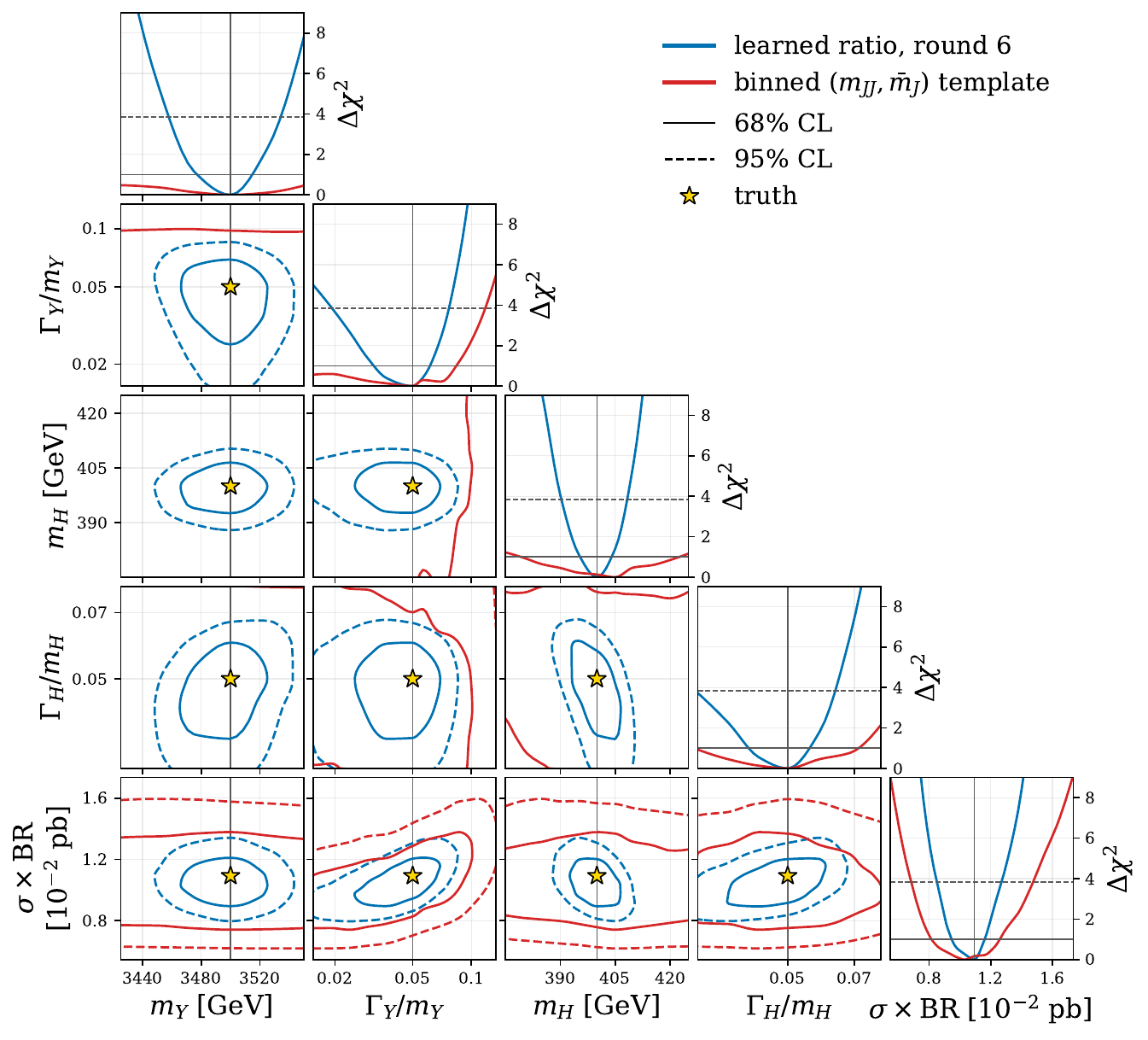}
    \caption{The learned ratio at round~6 against the binned $(m_{JJ},\bar m_J)$ template, on the same events with the same efficiency, calibration and lattice, at $68\%$ (solid) and $95\%$ (dashed) CL.}
    \label{fig:ytohh4t_hist_corner}
\end{figure}

\Cref{fig:ytohh4t_hist_corner} compares the two. The binned baseline's $68\%$ intervals span the full scanned range in $m_Y$ and $[379,422]$~GeV in $m_H$, compared with $[3479,3511]$ and $[396,404]$~GeV, respectively, from sequential SBI. It constrains $\Gamma_Y/m_Y$ only from above, at $0.08$, and gives a $\sigma\times\text{BR}$ interval almost $2$ times wider.

%%%%%%%%%%%%%%%%%% figure %%%%%%%%%%%%%%%%%%%
%%%%%%%%%%%%%%%%%% figure %%%%%%%%%%%%%%%%%%%

%%%%%%%%%%%%%%
\FloatBarrier
\section{Background fluctuations in the observed sample}
\label{sec:app_bkgfluct}

We obtain the limits in this paper by replacing the event sum in \cref{eq:full_lr_bonly} by its expectation under the injected hypothesis instead of drawing a pseudo-dataset. Writing the per-event contribution as
\begin{align}
    f_\theta(x) = \log\left[1 + \frac{S(\theta)}{B}\,r(x|\theta)\right] \;,
\end{align}
the log-likelihood ratio becomes
\begin{align}
    \log\Lambda_\text{full}(\theta,B)
    = -S(\theta)
    + N^{\text{exp}}_{S} \left\langle f_\theta \right\rangle_{p_S(x|\theta_\text{true})}
    + N^{\text{exp}}_{B} \left\langle f_\theta \right\rangle_{p_B(x)} \;.
    \label{eq:asimov_llr}
\end{align}
Neither expectation is available in closed form. Both are averages over finite simulated samples. We can estimate the introduced numerical noise via
\begin{align}
    \left\langle f_\theta \right\rangle_{p_B(x)}
    \;\longrightarrow\;
    \frac{1}{N_b}\sum_{j=1}^{N_b} f_\theta(x_j)
    \;\equiv\;
    \left\langle f_\theta \right\rangle_{p_B(x)} + \delta_\theta \;,
    \qquad
    \left\langle \delta_\theta \right\rangle = 0 \;,
    \qquad
    \left\langle \delta_\theta^2 \right\rangle = \frac{\sigma_\theta^2}{N_b} \;,
    \label{eq:delta_def}
\end{align}
where $N_b$ is the number of simulated background events and
\begin{align}
    \sigma_\theta^2
    = \left\langle f_\theta^2 \right\rangle_{p_B(x)} - \left\langle f_\theta \right\rangle^2_{p_B(x)}
    \label{eq:sigma_theta}
\end{align}
is the variance of the per-event contribution $f_\theta$ over the background distribution, so $\sigma_\theta$ measures its spread. The estimator is unbiased, so on average the evaluation results in the correct likelihood. What matters is how far a given simulated sample scatters around it. \Cref{eq:asimov_llr} does not use $\left\langle f_\theta \right\rangle_{p_B}$ on its own, it multiplies it by the expected background yield, so the estimation error enters the log-likelihood ratio as $N^{\text{exp}}_{B}\,\delta_\theta$. The quantity this replaces in a real experiment is the fluctuation of the actual event sum $\sum_j^{N^{\text{exp}}_{B}} f_\theta(x_j)$ around its mean, which for $N^{\text{exp}}_{B}$ independent events is a Poisson-like fluctuation of size $\sqrt{N^{\text{exp}}_{B}}\,\sigma_\theta$ and is part of what the experiment genuinely measures. The two involve the same per-event spread $\sigma_\theta$ and differ only in how many events they are averaged over,
\begin{align}
    \underbrace{N^{\text{exp}}_{B}\,\frac{\sigma_\theta}{\sqrt{N_b}}}_\text{simulation noise}
    \qquad \text{against} \qquad
    \underbrace{\sqrt{N^{\text{exp}}_{B}}\,\sigma_\theta}_\text{statistical noise}
    \qquad \Longrightarrow \qquad
    \frac{\text{simulation noise}}{\text{statistical noise}} = \sqrt{\frac{N^{\text{exp}}_{B}}{N_b}} \;.
    \label{eq:mc_noise_ratio}
\end{align}
Our evaluation is meant to describe the experiment with its statistical fluctuations switched off. Consequently, it is only reliable once the simulated sample is much larger than the expected yield, $N_b \gg N^{\text{exp}}_{B}$. For the four-top channel, $N^{\text{exp}}_{B} = 6.82\times10^5$ against $N_b = 1.5\times10^6$, the ratio is $0.67$: the simulated background adds about two thirds of the statistical fluctuation of the experiment it emulates. For the resolved cascade of \cref{sec:zhllbb}, $N^{\text{exp}}_{B} = 1.20\times10^5$ against $N_b = 1.3\times10^6$, it is $0.30$ and the effect is a correction. The signal term of \cref{eq:asimov_llr} carries the same structure, but its prefactor $N^{\text{exp}}_{S}$ is smaller by two to three orders of magnitude, so we can ignore it.

The size of $\delta_\theta$ is, however, not the whole story. The same simulated events are used at every parameter point, so $\delta_\theta$ is not independent noise from point to point, but a single smooth but random function of $\theta$, drawn once and then held fixed for the entire analysis. It therefore does not average out when we scan the grid or profile a parameter. Instead, it tilts the whole likelihood surface coherently and displaces the position of its minimum. To see this, note that by \cref{eq:asimov_llr} the noise enters the log-likelihood ratio linearly, $-\log\Lambda_\text{full}(\theta) = -\log\bar\Lambda_\text{full}(\theta) - N^{\text{exp}}_{B}\,\delta_\theta$, where the bar denotes the noiseless $N_b\to\infty$ limit. Writing the minimum of $-\log\bar\Lambda_\text{full}$ as $\bar\theta$ and the perturbed one as $\bar\theta + \Delta\theta$, the stationarity condition for the noisy surface reads $\nabla_\theta \left[ -\log\bar\Lambda_\text{full} \right](\bar\theta + \Delta\theta) = N^{\text{exp}}_{B} \nabla_\theta \delta_\theta$. Expanding the left-hand side around $\bar\theta$, where its leading term vanishes by construction, and keeping the first order in the noise, the displacement is
\begin{align}
    \Delta\theta \simeq N^{\text{exp}}_{B}\, H^{-1} \nabla_\theta \delta_\theta \;,
    \label{eq:llr_displacement}
\end{align}
where $H = \nabla_\theta \nabla_\theta \left[ -\log\bar\Lambda_\text{full} \right](\bar\theta)$ is the Hessian at the unperturbed minimum --- the same matrix that determines the width of the confidence region. The two therefore scale together: once the noise ratio in \cref{eq:mc_noise_ratio} is of order one, the displacement is of order the quoted uncertainty. On a single dataset with a single simulated background sample such a shift is indistinguishable from a genuine bias of the method. Separating them requires repeating the analysis with independent background samples. Residual offsets that survive the iteration are of this kind: chains that differ in their training data and their networks still fit the same realisation of $\delta_\theta$.

The calibration of \cref{eq:ratio_calibration} feeds the same kind of fluctuation into the likelihood through a much narrower channel. Its normalization $\left\langle \hat r \right\rangle_{p_B}$ is also a finite-sample mean over background events, but of $e^{\ell}$ rather than of $f_\theta$, so it is an importance-sampling mean dominated by the handful of background events which the classifier ranks most signal-like. The number of events that effectively contribute is not $N_b$, but the Kish effective sample size
\begin{align}
    N_\text{eff}(\theta) = \frac{\left( \sum_j e^{\ell_j} \right)^2}{\sum_j e^{2\ell_j}} \;,
\end{align}
In our runs, the calibration is computed from $N_b = 1.5\times10^6$ background events for the four-top channel and from $4\times10^5$ for the resolved cascade. Over the evaluation grids, $N_\text{eff}$ ranges from about $7\times10^3$ to $2.5\times10^4$ for the four-top channel and from about $7\times10^2$ to $6\times10^3$ per network for the resolved cascade, i.e.\ $0.2\%$ to $1.7\%$ of the background events. If we call $\eta_\theta$ the resulting error on $\log\left\langle \hat r \right\rangle_{p_B}$, with $\left\langle \eta_\theta^2 \right\rangle \simeq 1/N_\text{eff}(\theta)$, it shifts \emph{every} event's calibrated log-ratio at that parameter point by the same amount $-\eta_\theta$. This is exactly the coherent, $\theta$-dependent tilt that \cref{eq:llr_displacement} responds to, only now with an effective sample size about two orders of magnitude below $N_b$.

\clearpage
%%%%%%%%%%%%%%%%%%%%%%%%%%%%%%%%%%%%%%%%%%%%%%%%%%
\bibliography{tilman,refs}
\end{document}

%% file: incl_settings.tex
\usepackage[utf8]{inputenc} % input encodings (allow UTF-8 input)
\usepackage[T1]{fontenc} 	% selecting font encodings (use 8-bit T1 fonts)
\usepackage[english]{babel} % multilingual support (English language/hyphenation)

\usepackage[bitstream-charter]{mathdesign}
\usepackage{geometry} 		% flexible and complete interface to document dimensions
\usepackage{amsmath} 		% math package (American Mathematical Society)
\usepackage{mathtools} 		% math package (fixes various deficiencies of amsmath)
\usepackage{float} 			% floating objects such as figures and tables
\usepackage{graphicx} 		% enhanced support for graphics
\usepackage{tabularx} 		% tabulars with adjustable-width columns
\usepackage{booktabs} 		% professional-quality tables
\usepackage{color, xcolor} 	% foreground and background colour management
\usepackage{pdfpages} 		% inclusion of external multi-page PDF documents
\usepackage{extarrows} 		% extra arrows beyond those provided in amsmath
\usepackage{multirow} 		% create tabular cells spanning multiple rows
\usepackage{multicol} 		% intermix single and multiple columns
\usepackage{enumitem} 		% control layout of itemize, enumerate, description
\usepackage{xspace} 		% define commands that appear not to eat spaces
\usepackage{stackrel} 		% enhancement to the \stackrel command
\usepackage{tikz} 			% create PostScript and PDF graphics
\usepackage{braket} 		% Dirac bra-ket notation
\usepackage{bm} 			% access bold symbols in maths mode
\usepackage{tensor} 		% typeset tensors (tensor-style super- and subscripts)
\usepackage{slashed} 		% slash through characters (Feynman slash notation)
\usepackage{siunitx} 		% SI units package (typesetting values with units)
\usepackage{lastpage} 		% reference last page
\usepackage{cite} 			% improved citation handling
\usepackage[normalem]{ulem} % package for underlining
\usepackage{fontawesome} 	% access to web-related icons
\usepackage{tocloft} 		% control over the typography of the TOC, LOF, and LOT
\usepackage{titlesec} 		% interface to sectioning commands (various title styles)
\usepackage{doi} 			% create correct hyperlinks for DOI numbers
\usepackage{hyperref} 		% hypertext links (handel cross-referencing commands)
\usepackage[most]{tcolorbox} 					% coloured and framed text boxes
\usepackage[nameinlink, capitalize]{cleveref} 	% intelligent cross-referencing
\usepackage[nottoc, notlot, notlof]{tocbibind} 	% add references/index/contents to TOC
\usepackage[ruled, vlined]{algorithm2e} 		% floating algorithm environment
\usepackage{makecell}
\usepackage[makeroom]{cancel}
\usepackage{feynmf}

\makeatletter
\def\BState{\State\hskip-\ALG@thistlm}
\makeatother

\makeatletter
\@ifundefined{pdfoutput}{}{\DeclareGraphicsRule{*}{mps}{*}{}}
\makeatother

\makeatletter
\DeclareRobustCommand*{\bfseries}{%
   \not@math@alphabet\bfseries\mathbf
   \fontseries\bfdefault\selectfont
   \boldmath
}
\makeatother

\hypersetup{
	pdftitle={ForecastingAmplification},
	pdfauthor={Bahl et al.},
	colorlinks=true, 			% false: boxed links, true: colored links
	linkcolor={red!50!black}, 	% color of internal links (sections, pages, etc.)
	citecolor={blue!50!black}, 	% color of citation links (links to bibliography)
	urlcolor={blue!80!black} 	% color of URL links (external links)
} 
\DeclareSymbolFont{usualmathcal}{OMS}{cmsy}{m}{n}
\DeclareSymbolFontAlphabet{\mathcal}{usualmathcal}

\SetArgSty{textnormal}
\SetKwComment{Comment}{{\small\#}~}{}
\SetCommentSty{mycommfont}

\setitemize{itemsep=0pt, parsep=0pt} 				% adjust itemize environment
\setenumerate{itemsep=0pt, parsep=0pt} 				% adjust enumerate environment
\setitemize{itemsep=2pt,topsep=2pt,parsep=0pt,partopsep=0pt,leftmargin=*}
\setenumerate{itemsep=0pt,topsep=2pt,parsep=0pt,partopsep=0pt,labelindent=3pt,leftmargin=*}
\newlist{todolist}{itemize}{2}
\setlist[todolist]{label=$\square$}
\usepackage{pifont}

\usepackage{amsmath}
 
\usepackage{amsthm} 		% math package (typesetting theorems using AMS style)
\theoremstyle{definition}

%% file: incl_shortcuts.tex
\definecolor{red_cb}{HTML}{e41a1c}
\definecolor{blue_cb}{HTML}{377eb8}
\definecolor{green_cb}{HTML}{4daf4a}
\definecolor{purple_cb}{HTML}{984ea3}
\definecolor{orange_cb}{HTML}{ff7f00}

\definecolor{EmeraldGreen}{HTML}{1ea78d}
\definecolor{EnglishRed}{HTML}{b02427}
\hypersetup{colorlinks=true,urlcolor=EmeraldGreen,citecolor=EmeraldGreen,linkcolor=EnglishRed}

\newcommand\one{\leavevmode\hbox{\small1\normalsize\kern-.33em1}}
\newcommand{\arXiv}[2][]{%
	\ifthenelse{\equal{#1}{}}%
	{\href{http://arxiv.org/abs/#2}{arXiv:#2}}%
	{\href{http://arxiv.org/abs/#2}{arXiv:#2~[#1]}}}

\def\slashchar#1{\setbox0=\hbox{$#1$}           % set a box for #1
   \dimen0=\wd0                                 % and get its size
   \setbox1=\hbox{/} \dimen1=\wd1               % get size of /
   \ifdim\dimen0>\dimen1                        % #1 is bigger
      \rlap{\hbox to \dimen0{\hfil/\hfil}}      % so center / in box
      #1                                        % and print #1
   \else                                        % / is bigger
      \rlap{\hbox to \dimen1{\hfil$#1$\hfil}}   % so center #1
      /                                         % and print /
   \fi}

\newcommand{\tikznode}[2]{%
\ifmmode%
\tikz[remember picture,baseline=(#1.base),inner sep=0pt] \node (#1) {$#2$};%
\else
\tikz[remember picture,baseline=(#1.base),inner sep=0pt] \node (#1) {#2};%
\fi}

\def\mathswitchr#1{\relax\ifmmode{\mathrm{#1}}\else$\mathrm{#1}$\xspace\fi}
\def\mathswitch#1{\relax\ifmmode#1\else$#1$\xspace\fi}

%% file: paper.bbl
\providecommand{\href}[2]{#2}\begingroup\raggedright\begin{thebibliography}{10}

\bibitem{Plehn:2022ftl}
T.~Plehn, A.~Butter, B.~Dillon, T.~Heimel, C.~Krause, and R.~Winterhalder, {\it {Modern Machine Learning for LHC Physicists}},  \href{http://arxiv.org/abs/2211.01421}{{arXiv:2211.01421 [hep-ph]}}.

\bibitem{Feickert:2021ajf}
M.~Feickert and B.~Nachman, {\it {A Living Review of Machine Learning for Particle Physics}},  \href{http://arxiv.org/abs/2102.02770}{{arXiv:2102.02770 [hep-ph]}}.

\bibitem{Butter:2022rso}
S.~Badger {\em et al.}, {\it {Machine learning and LHC event generation}},  \href{http://dx.doi.org/10.21468/SciPostPhys.14.4.079}{SciPost Phys. {\bfseries 14} (2023) 4, 079}, \href{http://arxiv.org/abs/2203.07460}{{arXiv:2203.07460 [hep-ph]}}.

\bibitem{Cranmer:2015bka}
K.~Cranmer, J.~Pavez, and G.~Louppe, {\it {Approximating Likelihood Ratios with Calibrated Discriminative Classifiers}},
\href{http://arxiv.org/abs/1506.02169}{{arXiv:1506.02169 [stat.AP]}}.
%%CITATION = ARXIV:1506.02169;%%.

\bibitem{Cranmer:2019eaq}
K.~Cranmer, J.~Brehmer, and G.~Louppe, {\it {The frontier of simulation-based inference}},  \href{http://arxiv.org/abs/1911.01429}{{arXiv:1911.01429 [stat.ML]}}.

\bibitem{Rizvi:2023mws}
S.~Rizvi, M.~Pettee, and B.~Nachman, {\it {Learning Likelihood Ratios with Neural Network Classifiers}},  \href{http://arxiv.org/abs/2305.10500}{{arXiv:2305.10500 [hep-ph]}}.

\bibitem{Martini:2020rnb}
T.~Martini, M.~Kraus, S.~Peitzsch, and P.~Uwer, {\it {The Matrix Element Method as a tool for precision and accuracy}},  \href{http://dx.doi.org/10.22323/1.364.0673}{PoS {\bfseries EPS-HEP2019} (2020)  673}, \href{http://arxiv.org/abs/2009.14642}{{arXiv:2009.14642 [hep-ph]}}.

\bibitem{Butter:2022vkj}
A.~Butter, T.~Heimel, T.~Martini, S.~Peitzsch, and T.~Plehn, {\it {Two invertible networks for the matrix element method}},  \href{http://dx.doi.org/10.21468/SciPostPhys.15.3.094}{SciPost Phys. {\bfseries 15} (2023) 3, 094}, \href{http://arxiv.org/abs/2210.00019}{{arXiv:2210.00019 [hep-ph]}}.

\bibitem{Heimel:2023mvw}
T.~Heimel, N.~Huetsch, R.~Winterhalder, T.~Plehn, and A.~Butter, {\it {Precision-machine learning for the matrix element method}},  \href{http://dx.doi.org/10.21468/SciPostPhys.17.5.129}{SciPost Phys. {\bfseries 17} (2024) 5, 129}, \href{http://arxiv.org/abs/2310.07752}{{arXiv:2310.07752 [hep-ph]}}.

\bibitem{Brehmer:2020cvb}
J.~Brehmer and K.~Cranmer, {\it {Simulation-based inference methods for particle physics}},  \href{http://arxiv.org/abs/2010.06439}{{arXiv:2010.06439 [hep-ph]}}.

\bibitem{Brehmer:2018eca}
J.~Brehmer, K.~Cranmer, G.~Louppe, and J.~Pavez, {\it {A Guide to Constraining Effective Field Theories with Machine Learning}},  \href{http://dx.doi.org/10.1103/PhysRevD.98.052004}{Phys. Rev. D {\bfseries 98} (2018) 5, 052004}, \href{http://arxiv.org/abs/1805.00020}{{arXiv:1805.00020 [hep-ph]}}.

\bibitem{Brehmer:2018kdj}
J.~Brehmer, K.~Cranmer, G.~Louppe, and J.~Pavez, {\it {Constraining Effective Field Theories with Machine Learning}},  \href{http://dx.doi.org/10.1103/PhysRevLett.121.111801}{Phys. Rev. Lett. {\bfseries 121} (2018) 11, 111801},
\href{http://arxiv.org/abs/1805.00013}{{arXiv:1805.00013 [hep-ph]}}.
%%CITATION = ARXIV:1805.00013;%%.

\bibitem{Brehmer:2018hga}
J.~Brehmer, G.~Louppe, J.~Pavez, and K.~Cranmer, {\it {Mining gold from implicit models to improve likelihood-free inference}},  \href{http://dx.doi.org/10.1073/pnas.1915980117}{Proc. Nat. Acad. Sci. (2018)  201915980},
\href{http://arxiv.org/abs/1805.12244}{{arXiv:1805.12244 [stat.ML]}}.
%%CITATION = ARXIV:1805.12244;%%.

\bibitem{Chatterjee:2021nms}
S.~Chatterjee, N.~Frohner, L.~Lechner, R.~Sch\"ofbeck, and D.~Schwarz, {\it {Tree boosting for learning EFT parameters}},  \href{http://arxiv.org/abs/2107.10859}{{arXiv:2107.10859 [hep-ph]}}.

\bibitem{Chatterjee:2022oco}
S.~Chatterjee, S.~Rohshap, R.~Sch{\"o}fbeck, and D.~Schwarz, {\it {Learning the EFT likelihood with tree boosting}},  \href{http://arxiv.org/abs/2205.12976}{{arXiv:2205.12976 [hep-ph]}}.

\bibitem{Kong:2022rnd}
K.~Kong, K.~T. Matchev, S.~Mrenna, and P.~Shyamsundar, {\it {New Machine Learning Techniques for Simulation-Based Inference: InferoStatic Nets, Kernel Score Estimation, and Kernel Likelihood Ratio Estimation}},  \href{http://arxiv.org/abs/2210.01680}{{arXiv:2210.01680 [stat.ML]}}.

\bibitem{Schofbeck:2024zjo}
R.~Sch{\"o}fbeck, {\it {Refinable modeling for unbinned SMEFT analyses}},  \href{http://dx.doi.org/10.1088/2632-2153/ad9fd1}{Mach. Learn. Sci. Tech. {\bfseries 6} (2025) 1, 015007}, \href{http://arxiv.org/abs/2406.19076}{{arXiv:2406.19076 [hep-ph]}}.

\bibitem{Bahl:2024meb}
H.~Bahl, V.~Bres{\'o}, G.~De~Crescenzo, and T.~Plehn, {\it {Advancing Tools for Simulation-Based Inference}},  \href{http://dx.doi.org/10.21468/SciPostPhysCore.8.3.060}{SciPost Phys. Core {\bfseries 8} (2025)  060}, \href{http://arxiv.org/abs/2410.07315}{{arXiv:2410.07315 [hep-ph]}}.

\bibitem{Brehmer:2019xox}
J.~Brehmer, F.~Kling, I.~Espejo, and K.~Cranmer, {\it {MadMiner: Machine learning-based inference for particle physics}},  \href{http://dx.doi.org/10.1007/s41781-020-0035-2}{Comput. Softw. Big Sci. {\bfseries 4} (2020) 1, 3}, \href{http://arxiv.org/abs/1907.10621}{{arXiv:1907.10621 [hep-ph]}}.

\bibitem{Brehmer:2019gmn}
J.~Brehmer, S.~Dawson, S.~Homiller, F.~Kling, and T.~Plehn, {\it {Benchmarking simplified template cross sections in $WH$ production}},  \href{http://dx.doi.org/10.1007/JHEP11(2019)034}{JHEP {\bfseries 11} (2019)  034}, \href{http://arxiv.org/abs/1908.06980}{{arXiv:1908.06980 [hep-ph]}}.

\bibitem{Barman:2021yfh}
R.~K. Barman, D.~Gon\c{c}alves, and F.~Kling, {\it {Machine learning the Higgs boson-top quark CP phase}},  \href{http://dx.doi.org/10.1103/PhysRevD.105.035023}{Phys. Rev. D {\bfseries 105} (10, 2021)   035023}, \href{http://arxiv.org/abs/2110.07635}{{arXiv:2110.07635 [hep-ph]}}.

\bibitem{Bahl:2021dnc}
H.~Bahl and S.~Brass, {\it {Constraining CP-violation in the Higgs-top-quark interaction using machine-learning-based inference}},  \href{http://dx.doi.org/10.1007/JHEP03(2022)017}{JHEP {\bfseries 03} (10, 2021)   017}, \href{http://arxiv.org/abs/2110.10177}{{arXiv:2110.10177 [hep-ph]}}.

\bibitem{GomezAmbrosio:2022mpm}
R.~Gomez~Ambrosio, J.~ter Hoeve, M.~Madigan, J.~Rojo, and V.~Sanz, {\it {Unbinned multivariate observables for global SMEFT analyses from machine learning}},  \href{http://dx.doi.org/10.1007/JHEP03(2023)033}{JHEP {\bfseries 03} (2023)  033}, \href{http://arxiv.org/abs/2211.02058}{{arXiv:2211.02058 [hep-ph]}}.

\bibitem{Barrue:2023ysk}
R.~Barru{\'e}, P.~Conde-Mu{\'\i}{\~n}o, V.~Dao, and R.~Santos, {\it {Simulation-based inference in the search for CP violation in leptonic WH production}},  \href{http://dx.doi.org/10.1007/JHEP04(2024)014}{JHEP {\bfseries 04} (2024)  014}, \href{http://arxiv.org/abs/2308.02882}{{arXiv:2308.02882 [hep-ph]}}.

\bibitem{Mastandrea:2024irf}
R.~Mastandrea, B.~Nachman, and T.~Plehn, {\it {Constraining the Higgs potential with neural simulation-based inference for di-Higgs production}},  \href{http://dx.doi.org/10.1103/PhysRevD.110.056004}{Phys. Rev. D {\bfseries 110} (2024) 5, 056004}, \href{http://arxiv.org/abs/2405.15847}{{arXiv:2405.15847 [hep-ph]}}.

\bibitem{Ghosh:2025fma}
A.~Ghosh, M.~Griese, U.~Haisch, and T.~H. Park, {\it {Neural simulation-based inference of the Higgs trilinear self-coupling via off-shell Higgs production}},  \href{http://dx.doi.org/10.1140/epjc/s10052-026-15605-3}{Eur. Phys. J. C {\bfseries 86} (2026) 4, 415}, \href{http://arxiv.org/abs/2507.02032}{{arXiv:2507.02032 [hep-ph]}}.

\bibitem{Silva:2025hzo}
M.~Silva, R.~Barru{\'e}, I.~Ochoa, and P.~Conde~Mu{\'\i}{\~n}o, {\it {Searching for HWW anomalous couplings with simulation-based inference}},  \href{http://dx.doi.org/10.1103/29xq-b66r}{Phys. Rev. D {\bfseries 113} (2026) 5, 056011}, \href{http://arxiv.org/abs/2509.03307}{{arXiv:2509.03307 [hep-ph]}}.

\bibitem{Benato:2025rgo}
L.~Benato, C.~Giordano, C.~Krause, A.~Li, R.~Sch{\"o}fbeck, D.~Schwarz, M.~Shooshtari, and D.~Wang, {\it {Unbinned inclusive cross-section measurements with machine-learned systematic uncertainties}},  \href{http://dx.doi.org/10.1103/zwzt-1rrw}{Phys. Rev. D {\bfseries 112} (2025) 5, 052006}, \href{http://arxiv.org/abs/2505.05544}{{arXiv:2505.05544 [hep-ph]}}.

\bibitem{Bahl:2025mib}
H.~Bahl, T.~Plehn, and N.~Schmal, {\it {Unbinning global LHC analyses}},  \href{http://arxiv.org/abs/2509.05409}{{arXiv:2509.05409 [hep-ph]}}.

\bibitem{ATLAS:2024jry}
ATLAS, G.~Aad {\em et al.}, {\it {Measurement of off-shell Higgs boson production in the $H^*\rightarrow ZZ\rightarrow 4\ell$ decay channel using a neural simulation-based inference technique in 13{\,}TeV pp collisions with the ATLAS detector}},  \href{http://dx.doi.org/10.1088/1361-6633/adcd9a}{Rept. Prog. Phys. {\bfseries 88} (2025) 5, 057803}, \href{http://arxiv.org/abs/2412.01548}{{arXiv:2412.01548 [hep-ex]}}.

\bibitem{ATLAS:2025clx}
ATLAS, G.~Aad {\em et al.}, {\it {An implementation of neural simulation-based inference for parameter estimation in ATLAS}},  \href{http://dx.doi.org/10.1088/1361-6633/add370}{Rept. Prog. Phys. {\bfseries 88} (2025) 6, 067801}, \href{http://arxiv.org/abs/2412.01600}{{arXiv:2412.01600 [physics.data-an]}}.

\bibitem{Degrande:2012wf}
C.~Degrande, N.~Greiner, W.~Kilian, O.~Mattelaer, H.~Mebane, T.~Stelzer, S.~Willenbrock, and C.~Zhang, {\it {Effective Field Theory: A Modern Approach to Anomalous Couplings}},  \href{http://dx.doi.org/10.1016/j.aop.2013.04.016}{Annals Phys. {\bfseries 335} (2013)  21}, \href{http://arxiv.org/abs/1205.4231}{{arXiv:1205.4231 [hep-ph]}}.

\bibitem{Brivio:2017vri}
I.~Brivio and M.~Trott, {\it {The Standard Model as an Effective Field Theory}},  \href{http://dx.doi.org/10.1016/j.physrep.2018.11.002}{Phys. Rept. {\bfseries 793} (2019)  1}, \href{http://arxiv.org/abs/1706.08945}{{arXiv:1706.08945 [hep-ph]}}.

\bibitem{Chen:2023ind}
S.~Chen, A.~Glioti, G.~Panico, and A.~Wulzer, {\it {Boosting likelihood learning with event reweighting}},  \href{http://dx.doi.org/10.1007/JHEP03(2024)117}{JHEP {\bfseries 03} (2024)  117}, \href{http://arxiv.org/abs/2308.05704}{{arXiv:2308.05704 [hep-ph]}}.

\bibitem{D0:1998eiz}
D0, B.~Abbott {\em et al.}, {\it {Measurement of the Top Quark Mass in the Dilepton Channel}},  \href{http://dx.doi.org/10.1103/PhysRevD.60.052001}{Phys. Rev. D {\bfseries 60} (1999)  052001}, \href{http://arxiv.org/abs/hep-ex/9808029}{{arXiv:hep-ex/9808029}}.

\bibitem{CDF:2006nne}
CDF, A.~Abulencia {\em et al.}, {\it {Top quark mass measurement from dilepton events at CDF II with the matrix-element method}},  \href{http://dx.doi.org/10.1103/PhysRevD.74.032009}{Phys. Rev. D {\bfseries 74} (2006)  032009}, \href{http://arxiv.org/abs/hep-ex/0605118}{{arXiv:hep-ex/0605118}}.

\bibitem{Fiedler:2010sg}
F.~Fiedler, A.~Grohsjean, P.~Haefner, and P.~Schieferdecker, {\it {The Matrix Element Method and its Application in Measurements of the Top Quark Mass}},  \href{http://dx.doi.org/10.1016/j.nima.2010.09.024}{Nucl. Instrum. Meth. A {\bfseries 624} (2010)  203}, \href{http://arxiv.org/abs/1003.1316}{{arXiv:1003.1316 [hep-ex]}}.

\bibitem{Hollingsworth:2020kjg}
J.~Hollingsworth and D.~Whiteson, {\it {Resonance Searches with Machine Learned Likelihood Ratios}},
\href{http://arxiv.org/abs/2002.04699}{{arXiv:2002.04699 [hep-ph]}}.
%%CITATION = ARXIV:2002.04699;%%.

\bibitem{Cheng:2024yig}
C.~L. Cheng, G.~Singh, and B.~Nachman, {\it {Incorporating Physical Priors into Weakly Supervised Anomaly Detection}},  \href{http://dx.doi.org/10.1103/8259-wt5p}{Phys. Rev. Lett. {\bfseries 135} (2025) 2, 021801}, \href{http://arxiv.org/abs/2405.08889}{{arXiv:2405.08889 [hep-ph]}}.

\bibitem{Gutmann:2016bolfi}
M.~U. Gutmann and J.~Corander, {\it {Bayesian Optimization for Likelihood-Free Inference of Simulator-Based Statistical Models}},  J. Mach. Learn. Res. {\bfseries 17} (2016) 125, 1, \href{http://arxiv.org/abs/1501.03291}{{arXiv:1501.03291 [stat.ML]}}.

\bibitem{papamakarios2016fast}
G.~Papamakarios and I.~Murray, {\it {Fast $\varepsilon$-free Inference of Simulation Models with Bayesian Conditional Density Estimation}},  in {\em {Advances in Neural Information Processing Systems}}.
\newblock 2016.
\newblock \href{http://arxiv.org/abs/1605.06376}{{arXiv:1605.06376 [stat.ML]}}.

\bibitem{Lueckmann:2017snpe}
J.-M. Lueckmann, P.~J. Gon{\c{c}}alves, G.~Bassetto, K.~{\"O}cal, M.~Nonnenmacher, and J.~H. Macke, {\it {Flexible statistical inference for mechanistic models of neural dynamics}},  in {\em {Advances in Neural Information Processing Systems}}.
\newblock 2017.
\newblock \href{http://arxiv.org/abs/1711.01861}{{arXiv:1711.01861 [stat.ML]}}.

\bibitem{Papamakarios:2018zoy}
G.~Papamakarios, D.~C. Sterratt, and I.~Murray, {\it {Sequential Neural Likelihood: Fast Likelihood-free Inference with Autoregressive Flows}},  \href{http://arxiv.org/abs/1805.07226}{{arXiv:1805.07226 [stat.ML]}}.

\bibitem{Greenberg:2019apt}
D.~S. Greenberg, M.~Nonnenmacher, and J.~H. Macke, {\it {Automatic Posterior Transformation for Likelihood-Free Inference}},  in {\em {Proceedings of the 36th International Conference on Machine Learning}}.
\newblock 2019.
\newblock \href{http://arxiv.org/abs/1905.07488}{{arXiv:1905.07488 [cs.LG]}}.

\bibitem{Hermans:2019ioj}
J.~Hermans, V.~Begy, and G.~Louppe, {\it {Likelihood-free MCMC with Amortized Approximate Ratio Estimators}},  \href{http://arxiv.org/abs/1903.04057}{{arXiv:1903.04057 [stat.ML]}}.

\bibitem{Durkan:2020snre}
C.~Durkan, I.~Murray, and G.~Papamakarios, {\it {On Contrastive Learning for Likelihood-free Inference}},  in {\em {Proceedings of the 37th International Conference on Machine Learning}}.
\newblock 2020.
\newblock \href{http://arxiv.org/abs/2002.03712}{{arXiv:2002.03712 [stat.ML]}}.

\bibitem{Miller:2020hua}
B.~K. Miller, A.~Cole, G.~Louppe, and C.~Weniger, {\it {Simulation-efficient marginal posterior estimation with swyft: stop wasting your precious time}},  \href{http://arxiv.org/abs/2011.13951}{{arXiv:2011.13951 [astro-ph.IM]}}.

\bibitem{Miller:2021hys}
B.~K. Miller, A.~Cole, P.~Forr{\'e}, G.~Louppe, and C.~Weniger, \href{http://dx.doi.org/10.5281/zenodo.5043706}{{\it {Truncated Marginal Neural Ratio Estimation}}, } in {\em {35th Conference on Neural Information Processing Systems}}.
\newblock 7, 2021.
\newblock \href{http://arxiv.org/abs/2107.01214}{{arXiv:2107.01214 [stat.ML]}}.

\bibitem{Deistler:2022tsnpe}
M.~Deistler, P.~J. Gon{\c{c}}alves, and J.~H. Macke, {\it {Truncated proposals for scalable and hassle-free simulation-based inference}},  in {\em {Advances in Neural Information Processing Systems}}.
\newblock 2022.
\newblock \href{http://arxiv.org/abs/2210.04815}{{arXiv:2210.04815 [stat.ML]}}.

\bibitem{Lyu:2025vqk}
H.~Lyu, J.~Alvey, N.~Anau~Montel, M.~Pieroni, and C.~Weniger, {\it {Dynamic SBI: Round-free Sequential Simulation-Based Inference with Adaptive Datasets}},  \href{http://arxiv.org/abs/2510.13997}{{arXiv:2510.13997 [astro-ph.IM]}}.

\bibitem{Lueckmann:2021sbibm}
J.-M. Lueckmann, J.~Boelts, D.~S. Greenberg, P.~J. Gon{\c{c}}alves, and J.~H. Macke, {\it {Benchmarking Simulation-Based Inference}},  in {\em {Proceedings of the 24th International Conference on Artificial Intelligence and Statistics}}.
\newblock 2021.
\newblock \href{http://arxiv.org/abs/2101.04653}{{arXiv:2101.04653 [stat.ML]}}.

\bibitem{Cole:2021gwr}
A.~Cole, B.~K. Miller, S.~J. Witte, M.~X. Cai, M.~W. Grootes, F.~Nattino, and C.~Weniger, {\it {Fast and credible likelihood-free cosmology with truncated marginal neural ratio estimation}},  \href{http://dx.doi.org/10.1088/1475-7516/2022/09/004}{JCAP {\bfseries 09} (2022)  004}, \href{http://arxiv.org/abs/2111.08030}{{arXiv:2111.08030 [astro-ph.CO]}}.

\bibitem{Montel:2022fhv}
N.~A. Montel, A.~Coogan, C.~Correa, K.~Karchev, and C.~Weniger, {\it {Estimating the warm dark matter mass from strong lensing images with truncated marginal neural ratio estimation}},  \href{http://dx.doi.org/10.1093/mnras/stac3215}{Mon. Not. Roy. Astron. Soc. {\bfseries 518} (2022) 2, 2746}, \href{http://arxiv.org/abs/2205.09126}{{arXiv:2205.09126 [astro-ph.CO]}}.

\bibitem{Bhardwaj:2023xph}
U.~Bhardwaj, J.~Alvey, B.~K. Miller, S.~Nissanke, and C.~Weniger, {\it {Sequential simulation-based inference for gravitational wave signals}},  \href{http://dx.doi.org/10.1103/PhysRevD.108.042004}{Phys. Rev. D {\bfseries 108} (2023) 4, 042004}, \href{http://arxiv.org/abs/2304.02035}{{arXiv:2304.02035 [gr-qc]}}.

\bibitem{Amram:2025vqw}
O.~Amram and M.~Szewc, {\it {Data-Driven High-Dimensional Statistical Inference with Generative Models}},  \href{http://arxiv.org/abs/2506.06438}{{arXiv:2506.06438 [hep-ph]}}.

\bibitem{Cheng:2025ewj}
C.~L. Cheng, R.~Das, R.~Li, R.~Mastandrea, V.~Mikuni, B.~Nachman, D.~Shih, and G.~Singh, {\it {Generator Based Inference (GBI)}},  \href{http://arxiv.org/abs/2506.00119}{{arXiv:2506.00119 [hep-ph]}}.

\bibitem{Caron:2019xkx}
S.~Caron, T.~Heskes, S.~Otten, and B.~Stienen, {\it {Constraining the Parameters of High-Dimensional Models with Active Learning}},  \href{http://dx.doi.org/10.1140/epjc/s10052-019-7437-5}{Eur. Phys. J. {\bfseries C79} (2019) 11, 944},
\href{http://arxiv.org/abs/1905.08628}{{arXiv:1905.08628 [cs.LG]}}.
%%CITATION = ARXIV:1905.08628;%%.

\bibitem{deSa:2026hch}
R.~C.~L. de~Sa and J.~Sandesara, {\it {Finite Asimov Sample Construction in Unbinned Neural Simulation-Based Inference}},  \href{http://arxiv.org/abs/2609.14136}{{arXiv:2609.14136 [physics.data-an]}}.

\bibitem{deSa:2026awa}
R.~C.~L. de~Sa and J.~Sandesara, {\it {Simulation-Based Inference and Unbinned Asimov Construction with Hybrid Neural Density Estimation}},  \href{http://arxiv.org/abs/2609.30196}{{arXiv:2609.30196 [physics.data-an]}}.

\bibitem{Wilks:1938dza}
S.~S. Wilks, {\it {The Large-Sample Distribution of the Likelihood Ratio for Testing Composite Hypotheses}},  \href{http://dx.doi.org/10.1214/aoms/1177732360}{Annals Math. Statist. {\bfseries 9} (1938) 1, 60}.

\bibitem{Cowan:2010js}
G.~Cowan, K.~Cranmer, E.~Gross, and O.~Vitells, {\it {Asymptotic formulae for likelihood-based tests of new physics}},  \href{http://dx.doi.org/10.1140/epjc/s10052-011-1554-0}{Eur. Phys. J. C {\bfseries 71} (2011)  1554}, \href{http://arxiv.org/abs/1007.1727}{{arXiv:1007.1727 [physics.data-an]}}. [Erratum: Eur.Phys.J.C 73, 2501 (2013)].

\bibitem{Alwall:2014hca}
J.~Alwall, R.~Frederix, S.~Frixione, V.~Hirschi, F.~Maltoni, O.~Mattelaer, H.~S. Shao, T.~Stelzer, P.~Torrielli, and M.~Zaro, {\it {The automated computation of tree-level and next-to-leading order differential cross sections, and their matching to parton shower simulations}},  \href{http://dx.doi.org/10.1007/JHEP07(2014)079}{JHEP {\bfseries 07} (2014)  079}, \href{http://arxiv.org/abs/1405.0301}{{arXiv:1405.0301 [hep-ph]}}.

\bibitem{Ball:2012cx}
R.~D. Ball {\em et al.}, {\it {Parton distributions with LHC data}},  \href{http://dx.doi.org/10.1016/j.nuclphysb.2012.10.003}{Nucl. Phys. B {\bfseries 867} (2013)  244}, \href{http://arxiv.org/abs/1207.1303}{{arXiv:1207.1303 [hep-ph]}}.

\bibitem{Degrande:2011ua}
C.~Degrande, C.~Duhr, B.~Fuks, D.~Grellscheid, O.~Mattelaer, and T.~Reiter, {\it {UFO - The Universal FeynRules Output}},  \href{http://dx.doi.org/10.1016/j.cpc.2012.01.022}{Comput. Phys. Commun. {\bfseries 183} (2012)  1201}, \href{http://arxiv.org/abs/1108.2040}{{arXiv:1108.2040 [hep-ph]}}.

\bibitem{Alloul:2013bka}
A.~Alloul, N.~D. Christensen, C.~Degrande, C.~Duhr, and B.~Fuks, {\it {FeynRules 2.0 - A complete toolbox for tree-level phenomenology}},  \href{http://dx.doi.org/10.1016/j.cpc.2014.04.012}{Comput. Phys. Commun. {\bfseries 185} (2014)  2250}, \href{http://arxiv.org/abs/1310.1921}{{arXiv:1310.1921 [hep-ph]}}.

\bibitem{SOBOL}
I.~M. Sobol', {\it On the distribution of points in a cube and the approximate evaluation of integrals},  \href{http://dx.doi.org/10.1016/0041-5553(67)90144-9}{USSR Computational Mathematics and Mathematical Physics {\bfseries 7} (1967) 4, 86}.

\bibitem{Branco:2011iw}
G.~C. Branco, P.~M. Ferreira, L.~Lavoura, M.~N. Rebelo, M.~Sher, and J.~P. Silva, {\it {Theory and phenomenology of two-Higgs-doublet models}},  \href{http://dx.doi.org/10.1016/j.physrep.2012.02.002}{Phys. Rept. {\bfseries 516} (2012)  1}, \href{http://arxiv.org/abs/1106.0034}{{arXiv:1106.0034 [hep-ph]}}.

\bibitem{Biekotter:2023eil}
T.~Biek\"otter, S.~Heinemeyer, J.~M. No, K.~Radchenko, M.~O. Olea~Romacho, and G.~Weiglein, {\it {First shot of the smoking gun: probing the electroweak phase transition in the 2HDM with novel searches for $A\to ZH$ in $\ell^+\ell^- t\bar t$ and $\nu\nu b\bar b$ final states}},  \href{http://dx.doi.org/10.1007/JHEP01(2024)107}{JHEP {\bfseries 01} (2024)  107}, \href{http://arxiv.org/abs/2309.17431}{{arXiv:2309.17431 [hep-ph]}}.

\bibitem{ATLAS:2020gxx}
ATLAS, G.~Aad {\em et al.}, {\it {Search for a heavy Higgs boson decaying into a Z boson and another heavy Higgs boson in the $\ell \ell bb$ and $\ell \ell WW$ final states in $pp$ collisions at $\sqrt{s}=13$ $\text {TeV}$ with the ATLAS detector}},  \href{http://dx.doi.org/10.1140/epjc/s10052-021-09117-5}{Eur. Phys. J. C {\bfseries 81} (2021) 5, 396}, \href{http://arxiv.org/abs/2011.05639}{{arXiv:2011.05639 [hep-ex]}}.

\bibitem{CMS:2024yiy}
CMS, A.~Hayrapetyan {\em et al.}, {\it {Search for heavy neutral Higgs bosons A and H in the $\mathrm{t\bar{t}}$Z channel in proton-proton collisions at 13 TeV}},  \href{http://dx.doi.org/10.1016/j.physletb.2025.139568}{Phys. Lett. B {\bfseries 866} (2025)  139568}, \href{http://arxiv.org/abs/2412.00570}{{arXiv:2412.00570 [hep-ex]}}.

\bibitem{Butterworth:2015oua}
J.~Butterworth {\em et al.}, {\it {PDF4LHC recommendations for LHC Run II}},  \href{http://dx.doi.org/10.1088/0954-3899/43/2/023001}{J. Phys. G {\bfseries 43} (2016)  023001}, \href{http://arxiv.org/abs/1510.03865}{{arXiv:1510.03865 [hep-ph]}}.

\bibitem{Buckley:2014ana}
A.~Buckley, J.~Ferrando, S.~Lloyd, K.~Nordstr\"om, B.~Page, M.~R\"ufenacht, M.~Sch\"onherr, and G.~Watt, {\it {LHAPDF6: parton density access in the LHC precision era}},  \href{http://dx.doi.org/10.1140/epjc/s10052-015-3318-8}{Eur. Phys. J. C {\bfseries 75} (2015)  132}, \href{http://arxiv.org/abs/1412.7420}{{arXiv:1412.7420 [hep-ph]}}.

\bibitem{Sjostrand:2014zea}
T.~Sj\"ostrand, S.~Ask, J.~R. Christiansen, R.~Corke, N.~Desai, P.~Ilten, S.~Mrenna, S.~Prestel, C.~O. Rasmussen, and P.~Z. Skands, {\it {An introduction to PYTHIA 8.2}},  \href{http://dx.doi.org/10.1016/j.cpc.2015.01.024}{Comput. Phys. Commun. {\bfseries 191} (2015)  159}, \href{http://arxiv.org/abs/1410.3012}{{arXiv:1410.3012 [hep-ph]}}.

\bibitem{deFavereau:2013fsa}
DELPHES 3, J.~de~Favereau, C.~Delaere, P.~Demin, A.~Giammanco, V.~Lemaître, A.~Mertens, and M.~Selvaggi, {\it {DELPHES 3, A modular framework for fast simulation of a generic collider experiment}},  \href{http://dx.doi.org/10.1007/JHEP02(2014)057}{JHEP {\bfseries 02} (2014)  057},
\href{http://arxiv.org/abs/1307.6346}{{arXiv:1307.6346 [hep-ex]}}.
%%CITATION = ARXIV:1307.6346;%%.

\bibitem{Cacciari:2008gp}
M.~Cacciari, G.~P. Salam, and G.~Soyez, {\it {The anti-$k_t$ jet clustering algorithm}},  \href{http://dx.doi.org/10.1088/1126-6708/2008/04/063}{JHEP {\bfseries 04} (2008)  063}, \href{http://arxiv.org/abs/0802.1189}{{arXiv:0802.1189 [hep-ph]}}.

\bibitem{Cacciari:2011ma}
M.~Cacciari, G.~P. Salam, and G.~Soyez, {\it {FastJet User Manual}},  \href{http://dx.doi.org/10.1140/epjc/s10052-012-1896-2}{Eur. Phys. J. C {\bfseries 72} (2012)  1896}, \href{http://arxiv.org/abs/1111.6097}{{arXiv:1111.6097 [hep-ph]}}.

\bibitem{FebresCordero:2009xzo}
F.~Febres~Cordero, L.~Reina, and D.~Wackeroth, {\it {W- and Z-boson production with a massive bottom-quark pair at the Large Hadron Collider}},  \href{http://dx.doi.org/10.1103/PhysRevD.80.034015}{Phys. Rev. D {\bfseries 80} (2009)  034015}, \href{http://arxiv.org/abs/0906.1923}{{arXiv:0906.1923 [hep-ph]}}.

\bibitem{Spinner:2024hjm}
J.~Spinner, V.~Bres{\'o}, P.~de~Haan, T.~Plehn, J.~Thaler, and J.~Brehmer, {\it {Lorentz-Equivariant Geometric Algebra Transformers for High-Energy Physics}},  \href{http://arxiv.org/abs/2405.14806}{{arXiv:2405.14806 [physics.data-an]}}.

\bibitem{Brehmer:2024yqw}
J.~Brehmer, V.~Bres{\'o}, P.~de~Haan, T.~Plehn, H.~Qu, J.~Spinner, and J.~Thaler, {\it {A Lorentz-Equivariant Transformer for All of the LHC}},  \href{http://arxiv.org/abs/2411.00446}{{arXiv:2411.00446 [hep-ph]}}.

\bibitem{Brehmer:2023gatr}
J.~Brehmer, P.~de~Haan, S.~Behrends, and T.~Cohen, {\it {Geometric Algebra Transformer}},  in {\em {Advances in Neural Information Processing Systems 36 (NeurIPS 2023)}}.
\newblock 2023.
\newblock \href{http://arxiv.org/abs/2305.18415}{{arXiv:2305.18415 [cs.LG]}}.

\bibitem{Petitjean:2025zjf}
A.~Petitjean, T.~Plehn, J.~Spinner, and U.~K{\"o}the, {\it {Economical Jet Taggers -- Equivariant, Slim, and Quantized}},  \href{http://arxiv.org/abs/2512.17011}{{arXiv:2512.17011 [hep-ph]}}.

\bibitem{Randall:1999ee}
L.~Randall and R.~Sundrum, {\it {A Large mass hierarchy from a small extra dimension}},  \href{http://dx.doi.org/10.1103/PhysRevLett.83.3370}{Phys. Rev. Lett. {\bfseries 83} (1999)  3370}, \href{http://arxiv.org/abs/hep-ph/9905221}{{arXiv:hep-ph/9905221}}.

\bibitem{Agashe:2007zd}
K.~Agashe, H.~Davoudiasl, G.~Perez, and A.~Soni, {\it {Warped Gravitons at the LHC and Beyond}},  \href{http://dx.doi.org/10.1103/PhysRevD.76.036006}{Phys. Rev. D {\bfseries 76} (2007)  036006}, \href{http://arxiv.org/abs/hep-ph/0701186}{{arXiv:hep-ph/0701186}}.

\bibitem{Das:2026spe}
R.~Das, M.~Hein, G.~Kasieczka, M.~Kr{\"a}mer, L.~Lang, R.~Mastandrea, L.~Moureaux, A.~M{\"u}ck, and D.~Shih, {\it {Kitchen Sink Anomaly Detection}},  \href{http://arxiv.org/abs/2604.20965}{{arXiv:2604.20965 [hep-ph]}}.

\bibitem{CMS:2024nsz}
CMS, V.~Chekhovsky {\em et al.}, {\it {Model-agnostic search for dijet resonances with anomalous jet substructure in proton{\textendash}proton collisions at $\sqrt{s}$ = 13 TeV}},  \href{http://dx.doi.org/10.1088/1361-6633/add762}{Rept. Prog. Phys. {\bfseries 88} (2025) 6, 067802}, \href{http://arxiv.org/abs/2412.03747}{{arXiv:2412.03747 [hep-ex]}}.

\bibitem{Kasieczka:2021xcg}
G.~Kasieczka {\em et al.}, {\it {The LHC Olympics 2020: A Community Challenge for Anomaly Detection in High Energy Physics}},  \href{http://arxiv.org/abs/2101.08320}{{arXiv:2101.08320 [hep-ph]}}.

\bibitem{Foreman-Mackey:2012any}
D.~Foreman-Mackey, D.~W. Hogg, D.~Lang, and J.~Goodman, {\it {emcee: The MCMC Hammer}},  \href{http://dx.doi.org/10.1086/670067}{Publ. Astron. Soc. Pac. {\bfseries 125} (2013)  306}, \href{http://arxiv.org/abs/1202.3665}{{arXiv:1202.3665 [astro-ph.IM]}}.

\bibitem{Baak:2014fta}
M.~Baak, S.~Gadatsch, R.~Harrington, and W.~Verkerke, {\it {Interpolation between multi-dimensional histograms using a new non-linear moment morphing method}},  \href{http://dx.doi.org/10.1016/j.nima.2014.10.033}{Nucl. Instrum. Meth. A {\bfseries 771} (2015)  39}, \href{http://arxiv.org/abs/1410.7388}{{arXiv:1410.7388 [physics.data-an]}}.

\bibitem{Neal:2003slice}
R.~M. Neal, {\it {Slice sampling}},  \href{http://dx.doi.org/10.1214/aos/1056562461}{Ann. Statist. {\bfseries 31} (2003) 3, 705}.

\bibitem{Virtanen:2019joe}
P.~Virtanen {\em et al.}, {\it {SciPy 1.0: Fundamental Algorithms for Scientific Computing in Python}},  \href{http://dx.doi.org/10.1038/s41592-019-0686-2}{Nature Meth. {\bfseries 17} (2020)  261}, \href{http://arxiv.org/abs/1907.10121}{{arXiv:1907.10121 [cs.MS]}}.

\end{thebibliography}\endgroup
